\documentclass{article}
\usepackage{geometry}
\usepackage{amsthm}
\usepackage{graphicx}
\usepackage[dvipsnames]{xcolor}
\usepackage{amsmath,mathtools,booktabs}
\usepackage{amssymb}

 \usepackage{multirow}
 \usepackage{geometry}
 \usepackage{caption} 
\usepackage{subcaption} 
\usepackage{nicefrac} 
\usepackage{colortbl}%
\usepackage{mathrsfs}%

\newcommand{\Vertices}{V}
\newcommand{\numag}{n}

\newcommand{\Weights}{W}

\newcommand{\Neig}{\mathcal{N}}

\newcommand{\FancyF}{\mathcal{F}}

\newcommand{\SetN}{\mathcal{W}}
\newcommand{\SetA}{\mathcal{A}}

\newcommand{\numques}{Q}
\newcommand{\newj}{l}
 
\newcommand{\Cost}{J}
\newcommand{\XSet}{\mathcal{R}}
\newcommand{\innertraits}{\psi}

\newcommand{\SetAA}{\mathscr{A}}
\newcommand{\SetNN}{\mathscr{N}}

\DeclareSymbolFont{symbolsC}{U}{pxsyc}{m}{n}
\DeclareMathSymbol{\coloneqq}{\mathrel}{symbolsC}{"42}

\DeclareMathOperator*{\argmin}{arg\,min}

\newcommand{\CAb}[1]{{\color{black} #1}}
\newcommand{\CAc}[1]{{\color{black} #1}}
\newcommand{\CAd}[1]{{\color{black} #1}}
\newcommand{\CAe}[1]{{\color{black} #1}}
\newcommand{\CAf}[1]{{\color{black} #1}}
\newcommand{\CAg}[1]{{\color{black} #1}}
\newcommand{\CAh}[1]{{\color{black} #1}}
\newcommand{\CA}[1]{{\color{black} #1}}

\newcommand{\CAbnew}[1]{{\color{black} #1}}
\newcommand{\CAnewcomment}[1]{{\color{black} #1}}

\title{Classification-Based Opinion Formation Model Embedding Agents' Psychological Traits}

\author{Carlos Andres Devia, Giulia Giordano}
\date{October 2022}

\begin{document}

\maketitle

\begin{abstract}
We propose an agent-based opinion formation model characterised by a two-fold novelty. First, we realistically assume that each agent cannot measure the opinion of its neighbours with infinite resolution and accuracy, and hence it can only classify the opinion of others as agreeing \emph{much more}, or \emph{more}, or \emph{comparably}, or \emph{less}, or \emph{much less} (than itself) with a given statement. This leads to a classification-based rule for opinion update. Second, we consider three complementary agent traits suggested by significant sociological and psychological research: \emph{conformism}, \emph{radicalism} and \emph{stubbornness}. We rely on World Values Survey data to \CAbnew{show that the proposed model has the potential} to predict the evolution of opinions in real life: the classification-based approach and complementary agent traits produce rich collective behaviours, such as polarisation, consensus, and clustering, which \CAbnew{can yield predicted opinions similar to survey results.}
\end{abstract}

\section{Introduction}

%
%

The development and analysis of opinion formation models has been an active field of research since the introduction of the first opinion formation models \cite{French1956,Harary1959,Harary1965,DeGroot1974}. Increasingly more sophisticated models have been developed by embedding different concepts such as \emph{susceptibility} \cite{Friedkin1986,Friedkin1999}, \emph{stubborness} \cite{Hegselmann2015,Masuda2015}, \emph{leaders} \cite{Kacperski1999,Kacperski2000}, \emph{emotions} \cite{Sobkowicz2010a,Chmiel2011}, \emph{trust} \cite{Yin2019,Krawczyk2010}, \emph{bounded confidence} \cite{Hegselmann2002}, \emph{coevolving networks} \cite{Su2014,Sobkowicz2009a}, \emph{biases} \cite{Sobkowicz2018},  \CAbnew{\emph{polarity} \cite{Lorenz2021Individual}, \emph{assimilation} \cite{Dandekar2013Biased,Fu2016,Lorenz2021Individual, Banisch2021Biased}, \emph{tolerance} \cite{Duggins2017A}, \emph{mass media} \cite{Chattoe2014Using}, \emph{controversy} \cite{Baumann2020Modeling}},   \CAbnew{\emph{weighted balance theory} \cite{Schweighofer2020Weighted}}, among others. Although there may be different reasons to construct mathematical models of opinion formation \cite{Epstein2008}, the ultimate goal is typically to capture the mechanisms behind opinion change in society and accurately predict the evolution of real-life opinions \cite{Thompson2009,Troitzsch2009}.
%
%

Agent-based models (ABMs), such as the French-DeGroot model \cite{DeGroot1974}, are very common in the opinion formation literature. In an ABM, every individual holds a different opinion (or vector of opinions) and interacts with the other agents according to a given function over a network that can be directed, weighted, or signed. Some notable examples of agent-based models are those by \cite{Hegselmann2006}, \cite{Salzarulo2006}, and \cite{Deffuant2002}, among many others \cite{Urbig2008,Afshar2010,Deffuant2006,Mckeown2006,Urbig2003}. An extensive literature \cite{Mastroeni2019} proposes and analyses opinion formation models for different types of agent interactions and network characteristics.

%
%

This paper proposes an agent-based model characterised by two novel features.
\begin{enumerate}
\item Even if the agents communicate, and openly express their real opinion, it is impossible for an agent to exactly measure and quantify the opinion of another. Therefore, the model introduces a classification-based approach,  \CAbnew{supported by the empirical finding that the assessment of the opinion of others depends on the perceived distance to those others \cite{Schweighofer2020Weighted}:}
each agent classifies its neighbours in different groups according to their \emph{perceived} opinion, distinguishing between those that agree \emph{much more}, or \emph{more}, or \emph{comparably}, or \emph{less}, or \emph{much less} (than itself) with a given statement. 

The fact that agents don't have access to the exact opinion of their neighbours with infinite resolution and accuracy has been taken into account by models with quantised opinions  \cite{Guo2013,Francesca2018}, while  threshold models \cite{Granovetter1978ThresholdMO,GRANOVETTER198683} could be seen as adopting a classification approach because  the opinion update law depends on the number of neighbours expressing a particular opinion or action. Our classification-based approach is based not on the opinion of an agent's neighbours, but on the \CAbnew{weighted} difference between the opinion of the agent and of its neighbours, accounting for the finite resolution with which agents perceive the opinions of their neighbours.
\CAb{Also in opinion formation models with private and public opinions   \CAbnew{\cite{YE2019371,Anderson2019129,Shang2021318,Duggins2017A, Banisch2019Opinion}} the agents cannot have perfect access to the real opinion of their neighbours. However, there is a critical difference. In these models, the agents can choose which public opinion they show, with certainty that it will be the opinion perceived by others, and hide their true private opinion: the misperception is intentional. Conversely, in our model, the misperception is unintentional and unavoidably caused by the inability to communicate with infinite accuracy: the agents wish to show as openly as possible their opinion, which still cannot be perceived with infinite resolution, and the other agents can only perceive the range in which the opinion falls, which depends on both the agent that expresses the opinion and the one that assesses it. Therefore an agent cannot know with certainty how its opinion is perceived by others.}
 \CAbnew{Also in the Continuous Opinions and Discrete Actions model \cite{Martins2008Continuous}, the mismatch between real and perceived opinions is intentional and due to the agents purposefully hiding their actual opinion to others (each agent controls the action it takes and consequently how its opinion is perceived by its neighbours), while in the classification-based model the mismatch is due to the intrinsically imperfect perception mechanism.}

\CAbnew{To reflect imperfect communication in the model, our proposed solution of classifying the opinion of others in one of five categories is inspired by the field of psychometrics: in questionnaires, the responses quantify opinions according to discrete scales.} Likert scales are a standard psychometric scale used to analyse surveys, which in turn are the typical approach to measure the opinions of individuals in a population. In our model, the process of agent $i$ assessing the opinion of agent $j$ yields, at each time step, the answer to a five-point Likert question, which asks how much agent $j$ agrees with a statement, compared with agent $i$,
where the possible answers are: \emph{Agrees much more}, \emph{Agrees more}, \emph{Agrees the same}, \emph{Agrees less}, and \emph{Agrees much less}. \CAbnew{For certain specific questions and specific social groups and connections, the perception may be sharper, while in other cases it may be less sharp; also, some agents may have a sharper perception than others. Five levels are chosen as a compromise resolution to account for the perception skills of the \emph{average agent} interacting with an \emph{average neighbour}. Still, the model could be modified to consider more than five levels, thus accounting for agents with a sharper average perception, and differentiating the sharpness of perception for different agents could also be interesting; however, this goes beyond the scope of the manuscript and is left for future work.}

\item Each agent behaves according to a combination of three \emph{internal traits} based on well studied sociological and psychological concepts: conformisms, radicalism, and stubbornness.
\begin{itemize}
\item Conformism: agents tend to agree with their neighbours. This behaviour was first shown in the conformity experiment by \cite{Asch1961,Asch1955,Asch1956} and evolved into social conformity theory \cite{Larsen1974}. A similar behaviour is supported by the cognitive dissonance theory \cite{festinger1957theory,matz2005cognitive}.
\item Radicalism: agents do not care if their opinion is different from their neighbours'. On the contrary, their opinion is strengthened by the presence of agents with a similar opinion, which reinforce their beliefs; this is known as the persuasive argument theory, which supports several polarisation models \cite{Mas2013,LaRocca2014,Liu2015,Fu2016,Pinasco2017}.
\item Stubbornness: agents refuse to change their opinion; this type of behaviour has been often present in opinion formation models starting from those by \cite{Friedkin1999,Friedkin2011}.
\end{itemize}
In the model, the behaviour of each agent is determined by a \emph{combination} of these three traits: in fact, actual people are not completely conformist, radical, or stubborn, but everyone is characterised by a peculiar blend of these three traits.
The inclusion of the radical trait can be seen as an extension of the model by \cite{Friedkin1999,Friedkin2011}, which includes both conformist and stubborn traits. 
\end{enumerate} 
%
%

 \CAbnew{The proposed model evolves over an invariant, directed, signed and unweighted network. Signed edges are interpreted as in structural balance theory: an edge from agent $j$ to agent $i$ is positive if agent $i$ approves, trusts, or follows agent $j$, whereas it is negative if agent $i$ disapproves, distrusts, or antagonises agent $j$ \cite{Altafini2013Consensus,Xia2016Structural,Cartwright1956Structural}.}

Despite significant research efforts in the development and analysis of opinion formation models, empirical validation is often lacking,  \CAbnew{and has been identified as one of the frontiers of opinion modelling \cite{Flache2017Models}}. In most cases, just an analytical or numerical characterisation of possible opinion evolutions is provided and, with some exceptions (most notably the model by \cite{Friedkin1999,Friedkin2011}), there are no systematic comparisons with real world behaviours.

\CAbnew{The problem of identifying individual-level parameters (in our case, agent inner traits) from population-level data (in our case, survey results) is known as the \emph{inverse problem} \cite{Kandler2018Generative} and arises, in the context of opinion dynamics, for any agent-based model, also when estimating agent interactions \cite{Lu2021Learning} and underlying networks \cite{Hassanibesheli2019Network} from data.} 
 \CAbnew{An approach to solve the inverse problem using survey results relies on evolutionary algorithms  \cite{Duggins2017A}; other papers taking into account survey results or empirical data in the study of opinion formation models include those by \cite{Banisch2021Biased,Chattoe2014Using,Baumann2020Modeling,Martins2008Continuous}.}

Here, we assess \CAbnew{the potential} of our model to predict opinion evolution in real-life settings using data from the World Values Survey \cite{WVS5,WVS6}, a global research project that studies people's values and beliefs over time, conducting surveys every five years. The results of these surveys are classified by `waves'. We use the results from wave 5 (year 2010) and wave 6 (year 2015). The answers of wave 5 are used as initial opinions that are evolved, according to the model dynamics, so as to produce \emph{predicted} opinions that are compared with the survey results from wave 6.
 \CAbnew{Our main purpose is to present a new opinion formation model; through the comparison with real data, we identify parameter choices showing that the model has the \emph{potential} to accurately predict real opinions starting from a variety of different initial opinions, but this does not fully or univocally solve the inverse problem \cite{Kandler2018Generative}.}

The paper is structured as follows. \CA{First, it introduces the model and its key parameters. Then, five types of simulation results are presented: simulations with simple parameters and digraphs, to gain intuition on the model behaviour; parameter sensitivity analysis, to explore the effect of different parameters on the opinion evolution; model validation with real data, to assess the predictive potential of the model by choosing the parameters through either a free or a constrained optimisation problem; a comparison with the Friedkin-Johnsen model \cite{Friedkin1999,Friedkin2011}; and qualitative model outcomes, illustrating the rich and varied opinion distributions that the model can yield.}

\section{The Classification-Based Model}\label{Sec:Model}
In our proposed classification-based (CB) model, the set $\Vertices=\{1, 2, \dots, \numag\}$ indexes the agents. The \emph{opinion} of agent $i\in\Vertices$  at time $k$, representing its level of agreement with a statement, is denoted by $x_i[k]\in[-1,1]$. The opinions $x_i=1$, $x_i=0$, and $x_i=-1$ represent complete agreement, indifference, and complete disagreement respectively. The vector of all opinions at time $k$ is denoted by $x[k]$. 

The agent opinions evolve in time due to opinion exchanges occurring over a signed digraph, represented by the matrix $\Weights\in\{-1,0,1\}^{\numag\times\numag}$, \CAbnew{whose entries are constant and, in particular, not opinion-dependent.} 
The self-confidence of each agent is expressed by $w_{ii}=1$ for all $i$.
The coefficient $w_{ij}$ represents the influence of agent $j$ over agent $i$. 
If $w_{ij}=0$, then agent $i$ is not influenced by agent $j$. If $w_{ij}\neq0$, then agent $j$ is a neighbour of agent $i$:
 \CAbnew{$w_{ij}=1$ means that agent $i$ approves, trusts, or follows agent $j$, while $w_{ij}=-1$ means that agent $i$ disapproves, mistrusts, or antagonises agent $j$. Signed edges have been interpreted in the opinion formation literature in terms of either cooperative/antagonistic interactions \cite{Altafini2013Consensus}, trust/mistrust \cite{Xia2016Structural}, or approval/disapproval \cite{Cartwright1956Structural}.
In our model, if $w_{ij}=1$ (respectively $w_{ij}=-1$), then agent $i$ perceives the opinion of agent $j$ as $x_j$ (resp. $-x_j$).} The set of neighbours of agent $i\in\Vertices$ is
\begin{equation}
\Neig_i = \Big\{ j\in\Vertices \mid w_{ij}\neq0 \Big\}.
\end{equation} 

The agent opinions evolve in discrete time and the opinion update relies on the assumption that agents cannot determine their neighbours' opinions precisely. Instead, each agent can classify its neighbours according to how close their \emph{perceived} opinion is to its own opinion. For instance, if agent $j$ influences agent $i$, and $x_i = 0.61$ and $x_j = 0.34$, then it is unrealistic to expect agent $i$ to know \emph{exactly} the opinion of agent $j$, or to assume that agent $i$ knows that the opinion difference is \emph{exactly} $0.27$. However, agent $i$ can perceive that agent $j$ agrees less than itself. On the contrary, if $x_j = 0.89$, agent $i$ can perceive that agent $j$ agrees more than itself. 

\begin{figure}[!h]
\centering
\includegraphics[width=\textwidth]{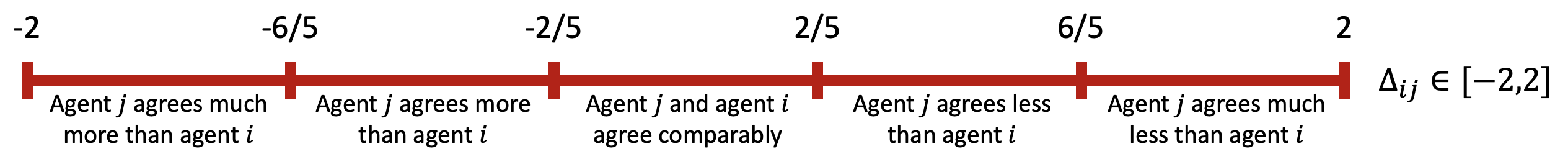}
\caption{Partition of the interval $[-2,2]$ in five equal subintervals. Depending on the interval to which the \CAbnew{weighted} opinion difference
$\Delta_{ij} = x_i - w_{ij} x_j$
belongs, agent $i$ will \emph{perceive} that agent $j$ agrees either: \emph{much more}; or \emph{more}; or \emph{comparably}; or \emph{less}; or \emph{much less}.}
\label{Fig:Cla}
\end{figure}

Therefore, agent $i$ can at most classify agent $j$ according to an estimation of $\Delta_{ij}$, which is the weighted difference between its opinion $x_i$ and the opinion of agent $j$, $x_j$: $\Delta_{ij}=x_i- w_{ij} x_j\in[-2,2]$. Let us divide the interval $[-2,2]$ in five equal subintervals. Then, depending on the subinterval to which $\Delta_{ij}$ belongs, agent $i$ can \emph{perceive} that agent $j$:  (1) agrees much more, (2) agrees more, (3) agrees comparably, (4) agrees less, or (5) agrees much less with the statement; see Figure \ref{Fig:Cla}. If $w_{ij}=-1$, then agent $i$ \CAbnew{disapproves/mistrusts/antagonises} agent $j$, therefore the \CAbnew{weighted} opinion difference is $\Delta_{ij}=x_i-(-x_j)\in[-2,2]$. If $w_{ij}=1$, then agent $i$ \CAbnew{approves/trusts/follows} agent $j$ and the \CAbnew{weighted} opinion difference is $\Delta_{ij}=x_i-x_j \in[-2,2]$.  \CAbnew{The combined effect of signed edges and neighbour classification leads to a three-step process: first, agent $i$ perceives the opinions of its neighbours; then, the opinions of neighbours that agent $i$ disapproves, mistrusts, or antagonises have the sign reversed; finally, the neighbours are classified according to the adjusted perceived opinion distance.}

The set $\Neig_i$ of all the neighbours of agent $i$ is thus partitioned into five time-dependent subsets: $D^+_i[k]$, $D_i[k]$, $N_i[k]$, $A_i[k]$, and $A^+_i[k]$, which contain the neighbours that agree much less, less, comparably, more, and much more, respectively. Mathematically these subsets are defined as 
\begin{align}
  & D^+_i[k] = \big\{ j\in\Neig_i \mid \nicefrac{6}{5}\leq  \Delta_{ij}[k] \leq2  \big\} \nonumber \\
  & D_i[k] = \big\{ j\in\Neig_i \mid \nicefrac{2}{5}\leq \Delta_{ij}[k] < \nicefrac{6}{5}  \big\} \nonumber \\
  & N_i[k] = \big\{ j\in\Neig_i \mid \nicefrac{-2}{5} < \Delta_{ij}[k] < \nicefrac{2}{5} \big\} \label{Eq:Subsets} \\
  & A_i[k] = \big\{ j\in\Neig_i \mid \nicefrac{-6}{5}< \Delta_{ij}[k] \leq\nicefrac{-2}{5}  \big\} \nonumber \\
  & A^+_i[k] = \big\{ j\in\Neig_i \mid -2\leq \Delta_{ij}[k] \leq\nicefrac{-6}{5}   \big\} \nonumber 
\end{align}
where $\Delta_{ij}[k] = x_i[k] - w_{ij} x_j[k]$. The cardinality of these sets has the following interpretation:
\begin{align}
|D^+_i[k]| & = \text{number of neighbours that agent $i$ \emph{perceives} as agreeing much less than itself at time $k$} \nonumber \\
|D_i[k]| & = \text{number of neighbours that agent $i$ \emph{perceives} as agreeing less than itself at time $k$} \nonumber \\
|N_i[k]| & = \text{number of neighbours that agent $i$ \emph{perceives} as agreeing the same as itself at time $k$} \nonumber \\
|A_i[k]| & = \text{number of neighbours that agent $i$ \emph{perceives} as agreeing more than itself at time $k$} \nonumber \\
|A^+_i[k]| & = \text{number of neighbours that agent $i$ \emph{perceives} as agreeing much more than itself at time $k$} \nonumber
\end{align} 

The overall behaviour of each agent results from the combination of three complementary inner traits: \emph{conformism}, leading the agent to agree with its neighbours; \emph{radicalism}, driving the agent to reinforce its opinion; and \emph{stubbornness}, anchoring the agent to its current opinion. The conformism, radicalism and stubbornness degree of agent $i$ is respectively denoted by $\alpha_i$, $\beta_i$ and $\gamma_i$. The parameters $\innertraits_i = (\alpha_i,\beta_i,\gamma_i)$, quantifying the \emph{inner traits} of agent $i$, satisfy $\alpha_i, \beta_i, \gamma_i\in[0,1]$ and $\alpha_i+\beta_i+\gamma_i = 1$ for all $i$. We call \emph{inner traits assignation} the collection of inner traits of all agents, $\innertraits \coloneqq (\innertraits_i )_{i\in\Vertices}$. The model features are summarised in Figure \ref{Fig:ModelExplanation}.

The opinion change $\Delta x_i[k]$ of agent $i$ at time $k$ is thus the convex combination of the behaviour of a purely conformist, purely radical, and purely stubborn agent,
\begin{equation}
\label{Eq:Threef}
\Delta x_i[k] = \alpha_i f^\text{con}_i + \beta_i f^\text{rad}_i + \gamma_i f^\text{stb}_i,
\end{equation}
with $f^\text{con}_i$, $f^\text{rad}_i$, and $f^\text{stb}_i$ taken as
\begin{equation}
\label{Eq:f_con}
f^\text{con}_i = \frac{\lambda}{|\Neig_i|}\Big( \xi|A^+_i| + |A_i| - |D_i| - \xi|D^+_i| \Big),  \qquad
f^\text{rad}_i = \frac{\lambda}{|\Neig_i|}\mu|N_i|x_i[k],  \qquad
f^\text{stb}_i = 0,
\end{equation}
where $\lambda$, $\xi$, and $\mu$ are positive parameters: $\lambda$ weighs the overall opinion change magnitude, $\xi$ weighs the increased influence that neighbours with distant opinions have over conformist traits, and $\mu$ weighs the influence of the agent's own opinion in radical traits. We call these \emph{opinion evolution parameters}: $\Omega = (\lambda, \xi, \mu)$.

To better understand Equations \eqref{Eq:f_con} and choose reasonable values for the parameters, one can think of how an extreme agent ($\alpha_i=1$, or $\beta_i=1$, or $\gamma_i=1$) behaves.
\begin{itemize}
\item A purely conformist agent ($\alpha_i = 1$, $\beta_i=0$, $\gamma_i=0$) evolves towards an opinion comparable to that of its neighbours. For instance, if $N_i = \Neig_i$ (all the neighbours of agent $i$ agree comparably), then agent $i$ does not change its opinion. If $A_i = \Neig_i$ (all the neighbours of agent $i$ agree more), agent $i$ increases its opinion $x_i$ by $\lambda$; given that all the neighbours of agent $i$ are in the set $A_i$, a value $\lambda = 0.4$ guarantees that, if all the neighbour opinions remain unchanged, then at the next time step all the neighbours of agent $i$ will be in the set $N_i$, hence perceived as having a comparable opinion.
Instead, if $A^+_i = \Neig_i$, then the opinion of agent $i$ needs to increase $0.8=2\lambda$ in order to be perceived as comparable to its neighbours' at the next time step, and therefore a natural choice is $\xi=2$. The same reasoning can be applied to the sets $D_i$ and $D^+_i$.
\item A purely radical agent ($\alpha_i = 0$, $\beta_i=1$, $\gamma_i=0$) ignores neighbours with a different opinion and only cares about agents that think comparably to itself, hence it reinforces its current opinion $x_i[k]$ depending on the magnitude of its own opinion and on the fraction of its neighbours in the set $N_i$.  \CAbnew{To make sure that radical traits can affect the opinion change more strongly than conformist traits, we need $\mu>1$. In fact, if $\mu=1$, then $|f^\text{rad}_i|<|f^\text{con}_i|$ in general: the opinion change caused by the radical trait (which is proportional to $x_i[k]$, and $|x_i[k]|\leq1$) is smaller in magnitude than the one caused by the conformist trait. In our simulations, we set $\mu = 5$. The effect of different values of $\mu$ can be seen in Table \ref{table:var_mu}.}
\item A purely stubborn agent ($\alpha_i = 0$, $\beta_i=0$, $\gamma_i=1$) does not change its opinion under any circumstance.
\end{itemize} 

The new opinion of agent $i$ at time $k+1$ is the sum of the previous opinion $x_i[k]$ and the opinion change $\Delta x_i[k]$, modulated by the saturation function $\sigma$
\begin{equation}\label{Eq:Sigmoid}
\sigma(x) = 
\begin{cases}
x &\text{if}\quad |x|\leq1 \\
\text{sign}(x) &\text{if}\quad |x|>1
\end{cases}
\end{equation} 
so as to guarantee that the opinions remain in the interval $[-1,1]$.
The complete opinion update law is therefore 
\begin{equation}
\label{Eq:CompleteLaw}
x_i[k+1] = \sigma\Bigg( x_i[k] + \frac{\lambda}{|\Neig_i|} \Big( \alpha_i\xi \big(|A^+_i| - |D^+_i|\big)  + \alpha_i\big(|A_i| - |D_i|\big) + \beta_i\mu|N_i|x_i[k] \Big) \Bigg), \qquad \forall i\in\Vertices.
\end{equation}
%
%

\subsection{Model Parameters}

The Classification-Based (CB) model has three types of parameters: the signed digraph weights $w_{ij}$; the inner traits assignation $\innertraits_i = (\alpha_i,\beta_i,\gamma_i)$; and the opinion evolution parameters $\Omega = (\lambda, \xi, \mu)=(0.4,2,5)$ whose values are fixed, and chosen based on the model interpretation. Later, a parameter sensitivity analysis explores how the model evolution is affected by changes in opinion evolution parameters.

\begin{figure}[h!]
\centering
\includegraphics[width=0.9\textwidth]{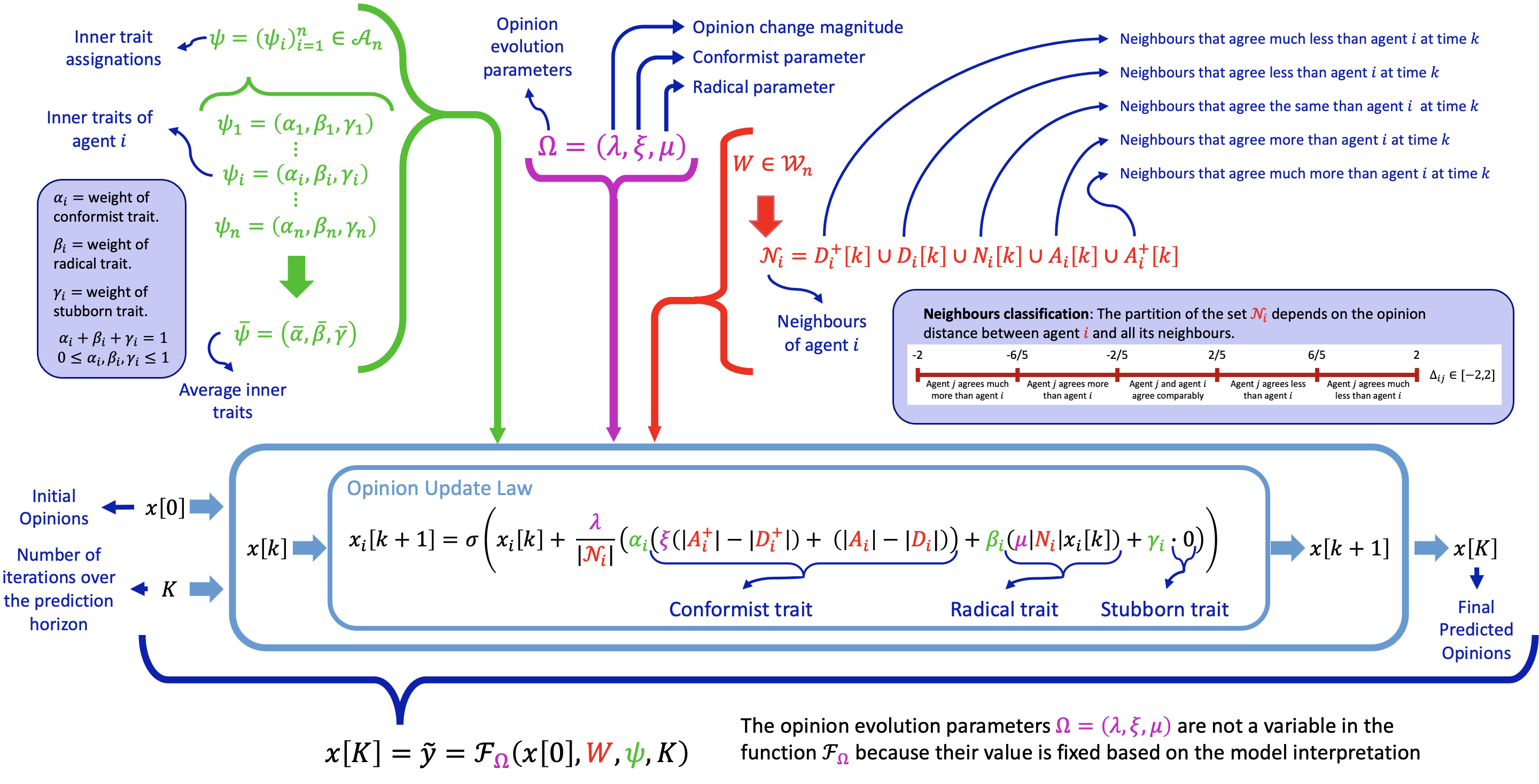}
\caption{Visualisation of the model features and parameters. The model has three parameter types: inner traits assignation $\innertraits$ (in green), opinion evolution parameters $\Omega$ (in magenta), and signed digraph weights $\Weights$ (in red). These parameters appear in the Opinion Update Law of Equation \eqref{Eq:CompleteLaw}, which is a convex combination of contributions by conformist, radical, and stubborn traits, with the opinions of neighbouring agents evaluated through a classification-based approach. Given the initial conditions $x[0]$ and the number $K$ of iterations over the prediction horizon, the Opinion Update Law produces the final predicted opinions $\tilde{y} = x[K]$. The opinion evolution parameters $\Omega$ can be fixed based on the model interpretation. Then, the final predicted opinions in each particular case are a function $x[K] = \FancyF_\Omega(x[0], \Weights, \innertraits, K)$ of the chosen initial opinions, signed digraph weights, inner traits assignation, and number of iterations in the prediction horizon.} 
\label{Fig:ModelExplanation}
\end{figure}

If the model has $\numag$ agents, then:
\begin{itemize}
\item The signed digraph has weight matrix $\Weights\in\SetN_\numag$. In general, $\SetN_\numag = \{-1,0,1\}^{\numag\times\numag}$, but we can focus for instance on small-world, or strongly connected, networks.
\item The inner traits assignation is $\innertraits \in\SetA_\numag$, where
\begin{equation}
\SetA_\numag = \Big\{\innertraits = (\innertraits_i)_{i\in\Vertices} = \big((\alpha_i, \beta_i, \gamma_i)\big)_{i=1}^\numag \mid \alpha_i,\beta_i,\gamma_i\in[0,1] \quad \mbox{and} \quad \alpha_i+\beta_i+\gamma_i = 1, \quad \forall i\in\Vertices\Big\}.
\end{equation}
\end{itemize}
We omit the subscript $\numag$ from the sets $\SetN$ and $\SetA$ for simplicity. 
Given $\numag$ agents, a signed digraph $\Weights\in\SetN$, an inner traits assignation $\innertraits\in\SetA$, and a vector of initial opinions $x[0]$, the opinion formation model evolves according to Equation \eqref{Eq:CompleteLaw}. The vector $x[K]$ of opinions after $K$ iterations can be explicitly represented as a function of $\Weights$, $\innertraits$, and $x[0]$ by the map $\FancyF_\Omega$ ($x[K]$ also depends on $\Omega$, whose value, given by the model interpretation, is fixed) as 
\begin{equation}
x[K] = \FancyF_\Omega(x[0], \Weights, \innertraits, K) 
\end{equation}
The value of $K$ depends on the type of statements and the prediction horizon. For statements related to core values or beliefs, opinions are not expected to change very fast and one could consider roughly 10 changes per year. Therefore, if the model is used to predict the opinions after 5 years, $K=50$. On the other hand, the opinions on more superficial topics could change faster and, over the same 5-year timespan, it could be $K=500$. See Figure \ref{Fig:ModelExplanation} for a summary of the model parameters and features.

To validate the model -- namely, assess its potential to closely reproduce the evolution of opinions in real life with suitably chosen parameters -- we consider \emph{real} initial and final opinions, denoted by $x$ and $y$  respectively, taken from survey data. Assuming that $y$ are the real opinions $K$ iterations after the real initial opinions $x$, these data can be used to  \CAbnew{find values of the model parameters (edge weights $\Weights$ and inner traits $\innertraits$) that match as closely as possible} the real opinion evolution, through the minimisation problem
\begin{equation}
\label{Eq:FirstMP}
(\widehat{\Weights}, \widehat{\innertraits}) = \argmin_{\substack{\Weights\in\SetN \\ \innertraits\in\SetA}}\Cost(y,\tilde{y}) \qquad \mbox{such that} \qquad \tilde{y} = \FancyF_\Omega(x, \Weights, \innertraits, K),
\end{equation}
where the cost function $\Cost(y,\tilde{y})= \sum_{i = 1}^\numag | y_i - \tilde{y}_i |$ quantifies the mismatch between opinion vectors $y$ and $\tilde{y}$.

If the same population is asked to quantify the agreement with $\numques$ different statements, the signed digraph cannot change. However, the inner traits assignation can vary depending on the statement, since each individual may have different attitudes towards different topics. Therefore, if $\innertraits^{(\newj)}$ represents the inner traits assignation associated with statement $\newj$,  \CAbnew{values for the parameters $\Weights$ and $(\innertraits^{(\newj)})_{\newj=1}^\numques$  that produce predicted opinions as similar as possible to the real ones} can be found through the \emph{free optimisation problem}
\begin{equation}
\label{Eq:LargeOP}
\big(\widehat{\Weights}, (\widehat{\innertraits^{(\newj)}})_{\newj=1}^\numques\big) = \argmin_{\substack{\Weights\in\SetN \\ \innertraits^{(\newj)}\in\SetA}} \sum_{\newj=1}^\numques\Cost(y_\newj,\tilde{y}_\newj) \qquad\qquad \tilde{y}_\newj = \FancyF_\Omega(x_\newj, \Weights, \innertraits^{(\newj)}, K) 
\end{equation}
where $x_\newj$ and $y_\newj$ are the known initial and final opinions related to statement $\newj$.

\CAg{If instead all the inner traits assignations are constrained to be the same for every question, we consider the \emph{constrained optimisation problem}}
\begin{equation}
\label{Eq:LargeOP_mod}
\CAg{(\widehat{\Weights}, \widehat{\innertraits}) =  \argmin_{\substack{\Weights\in\SetN \\ \innertraits\in\SetA}} \sum_{\newj=1}^\numques\Cost(y_\newj,\tilde{y}_\newj) \qquad\qquad \tilde{y}_\newj = \FancyF_\Omega(x_\newj, \Weights, \innertraits, K) }
\end{equation}
\CAg{The free optimisation problem, where the inner assignations can change, allows for a more thorough study of the behaviour of a population, while the constrained optimisation problem allows for a more rigorous testing of the prediction capabilities of the model in the form of cross-validation: the answers to some questions can be used as training datasets to  \CAbnew{choose the model parameters}, and the model performance can then be tested on the remaining questions.}

\section{Simulation Results}\label{Sec:Sim}

To gain insight into the classification-based (CB) model, this section presents five different types of simulation results:
1) \textbf{Simulations in Simple Cases} evolve the model in simple, special cases to gain intuition into its behaviour; 2) \textbf{Parameter Sensitivity Analysis} studies how changes in each of the model parameters (inner traits assignation, signed digraph, opinion evolution parameters) affect the model behaviour; 3) \textbf{Model Validation with Real Data} leverages real data from the WVS to show that the CB model has the potential to reproduce the time evolution of real opinions in society (with parameters chosen through the \emph{free} and the \emph{constrained} optimisation problems of Equations \eqref{Eq:LargeOP} and \eqref{Eq:LargeOP_mod} respectively) and presents \CAbnew{the transitions between different qualitative types of opinion distributions, such as Perfect Consensus, Consensus, Polarization, Clustering, Dissensus, based on the recently proposed transition tables \cite{Devia2022A}};
4) \textbf{Comparison with the Friedkin-Johnsen (FJ) Model} investigates the relation between the two models and their predictive capabilities (first, evolving equivalent populations; second, solving the optimisation problems  \eqref{Eq:LargeOP} and \eqref{Eq:LargeOP_mod}; and third, computing the corresponding transition tables);
5) \textbf{Model Outcome Capabilities} explores the rich variety of opinion vectors that the CB model can produce.\\
 \CAbnew{Due to the deterministic nature of the model, running it with the same initial opinions, inner traits, and interconnection network always produces the same results. Given a parameter constellation and a network, the model evolution could only change due to different initial conditions (see the repeated model runs in Figure \ref{Fig:BatchFig}).}

\begin{figure}[!h]
\centering
	\begin{subfigure}[t]{0.30\textwidth}
         \centering
         \includegraphics[width=\textwidth]{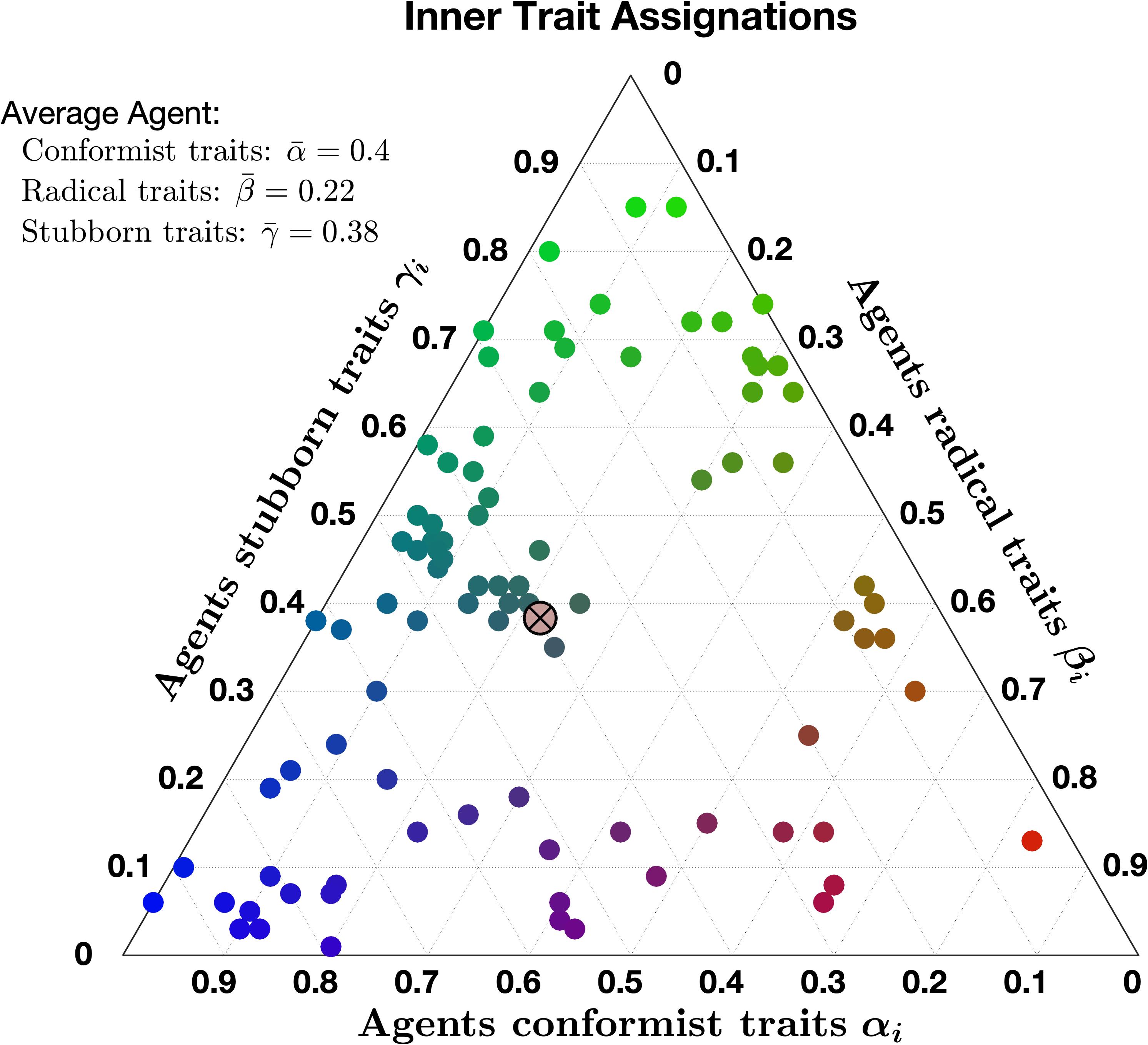}
         \caption{\CA{Each dot represents the inner traits of an agent; its RGB colour reflects the weight of each trait (blue: conformist; red: radical; green: stubborn). The crossed dot represents the average inner traits.}}
         \label{Fig:TernaryDiagramExample}
     \end{subfigure}
     \hfill
     \begin{subfigure}[t]{0.30\textwidth}
         \centering
         \includegraphics[width=\textwidth]{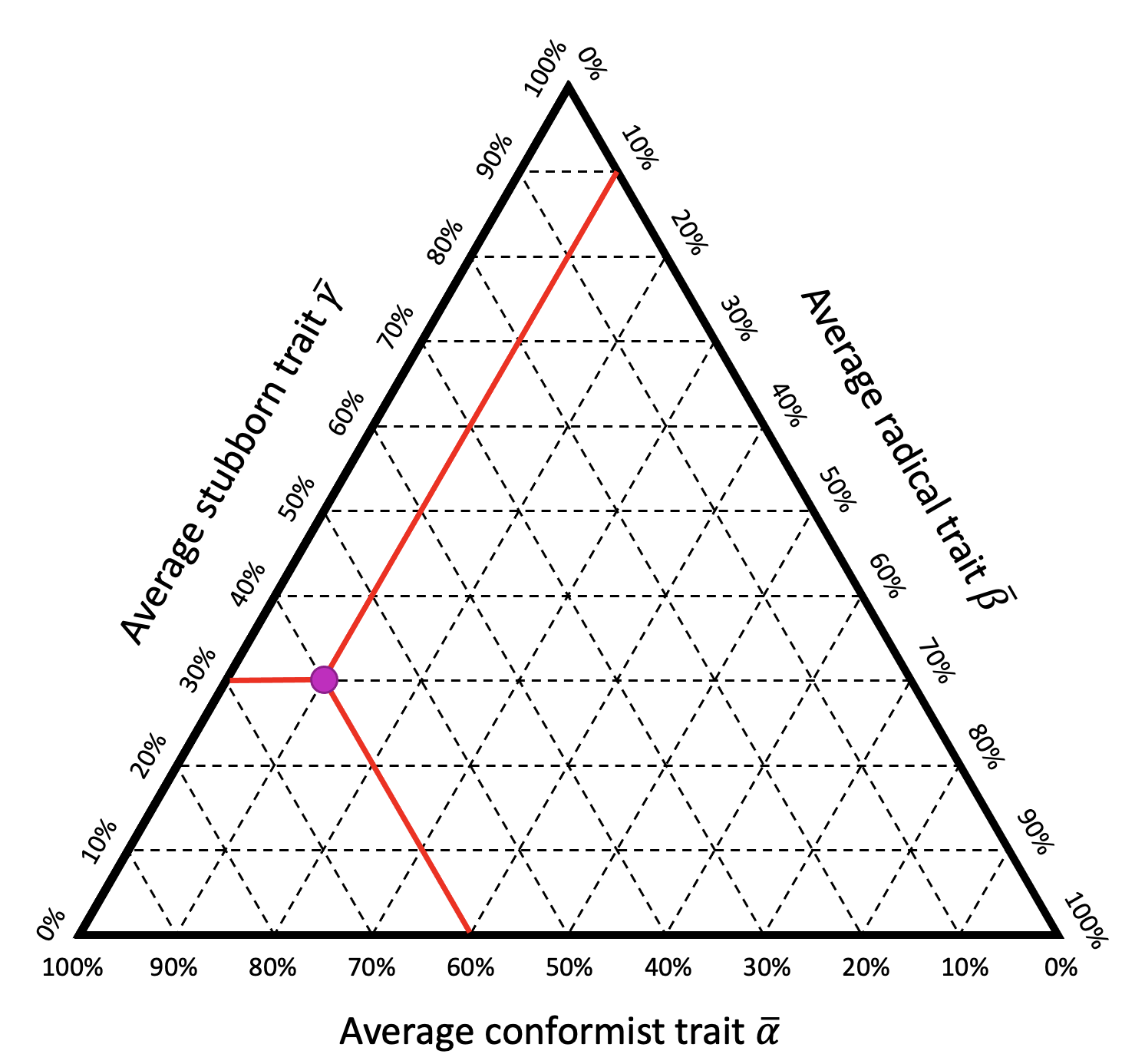}
         \caption{\CA{Example of average inner traits in the ternary diagram: $60\%$ conformist, $10\%$ radical, $30\%$ stubborn.}} 
         \label{Fig:TD1c}
     \end{subfigure}
     \hfill
     \begin{subfigure}[t]{0.38\textwidth}
         \centering
         \includegraphics[width=\textwidth]{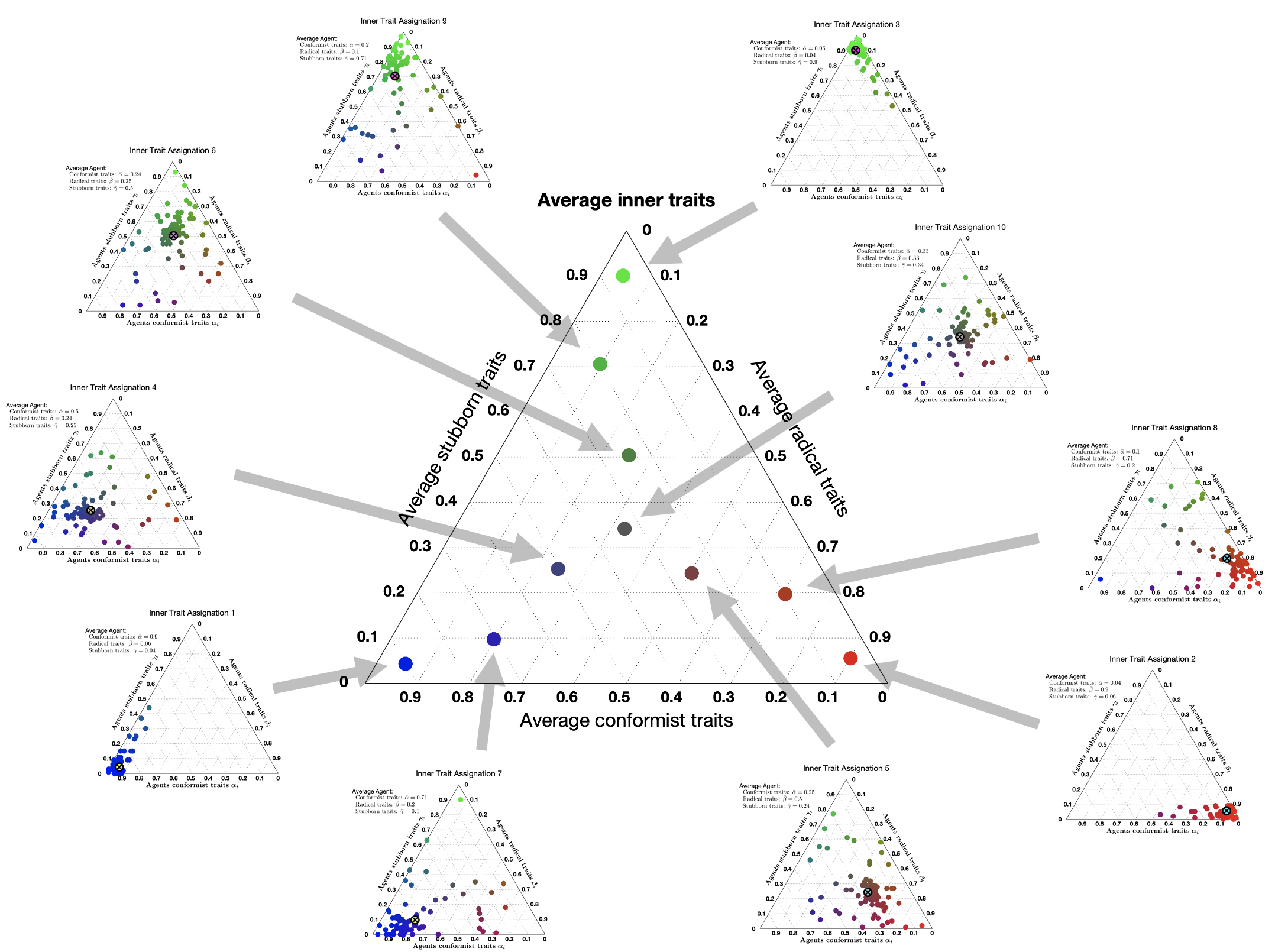}
         \caption{ \CAbnew{The ten inner traits assignations used in the simulations in Figure \ref{Fig:EvMultipleTop}, along with the corresponding average inner traits in the ternary diagram.}} 
         \label{Fig:TenInnerTraits}
     \end{subfigure}
\label{Fig:TernaryDiagramExample_main}
\caption{\CA{Ternary diagrams visualising inner traits assignations $\innertraits$ and average inner traits $\bar{\innertraits}$.} Panel \ref{Fig:TernaryDiagramExample} shows the whole inner traits assignation (along with its average), while panel \ref{Fig:TD1c} only shows the average inner traits. The relation between the two types of plots is visualised in panel \ref{Fig:TenInnerTraits}, showing that each of the ten inner traits assignations (plotted in a ternary diagram) can be represented by a single point, i.e. the average inner traits, which can also be plotted in a ternary diagram, with different axis labels.}
\end{figure}

To facilitate the interpretation of simulation results, we introduce some definitions. Given the inner traits assignation $\innertraits = (\innertraits_i)_{i\in\Vertices} = \big( (\alpha_i,\beta_i,\gamma_i) \big)_{i\in\Vertices}$, the associated \emph{average inner traits}
\begin{equation}
\bar{\innertraits} = (\bar{\alpha}, \bar{\beta}, \bar{\gamma}) \quad \text{where} \quad
\bar{\alpha} = \frac{1}{\numag}\sum_{i\in\Vertices}\alpha_i \qquad 
\bar{\beta} = \frac{1}{\numag}\sum_{i\in\Vertices}\beta_i \qquad 
\bar{\gamma} = \frac{1}{\numag}\sum_{i\in\Vertices}\gamma_i,
\end{equation}
represent the traits of an average agent in the considered society or population \CAbnew{with $n = |\Vertices|$ agents}. \CA{inner traits assignation $\innertraits$ and the corresponding average inner traits $\bar{\innertraits}$ can be plotted in a ternary diagram as shown in Figure \ref{Fig:TernaryDiagramExample}. Figure \ref{Fig:TD1c} explains how to interpret a point in the ternary diagram.}


\begin{figure}[!h]
\centering
	\begin{subfigure}[t]{0.39\textwidth}
         \centering
\includegraphics[width=\textwidth]{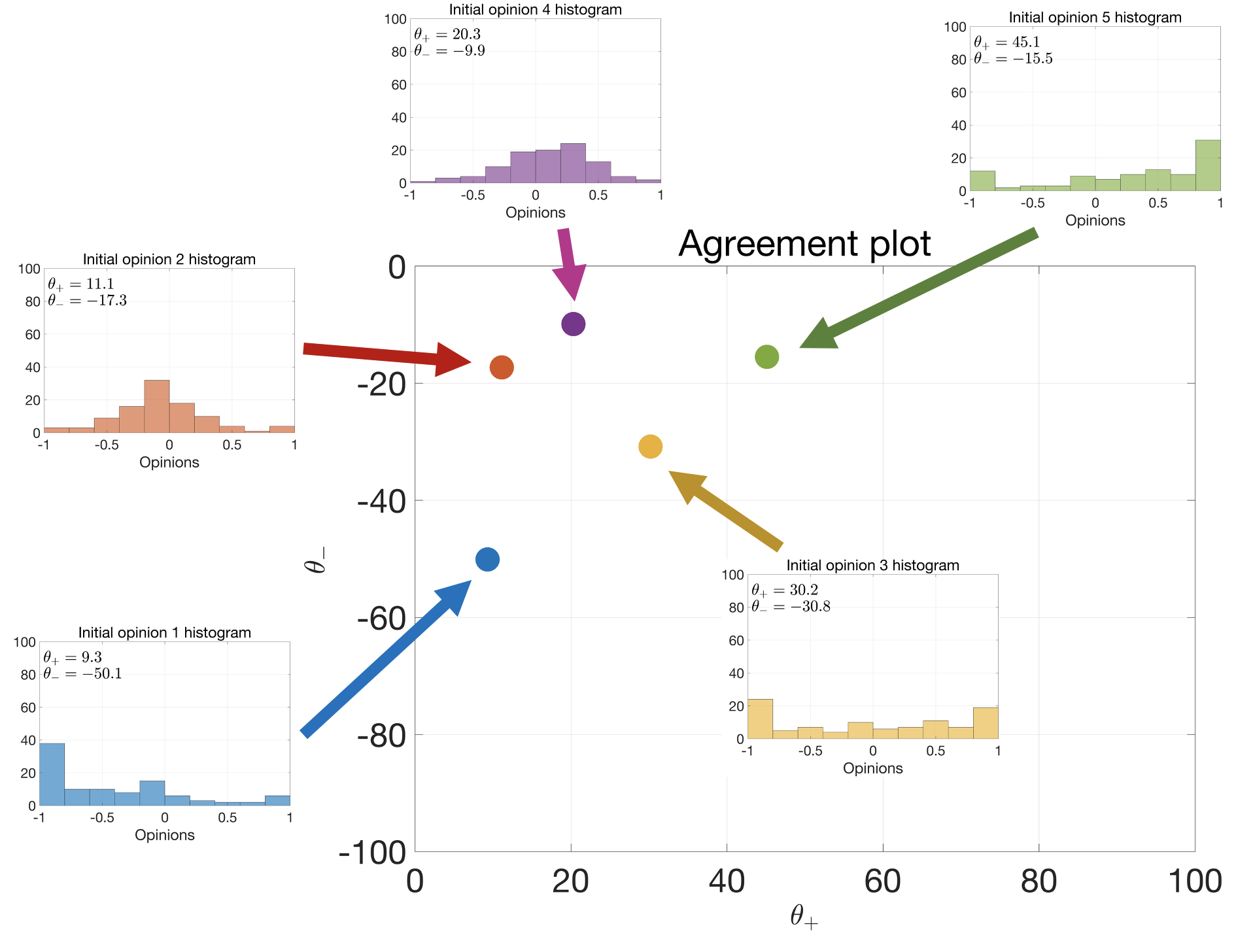}
\caption{\CAnewcomment{Agreement plot of 5 opinion vectors.}}
\label{Fig:AgPlotExpl}
     \end{subfigure}
     \hfill
     \begin{subfigure}[t]{0.59\textwidth}
         \centering
         \includegraphics[width=\textwidth]{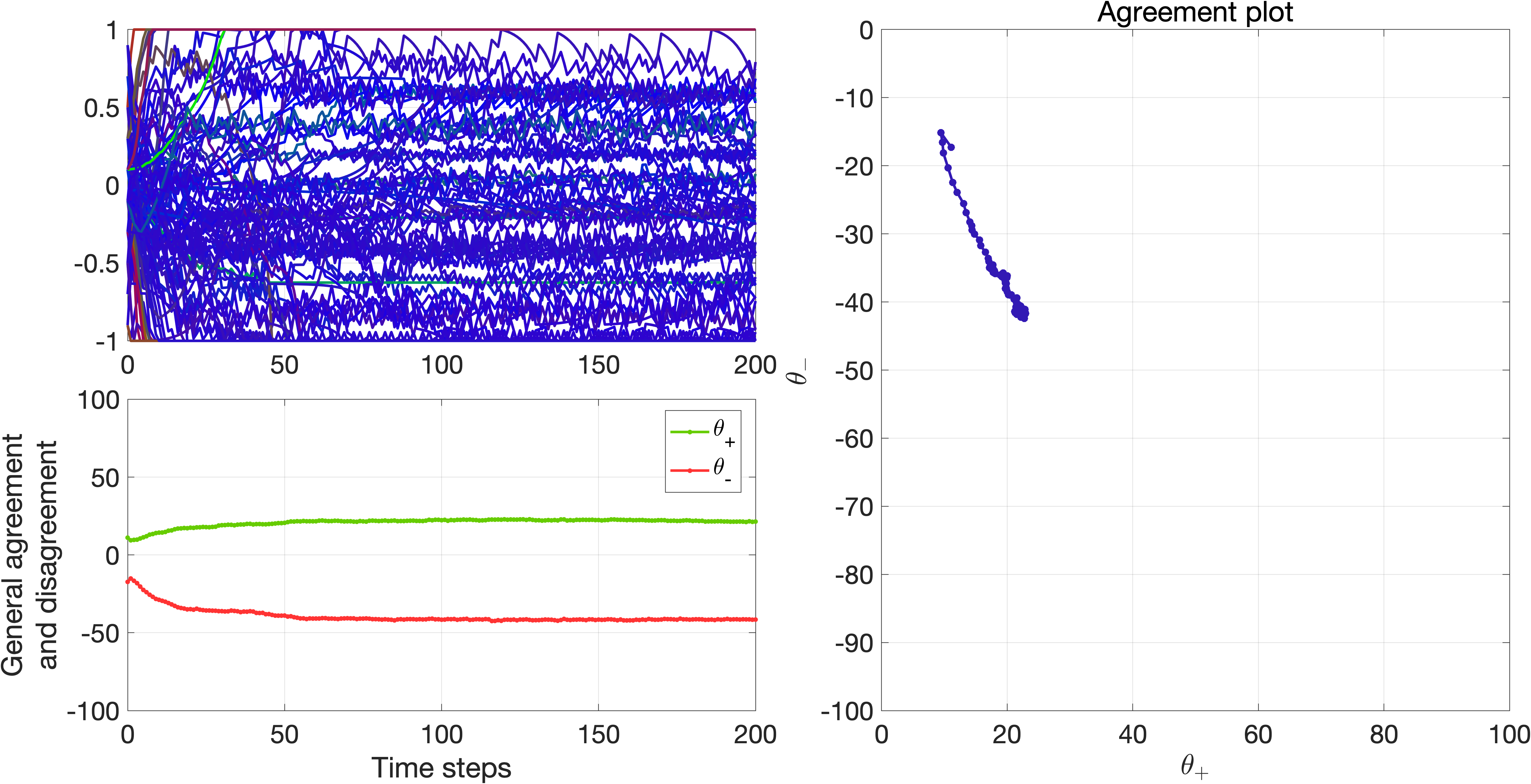}
         \caption{\CAnewcomment{Agreement plot of the opinion evolution (a parametric curve).}} 
\label{Fig:AgParamExpl}
     \end{subfigure}
\caption{\CAnewcomment{Panel \ref{Fig:AgPlotExpl} shows the general agreement of five different opinion vectors (represented in the figure by their corresponding histograms) plotted in the Cartesian plane. Panel \ref{Fig:AgParamExpl} shows a parametric curve in the the agreement plot: the opinion evolution of 100 agents and the corresponding values of $\theta_-$ and $\theta_+$ are shown as a function of time (left) and the corresponding general agreement evolution is plotted in the Cartesian plane (right), visualising information on the evolution of the opinions that may be difficult to grasp otherwise. }} 
\end{figure}

\CAnewcomment{The \emph{general agreement} of an opinion vector $x = (x_i)_{i=1}^\numag$, quantified by the pair $(\theta_+, \theta_-)$ where}
\begin{equation}
\label{Eq:ThetaEquations}
\CAnewcomment{ \theta_- = \sum_{x_i<0}x_i \qquad \mbox{and} \qquad \theta_+ = \sum_{x_i>0}x_i},
 \end{equation}
\CAnewcomment{is the overall level of agreement and disagreement in the whole society. Figure \ref{Fig:AgPlotExpl} shows the histograms corresponding to 5 different opinion vectors and their corresponding \emph{agreement plot}. The agreement plot, (i.e., the plot in the Cartesian plane of one or more general agreements) can be used to represent not only single opinion vectors, but also sequences of opinion vectors, resulting in a parametric curve of the opinion evolution, as shown in Figure \ref{Fig:AgParamExpl}.}


 \CAbnew{All the simulations involve a population of 100 agents.}

 \CAbnew{All the digraphs used in both Parameter Sensitivity Analysis and Model Validation with Real Data have a small-world network topology, with an assigned probability for positive and negative edges, and are strongly connected. We consider small-world networks because they have a high clustering coefficient (neighbours of neighbours of agent $i$ are likely also neighbours of agent $i$) and low diameter (maximum distance between two agents of the network), which are believed to be characteristics of real-life social networks \cite{Elgazzar2003Applications,Watts1998Collective}. The directed small-world networks were built based on the Watts-Strogatz algorithm. Appendix A describes the computation of network metrics. The signed digraphs are not restricted to be structurally balanced, to account for the fact that also non-structurally-balanced networks have been considered in the literature when modelling social dynamics \cite{Estrada2019Rethinking,Leinhardt1977Social,Opp1984Balance}. }

\CAbnew{In all the considered simulations, the initial opinions, traits and networks are assigned independently. A different approach -- which is left for future work -- could be to assign them in some correlated way: e.g., initial opinions and network could be correlated by assigning the initial opinions such that two vertices connected by an edge have a very similar (or very distant) initial opinion; traits and network could be correlated by assigning the agent parameters with a probability that depends on the corresponding vertex characteristics, for example assuming that vertices with higher out-degree have a higher probability of being completely conformist, or radical. Correlations between initial opinions, traits, and network characteristics can reproduce different types of societies present in real life (for instance, in a society that values tradition, highly stubborn agents may be more influential than others, and hence the corresponding vertices may have a higher out-degree).}

\subsection{Simulations in Simple Cases}

\CAd{To better understand the model behaviour, we simulate the model evolution in special simple cases. First, for the same digraph with a lattice topology, we vary the inner traits assignations (Figure \ref{Fig:NewFig_1}). Then, for the same inner traits assignation, we consider different digraph topologies (complete, lattice, ring, small-world) and different, randomly chosen, initial opinions (Figure \ref{Fig:BatchFig}).}  \CAbnew{Finally, for the five initial opinions shown in Figure \ref{Fig:AgPlotExpl}, we consider 12 different networks (4 topologies, each with 3 different probabilities for positive and negative edges) and 10 different inner traits assignations (Figure \ref{Fig:EvMultipleTop}).}

\subsubsection{Different Inner Traits Assignations}

\begin{figure}[!h]
\centering
\includegraphics[width=\textwidth]{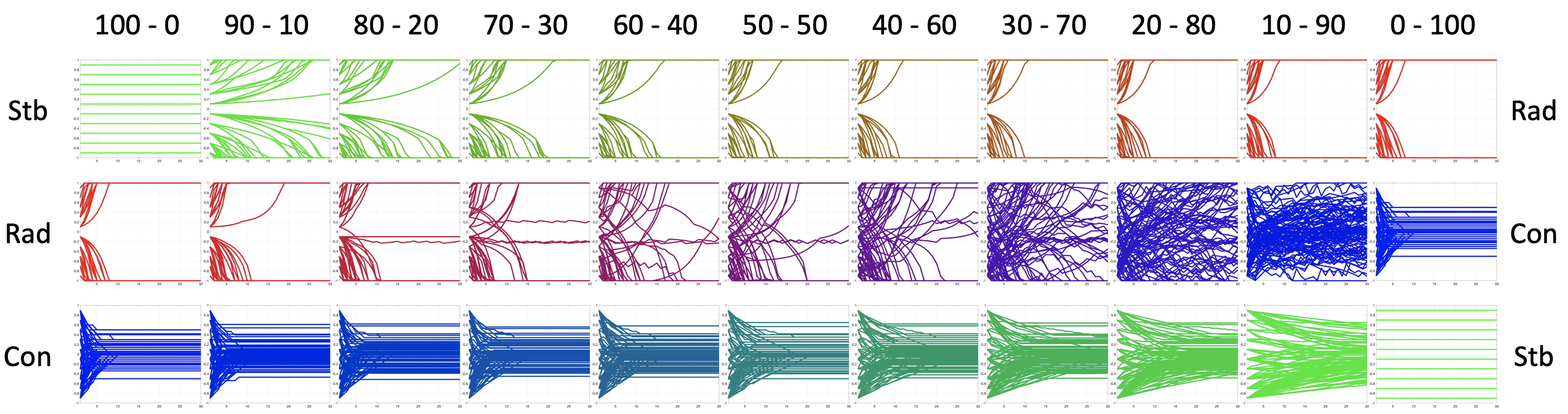}
\caption{\CAd{Evolution of the CB model, starting from the same initial opinions, over a signed lattice digraph where the edges are positive with probability $0.77$. In each simulation, all $100$ agents have the same inner traits that are a combination of only two of the possible traits: stubbornness (Stb), radicalism (Rad), and conformism (Con). The top labels show the proportion of each trait: the upper left (respectively, right) graph corresponds to a simulation where all agents are purely stubborn (resp. radical). The colour of the lines is the RGB representation of the inner traits assignations (blue: conformist; red: radical; green: stubborn).}}
\label{Fig:NewFig_1}
\end{figure}

\CAd{We consider a signed lattice digraph, where each agent has 4 in-neighbours and the edges are positive with probability $0.77$. All the agents have the same inner traits, combining only two inner traits: stubbornness and radicalism; radicalism and conformism; conformism and stubbornness. Starting from the same initial opinions, Figure \ref{Fig:NewFig_1} shows the opinion evolution over 30 time steps. Radicalism tends to form polarisation by driving the agents to extreme opposite views. Conformism tends to create consensus; however, because of the classification approach, the agents do not converge to the very same opinion (close enough agents are unable to perceive their opinion difference, because opinions are assessed with finite resolution). Stubbornness slows down the effect of the other two traits; only in a fully stubborn population everyone keeps its initial opinion. Among the three traits, radicalism appears to have the greatest effect: even a small amount of radicalism can prevent conformism from forming consensus, and can yield polarisation in a very stubborn society.}

\subsubsection{Different Digraph Topologies}

\begin{figure}[!h]
\centering
\includegraphics[width=0.8\textwidth]{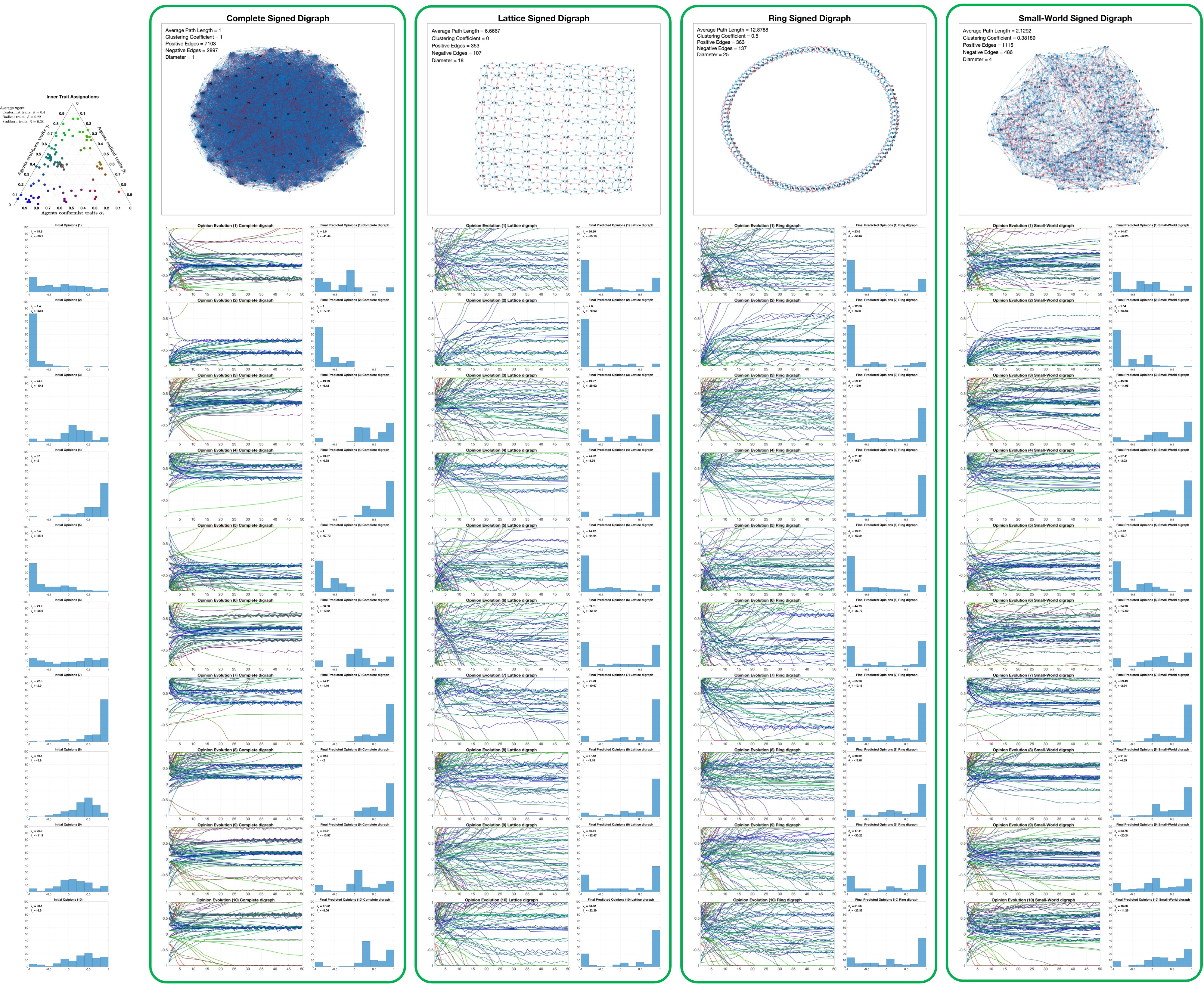}
\caption{\CAd{Evolution of the CB model, with 100 agents, over four different digraph topologies and starting from ten different, randomly chosen, initial opinions, shown on the left-most column. Each of the four columns framed in green shows the opinion evolution and the final opinion distribution for each of the initial opinions, for a different choice of the digraph (Complete, Lattice, Ring, Small-World), shown at the top of the column along with its characteristics. The inner traits assignation is presented at the top left and has average traits: $\overline{\alpha} = 0.4$ (conformist), $\overline{\beta} = 0.22$ (radical), and $\overline{\gamma} = 0.38$ (stubborn). The ternary diagram of these inner traits can also be seen in Figure \ref{Fig:TernaryDiagramExample}.}}
\label{Fig:BatchFig}
\end{figure}

\CAd{Additional intuition on the model behaviour can be gained by studying the effect of different initial opinions and different digraph topologies with fixed inner trait parameters. Figure \ref{Fig:BatchFig} shows 10 simulations starting from various, randomly chosen, initial opinions and evolved over four signed digraphs with Complete, Lattice, Ring, and Small-World topologies. The inner traits assignation for all these simulations is kept constant and is shown at the top left of Figure \ref{Fig:BatchFig} (it is the same also shown in Figure \ref{Fig:TernaryDiagramExample}). }\\
\CAd{Both the digraph topology (dictating how the agents communicate among them) and the initial opinions (providing the starting point of the evolution) have a significant effect on the opinion evolution and the final predicted opinions. For Lattice and Ring digraphs, there is a clear tendency towards consensus at one extreme opinion (completely agree or completely disagree), even when, as in this case, the average radical trait is relatively low. A possible explanation is that in both these topologies agents have less in-neighbours, so the radical trait can have a stronger effect. Another possible explanation is that both these types of networks have a larger average path length, and diameter, than Complete and Small-World networks, and therefore the `consensus effect' takes more time to act than in more connected networks.}\\
\CAd{Indeed, since they share common features, the Complete and Small-World digraphs (small diameter), as well as the Lattice and Ring digraphs (large diameter), showcase similar behaviours and similar final opinion distributions, across all the chosen initial opinion distributions. We have observed that this tendency is recurrent for several different choices of inner traits assignations.}

\CAd{As is apparent from Figure  \ref{Fig:NewFig_1} (and from all the opinion evolution simulations shown in the next subsection), the inner traits assignation has a tremendous effect on the opinion evolution. The simulations of Figure \ref{Fig:BatchFig} reinforce this idea by showing that, although the initial opinions and digraph topology do have an impact, keeping the same inner traits assignation restricts the final opinion distributions to some characteristic patterns.}

\subsubsection{Different Networks and Inner Traits}

\begin{figure}[!h]
\centering
	\begin{subfigure}[t]{0.49\textwidth}
         \centering
\includegraphics[width=\textwidth]{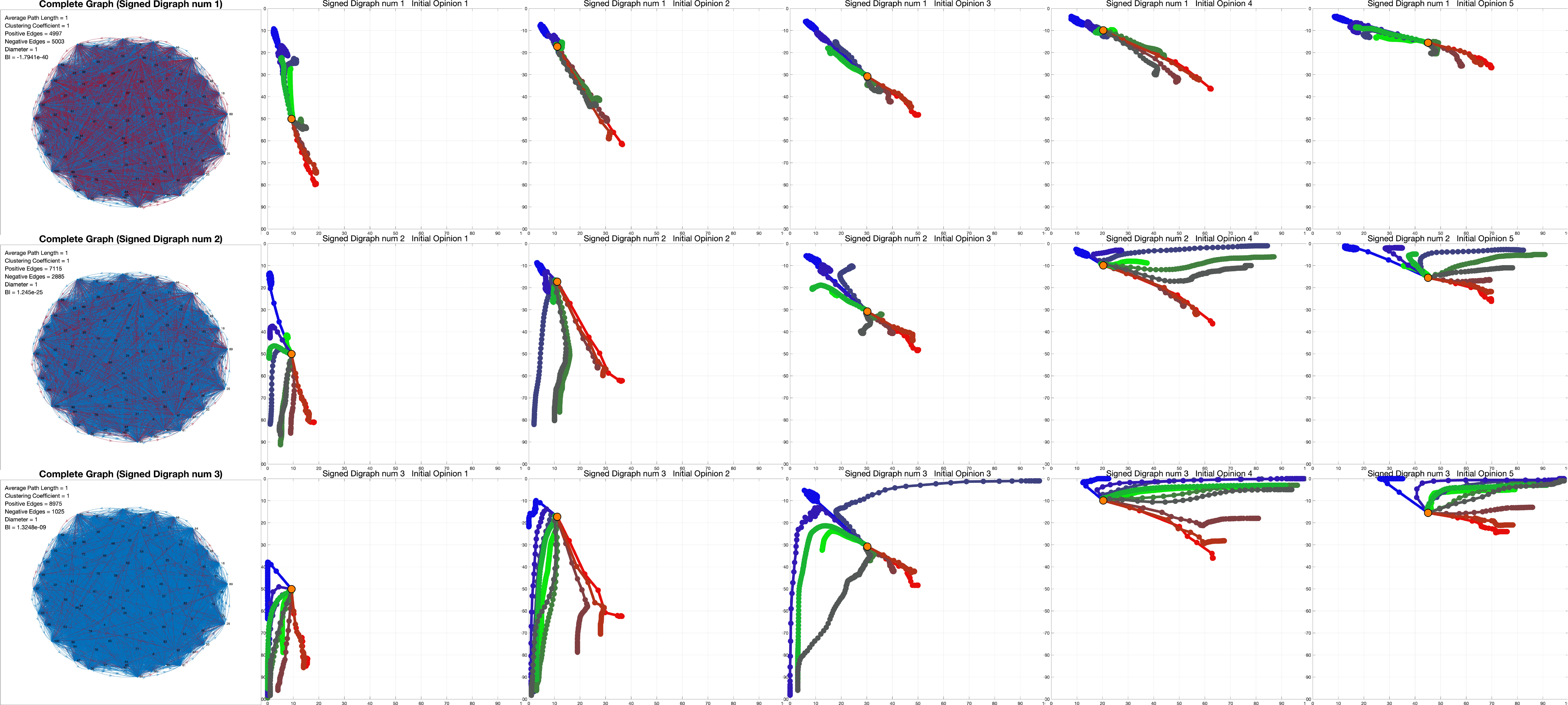}
\caption{ \CAbnew{Evolution over a Complete graph topology.}}
\label{Fig:EvCG}
     \end{subfigure}
     \hfill
     \begin{subfigure}[t]{0.49\textwidth}
         \centering
         \includegraphics[width=\textwidth]{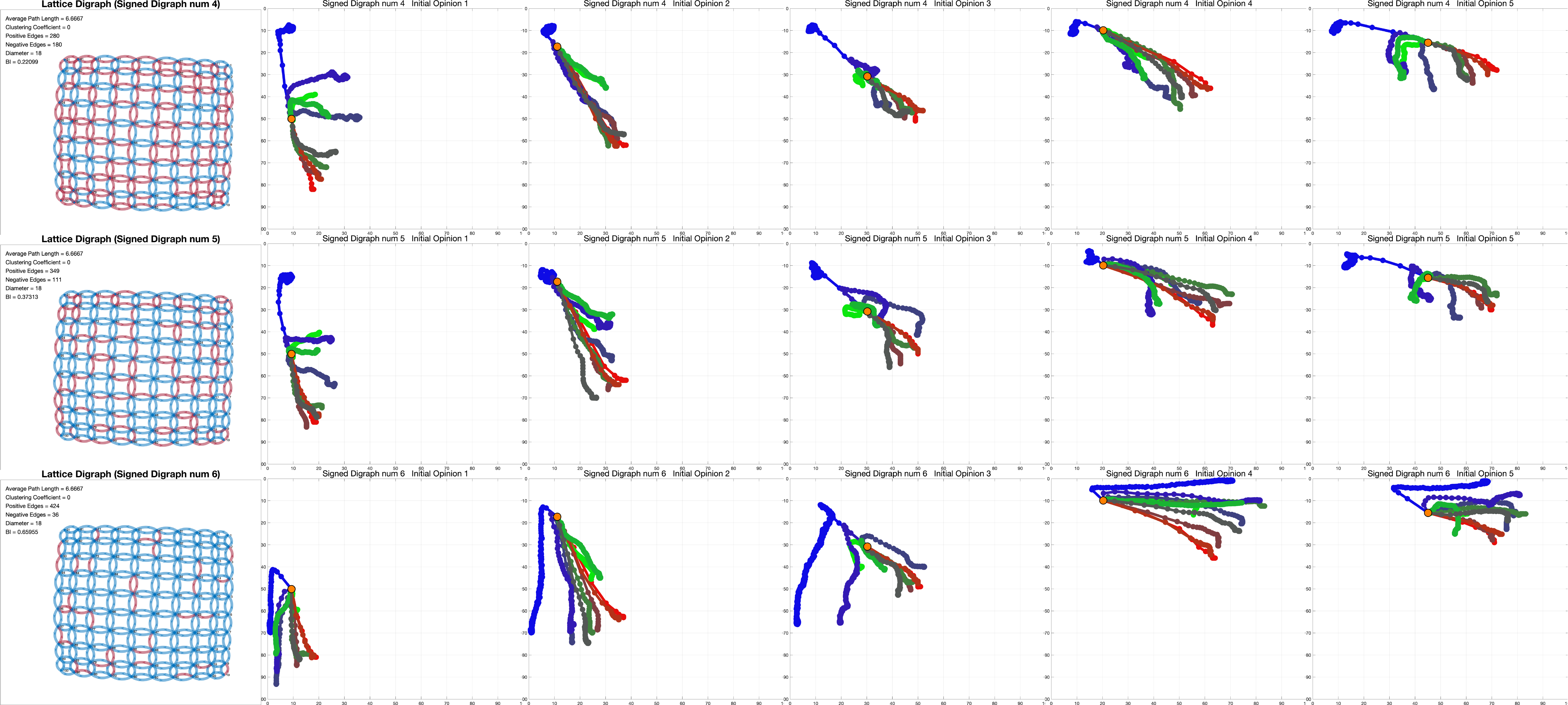}
\caption{ \CAbnew{Evolution over a Lattice digraph topology.}}
\label{Fig:EvLD} 
     \end{subfigure} \\
     \begin{subfigure}[t]{0.49\textwidth}
         \centering
\includegraphics[width=\textwidth]{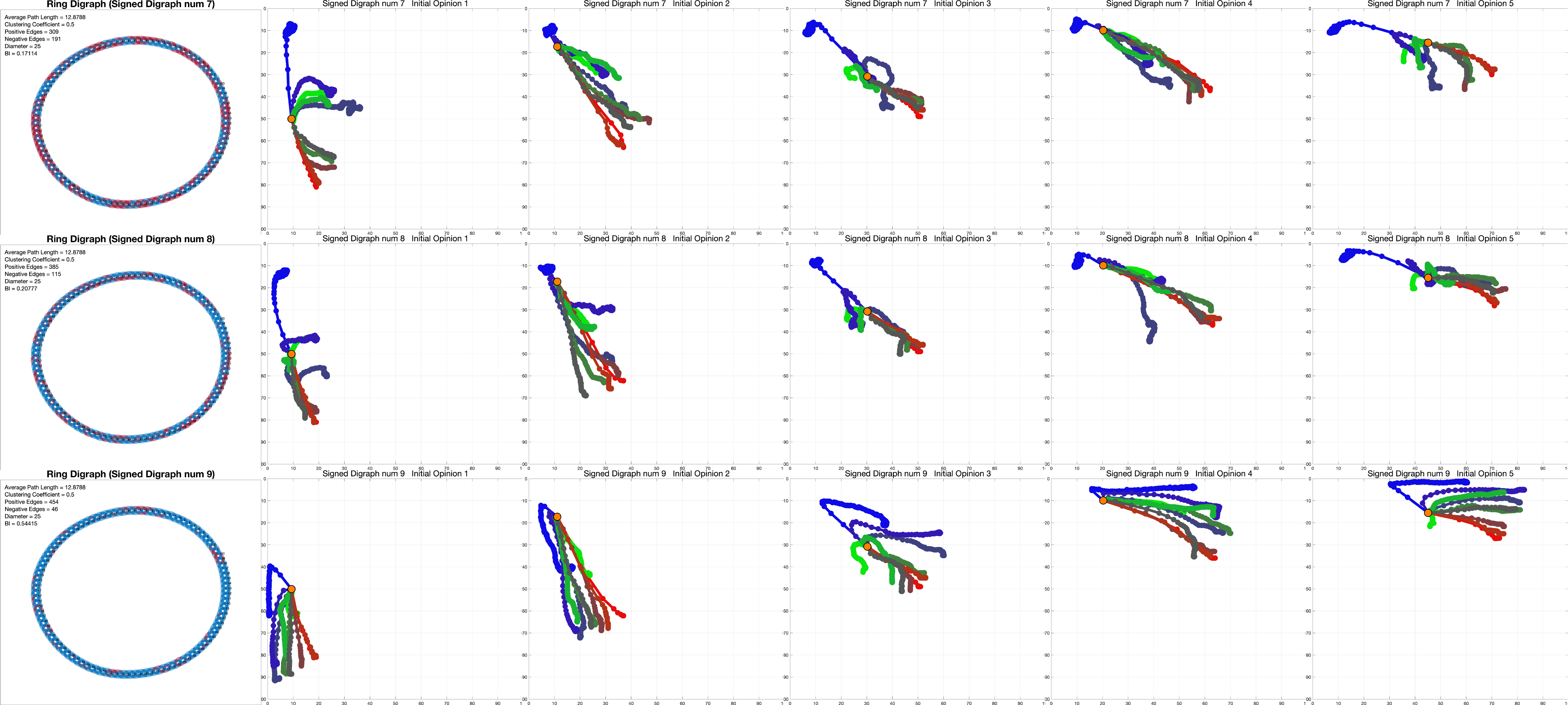}
\caption{ \CAbnew{Evolution over a Ring digraph topology.}}
\label{Fig:EvRD}
     \end{subfigure}
     \hfill
     \begin{subfigure}[t]{0.49\textwidth}
         \centering
         \includegraphics[width=\textwidth]{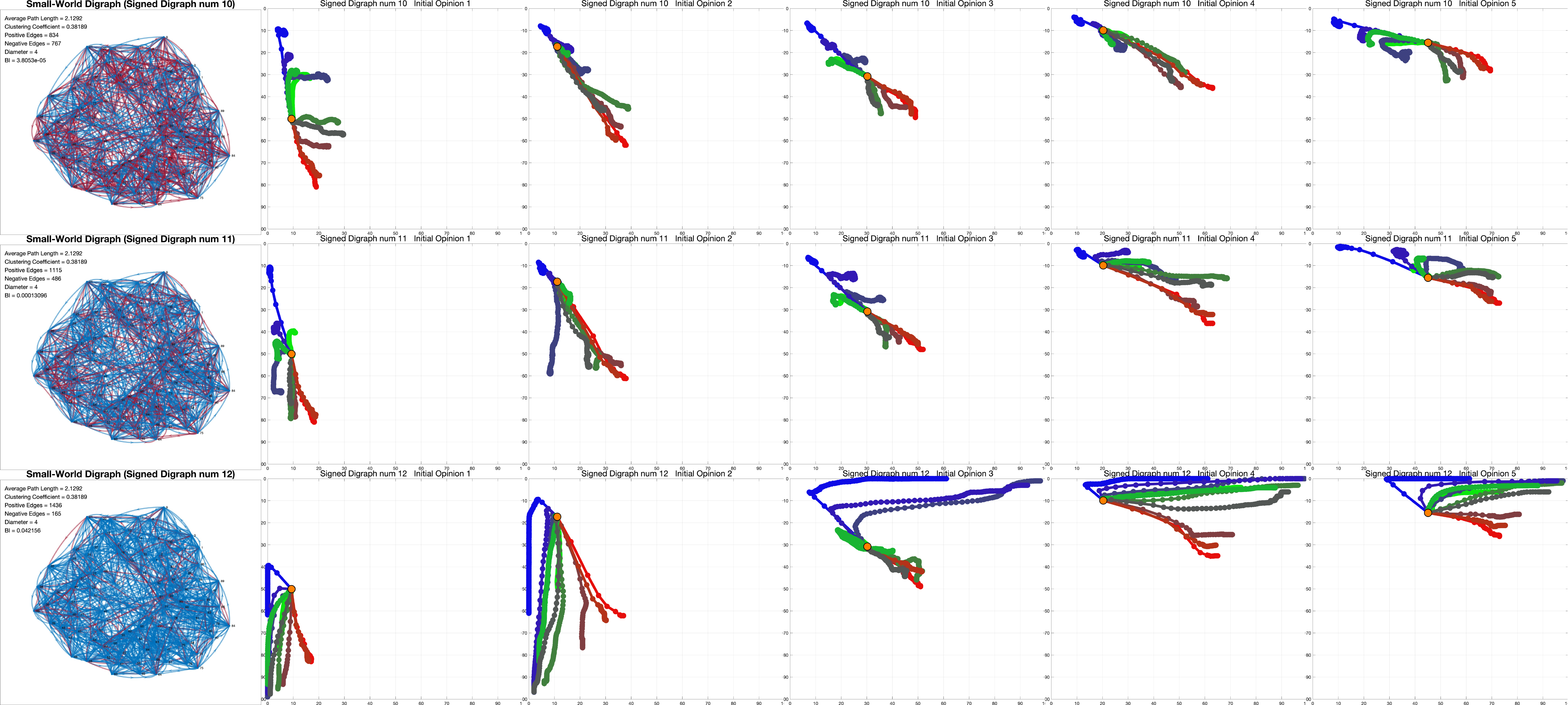}
         \caption{ \CAbnew{Evolution over a Small-World network topology. }}
\label{Fig:EvSW}
     \end{subfigure}
\caption{ \CAbnew{Each agreement plot shows 10 different opinion evolutions starting from the same initial opinions (identical for all plots along the same column), over the same digraph (identical for all plots along the same row and shown to the left) and with the inner traits shown in Figure \ref{Fig:TenInnerTraits}. All simulations involve 100 agents and have evolved for 1000 time steps.}} 
\label{Fig:EvMultipleTop}
\end{figure}

 \CAbnew{To provide an overview of the different behaviours that the model can produce, in relation to different initial opinions, signed digraphs, and inner traits assignations, Figure \ref{Fig:EvMultipleTop} shows the agreement plot of several opinion evolutions for complete, lattice, ring, and small-world graph topologies. In each panel, all the signed digraphs have the same topology, but the ratio of negative to positive edges changes from highest (row 1) to lowest (row 3). Simulations along the same row evolve over the same signed digraph, which is shown to the left together with digraph metrics. The simulations along the same column have the same initial opinion (which are the same as the initial opinion shown in Figure \ref{Fig:AgPlotExpl}). Each agreement plot contains 10 different opinion evolutions, each starting from the same initial opinions (given by the column), evolving over the same digraph (given by the row) and with the inner traits shown in Figure \ref{Fig:TenInnerTraits} (for every line the corresponding average inner traits are represented by the line colour; the average weight of conformist, radical, and stubborn traits are represented by blue, red, and green colours respectively).}
  

 \CAbnew{Figure \ref{Fig:EvCG} shows that increasing the ratio of positive to negative edges, corresponding to less antagonism and more balance, allows the opinions to be clearly expressed and reinforced, and thus ``propagate'' further, often reaching the top right or bottom left corners of the plot. To understand why this happens, think of a population of two agents, $a$ and $b$: if $a$ antagonises $b$ but $b$ does not antagonise $a$, then $a$ will tend to the opposite opinion than $b$ has, while $b$ will tend to the same opinion as $a$, and hence both their opinions converge to zero. An extreme example of how a decrease in this ratio makes opinions weaker (in fact, converge to zero) is the well-known model by \cite{Altafini2013Consensus}; the same phenomena is also present in the CB model, although less pronounced. Analogous trends can be seen irrespective of the graph topology.}
\CAbnew{Concerning the effect of inner traits assignations, the blue lines (prevalence of conformism) tend towards the top right and bottom left corners, indicating consensus with an extreme opinion, whereas the red lines (prevalence of radicalism) tend towards the bottom right corner, associated with polarization. }
%
%
%
%

\CAbnew{Comparing Figure \ref{Fig:EvLD}  with Figure \ref{Fig:EvCG} shows that, with a lattice topology, the opinions tend to be less extreme than with a complete graph topology: in fact, with a lattice graph, opinions take more time to propagate from one agent to the others, and become smaller in magnitude in the process. This also makes the effect of a different positive to negative edge ratio less prominent. However, it is still present: for instance, in the third column of Figure \ref{Fig:EvLD}, the blue line goes from moving to the origin, in the second row, to moving towards the bottom left corner, in the third row.}
%
%
\CAbnew{The results in Figures \ref{Fig:EvLD} and \ref{Fig:EvRD} look similar, but the evolution over the ring digraph seems to leave the opinions closer to the initial configuration for all the different inner traits assignations, perhaps because opinions take even more time to propagate on average due to less interaction among agents, and hence less opportunity to change the opinions of others.}
%
%
\CAbnew{Finally, simulations in Figure \ref{Fig:EvSW} show opinions that propagate considerably in the Cartesian plane, because of the higher connectivity of the Small-World topology (although the same effect is even more pronounced for opinions evolving over a complete graph). }


\begin{figure}[!h]
\centering
     \begin{subfigure}[t]{0.49\textwidth}
         \centering
\includegraphics[width=\textwidth]{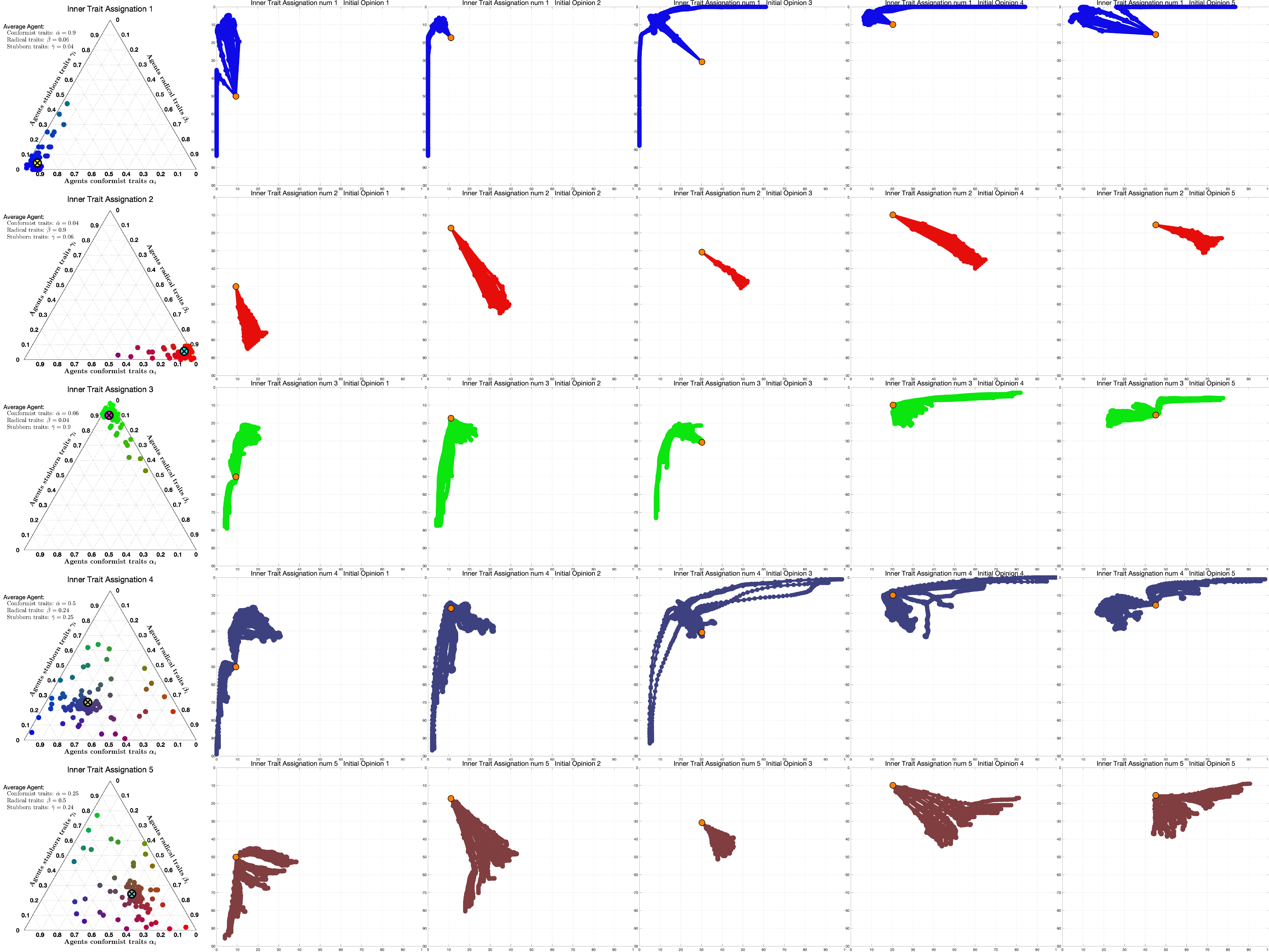}
     \end{subfigure}
     \hfill
     \begin{subfigure}[t]{0.49\textwidth}
         \centering
         \includegraphics[width=\textwidth]{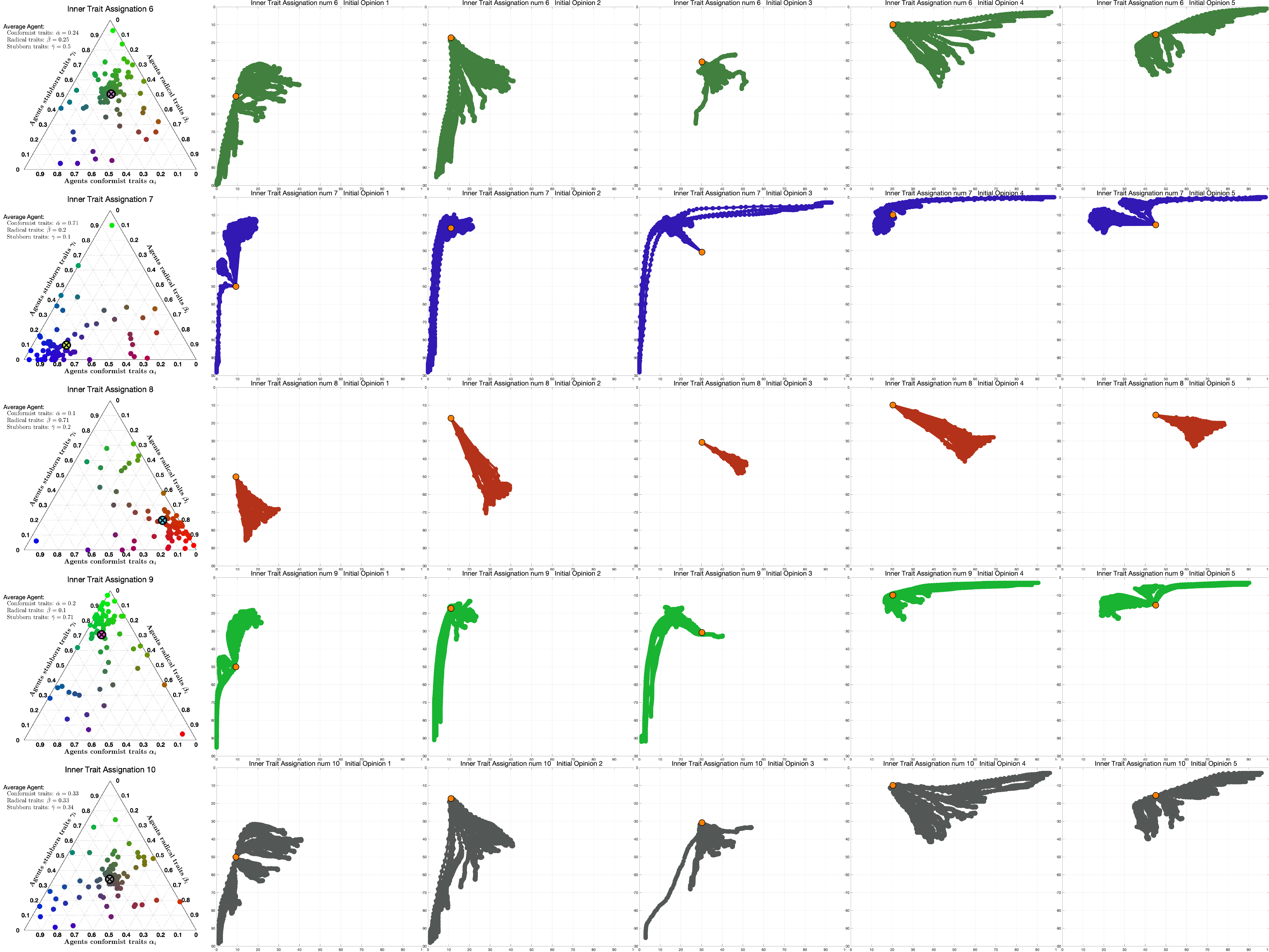}
              \end{subfigure}
\caption{ \CAbnew{Each agreement plot along the same column (respectively, row) corresponds to an opinion evolution with the same initial opinion (respectively, the same inner traits assignation). The corresponding inner traits assignation is shown to the left. Each agreement plot contains 12 curves, one for each of the 12 signed digraphs in Figure \ref{Fig:EvMultipleTop}.}}
\label{Fig:EvInnerTraits}
\end{figure}

 \CAbnew{Figure \ref{Fig:EvInnerTraits} shows the same opinion evolutions as in Figure \ref{Fig:EvMultipleTop}, but now the agreement plots are grouped by initial opinions and inner traits assignations. Evolutions shown in the same row have the inner traits shown to the left; the initial opinions are the same for all simulations along the same column; and each plot contains 12 lines corresponding to the 12 signed digraphs in Figure \ref{Fig:EvMultipleTop}. In addition to the effect of the network topology, each inner traits assignation leads to a characteristic behaviour for the opinion evolution. Highly conformist inner traits assignations tend to move towards the axis: most opinions are either positive or negative. On the contrary, predominantly radical inner traits assignations tend to move towards the bottom right corner, associated with polarisation, or at least with the presence of significant amount of agents with both positive and negative opinions. Inner traits assignations with a strong stubborn component can have either of the two behaviours. More heterogeneous inner traits assignations give rise to a wider variety of behaviours. }

\subsection{Parameter Sensitivity Analysis}

 \CAbnew{We select a set of \emph{nominal parameters} (which, for given initial conditions, produce \emph{nominal simulation results}) as a baseline with which other parameter choices can be compared. We choose a nominal inner traits assignation that leads to model outcomes that closely reproduce real data from the World Values Survey (in fact, it is close to some of the inner traits assignations resulting from the \emph{Free} optimisation problem \eqref{Eq:FirstMP}, see Figure \ref{Fig:TD1a}), and therefore has the potential to represent a realistic society; moreover, it allows us to showcase the wide range of different opinion evolutions that the model can produce.}
Then, we vary inner traits assignations, signed digraph and opinion evolution parameters, one by one, and study their effect on the simulated behaviour.

\subsubsection{Nominal Parameters and Nominal Results}

\begin{figure}[h!]
     \centering
     \begin{subfigure}[b]{0.32\textwidth}
         \centering
         \includegraphics[width=\textwidth]{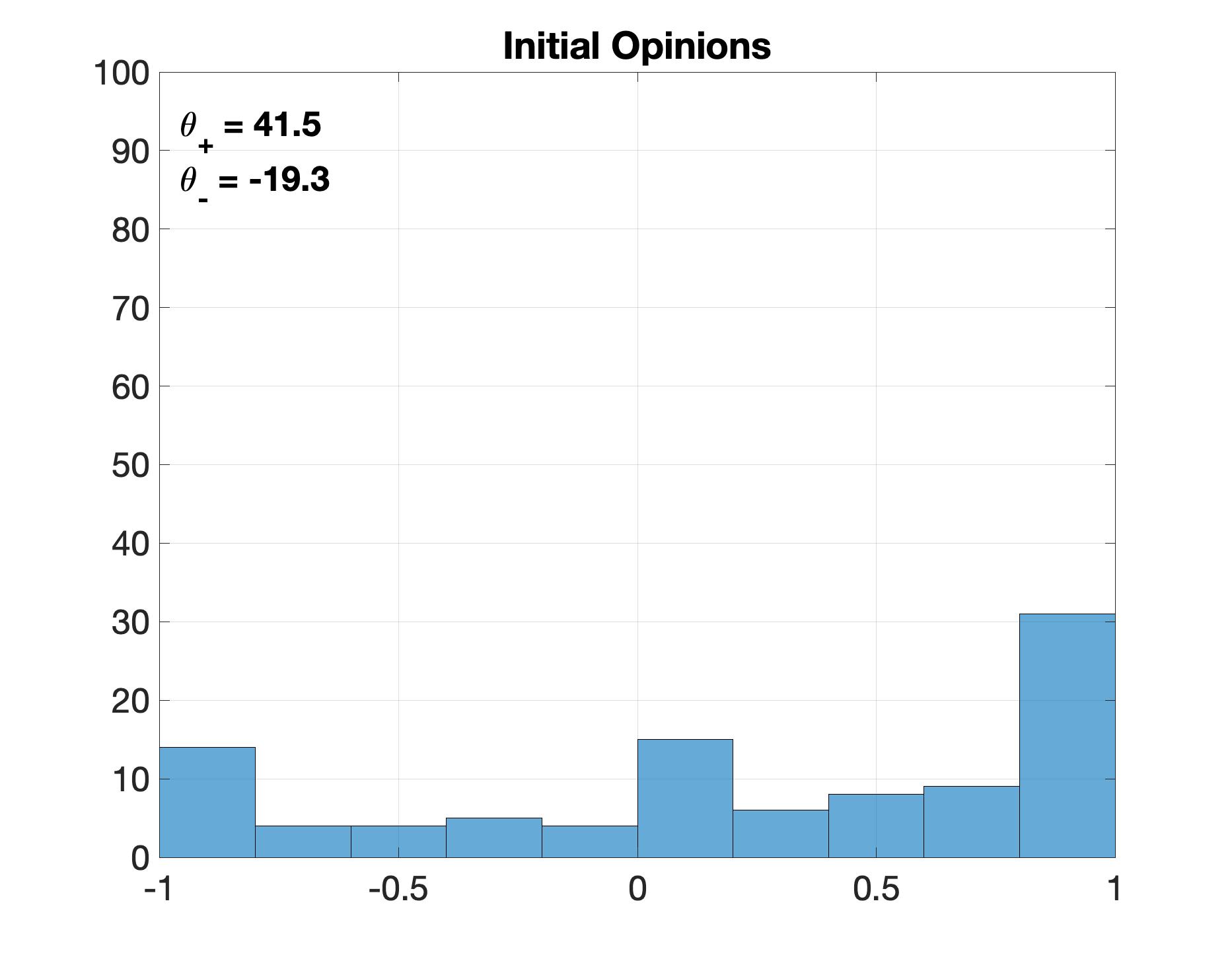}
         \caption{Initial Opinions Histogram}
         \label{SubFig:InitialOpinions}
     \end{subfigure}
     \hfill
     \begin{subfigure}[b]{0.32\textwidth}
         \centering
         \includegraphics[width=\textwidth]{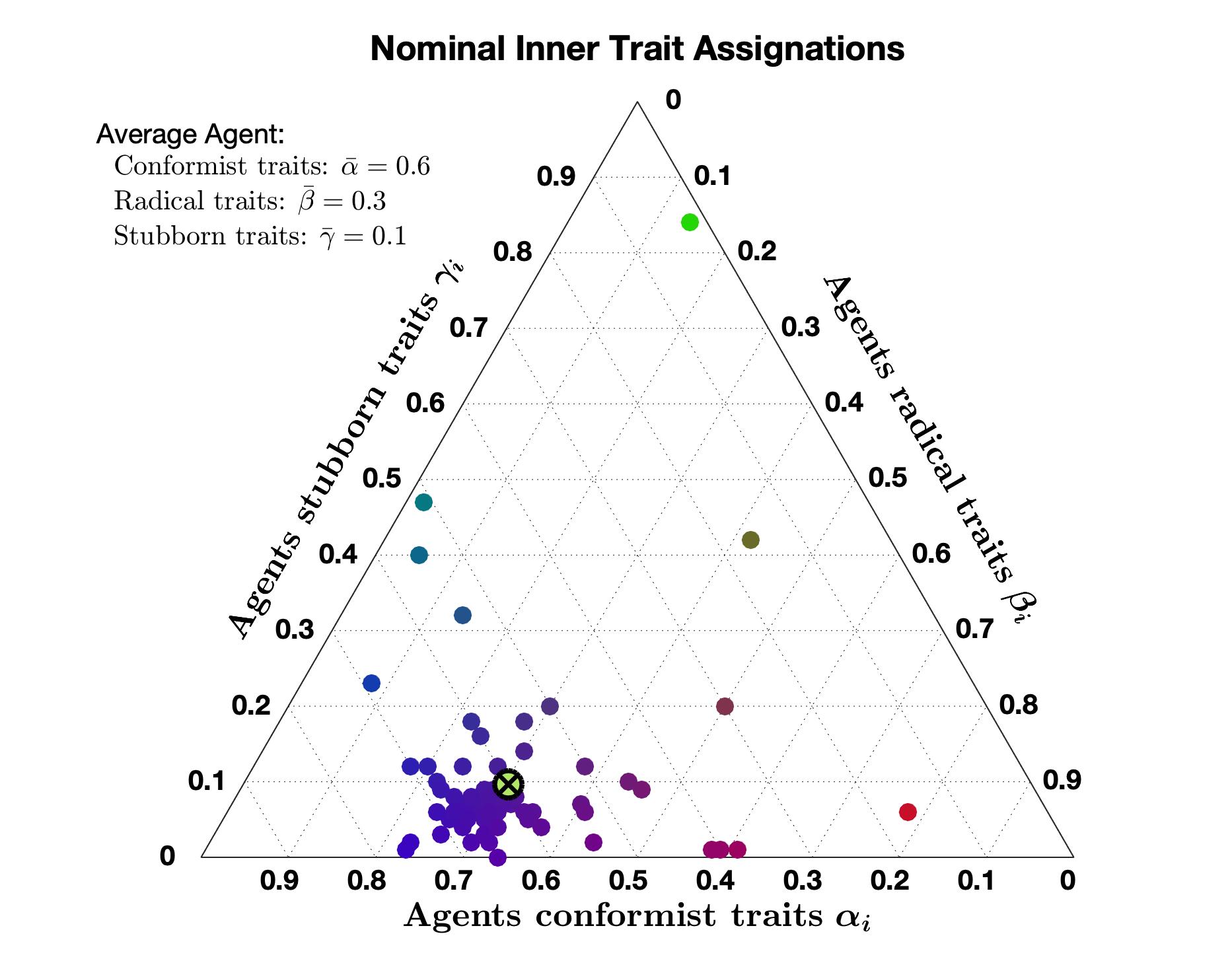}
         \caption{Nominal Inner Traits Assignation}
         \label{SubFig:NomInnerAssignations}
     \end{subfigure}
     \hfill
     \begin{subfigure}[b]{0.32\textwidth}
         \centering
         \includegraphics[width=\textwidth]{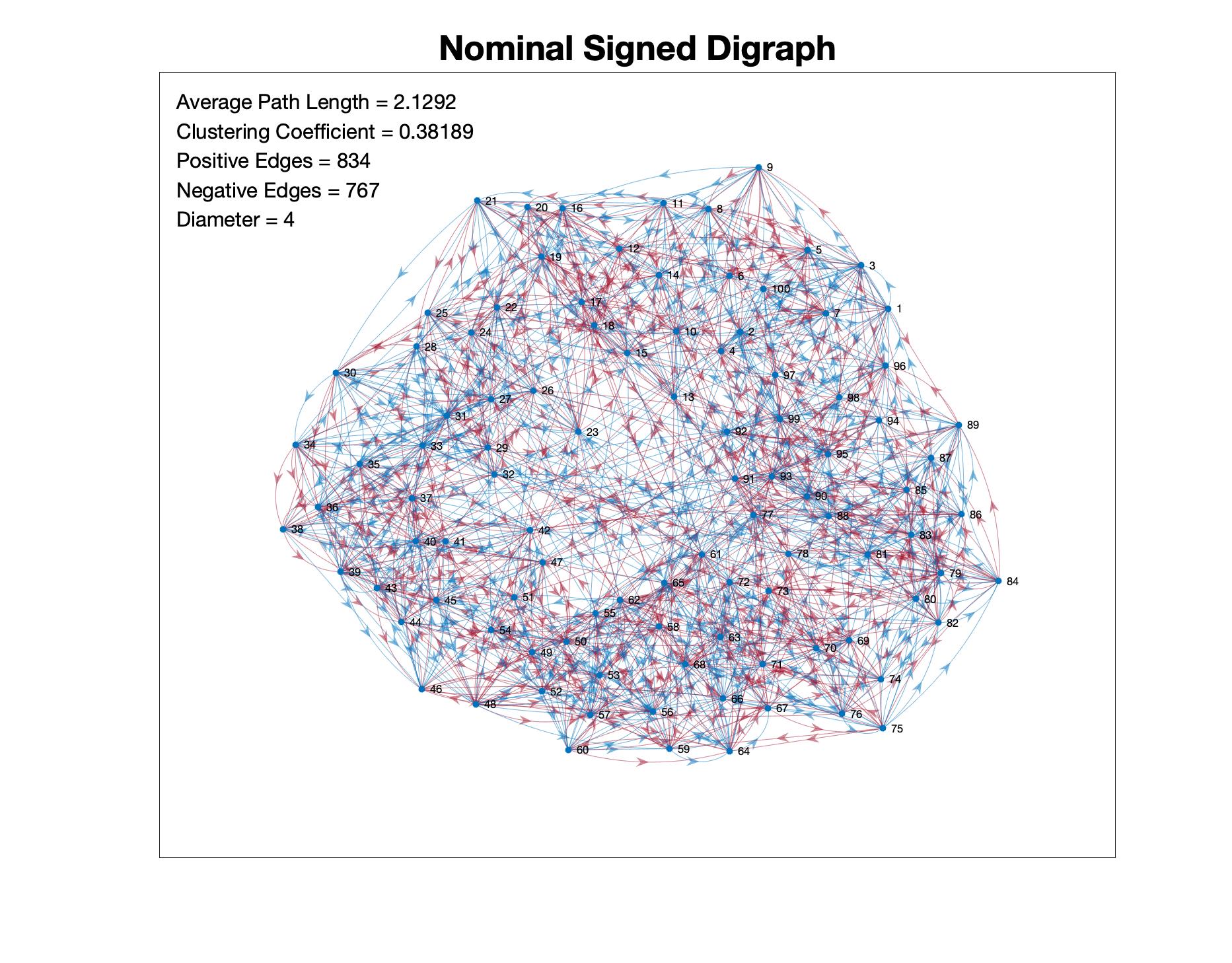}
         \caption{Nominal Signed Digraph}
         \label{SubFig:NomIntNetw}
     \end{subfigure} 
        \caption{Initial opinions and nominal parameters.} 
        \label{Fig:Examples_1}
\end{figure}

We consider the initial opinions shown in Figure \ref{SubFig:InitialOpinions}, which evolve according to the model with the nominal parameters: $\lambda = 0.4$, $\xi= 2$, $\mu = 5$, inner traits assignations in Figure \ref{SubFig:NomInnerAssignations}, and signed digraph in Figure \ref{SubFig:NomIntNetw}.

The initial opinions shown in Figure \ref{SubFig:InitialOpinions} have $\theta_- = -19.3$ and $\theta_+ = 41.5$, indicating a strong general agreement since $\theta_+>-\theta_-$. Figure  \ref{SubFig:NomInnerAssignations} shows that most agents have very strong conformist traits, with a notable percentage of radicalism, resulting in an average agent (crossed dot) with $60\%$ conformist traits, $30\%$ radical traits, and $10\%$ stubborn traits. The nominal signed digraph in Figure \ref{SubFig:NomIntNetw} is highly connected, with average path length $2.12$, clustering coefficient $0.38$, diameter $4$. It has $834$ positive edges and $767$ negative edges. 

\begin{figure}[h!]
     \centering
     \hspace{0.5cm}
     \begin{subfigure}[b]{0.56\textwidth} 
         \centering
         \includegraphics[width=\textwidth]{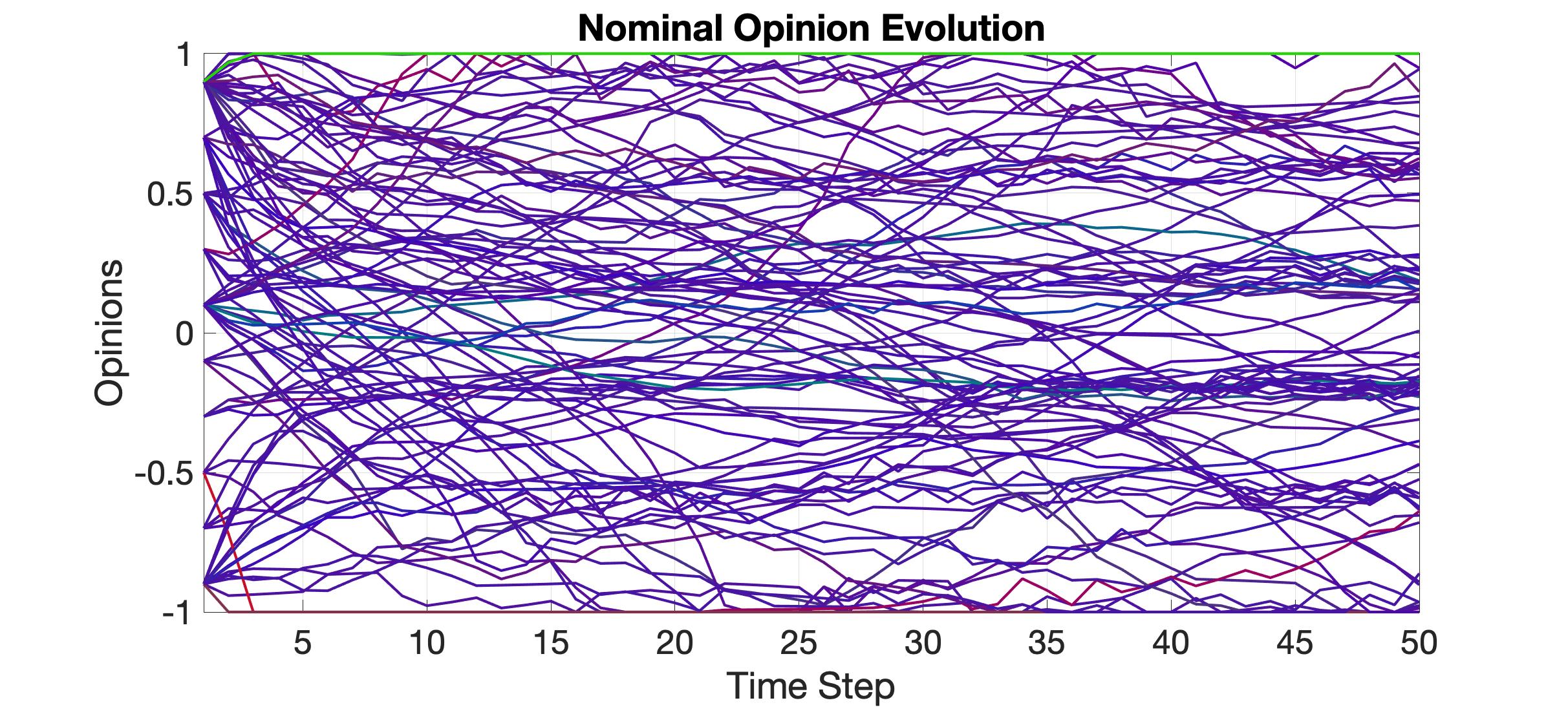}
         \caption{Nominal Opinion Evolution}
         \label{SubFig:NomOpEvo}
     \end{subfigure}
     \hfill
     \begin{subfigure}[b]{0.32\textwidth} 
         \centering
         \includegraphics[width=\textwidth]{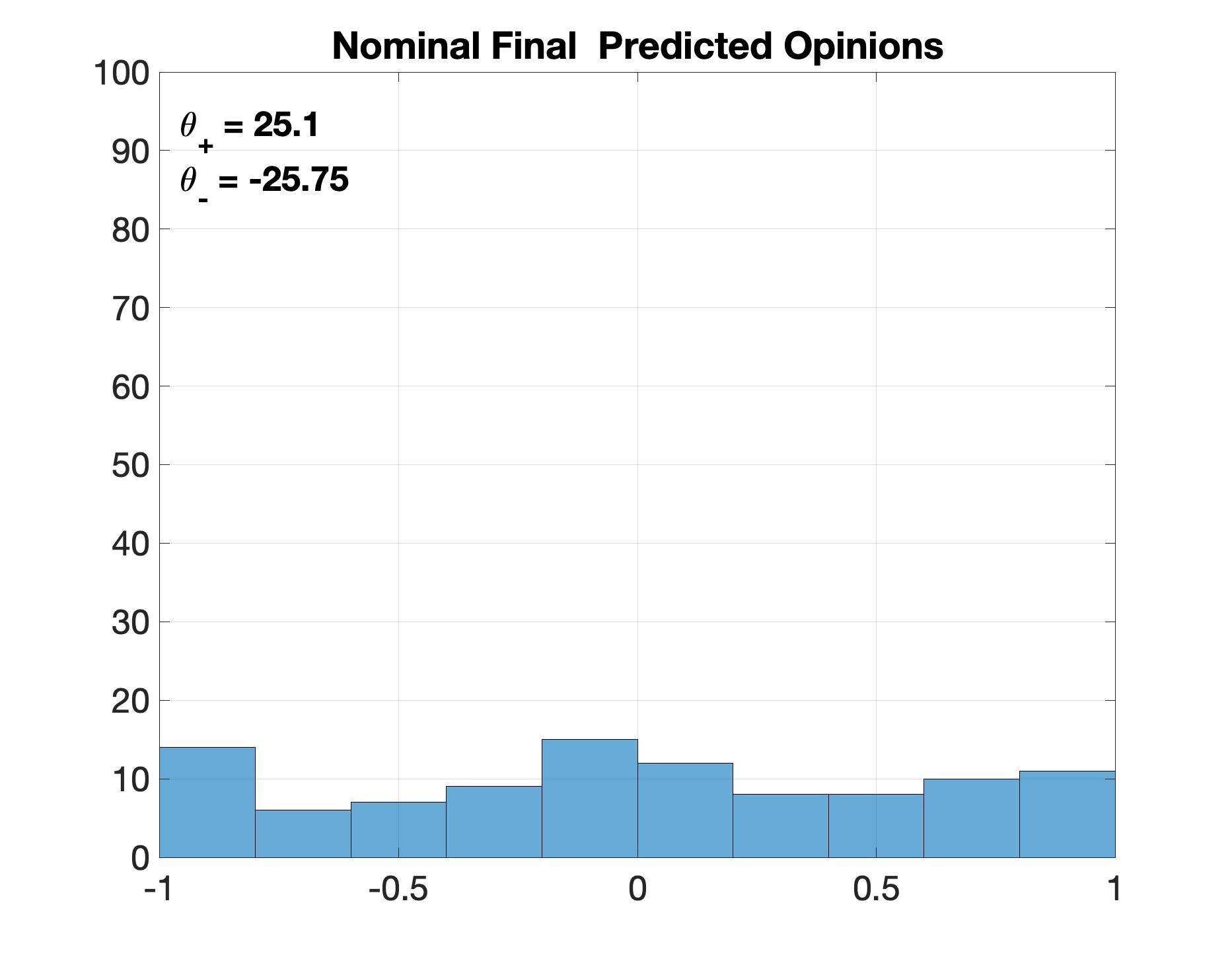}
         \caption{Nominal Final Opinion Histogram}
         \label{SubFig:NomFOH}
     \end{subfigure} \hspace{0.5cm}
        \caption{Simulation results with the nominal parameter values  \CAbnew{(evolving 100 agents)}.} 
        \label{Fig:Examples_NominalResults}
\end{figure}

The nominal results are shown in Figure \ref{Fig:Examples_NominalResults}. Figure \ref{SubFig:NomOpEvo} shows the opinion evolution of every agent. The line colour represents the percentage of conformist, radical, and stubborn agent traits (blue for conformist, red for radical, and green for stubborn). The purple colour of most lines corresponds to a combination of conformist and radical traits. The discontinuity in the opinion change is due to the classification process leading to a discontinuous opinion update law. The opinion evolution of the various agents shows a great variability in opinion changes, without a clear global tendency. 

Figure \ref{SubFig:NomFOH} shows the histogram of the nominal final opinions predicted by the model after 50 time steps. Compared with the initial opinions, the final opinions appear to have a more uniform distribution: in fact, for the nominal final opinions, $\theta_- = -25.75$ and $\theta_+ = 25.1$, hence $\theta_+ \approx -\theta_-$. The behaviour of the opinion evolution and the distribution of the final opinions is explained by the presence of two opposing forces that drive the opinion of all the agents: on one hand, the tendency to achieve consensus, due to the conformist traits, drives the agents towards the centre; on the other hand, the radical traits move the opinions towards extreme values.

\subsubsection{Varying the Inner Traits Assignations}

\begin{figure}[h!]
     \centering
     \begin{subfigure}[m]{0.27\textwidth}
         \centering
         \includegraphics[width=\textwidth]{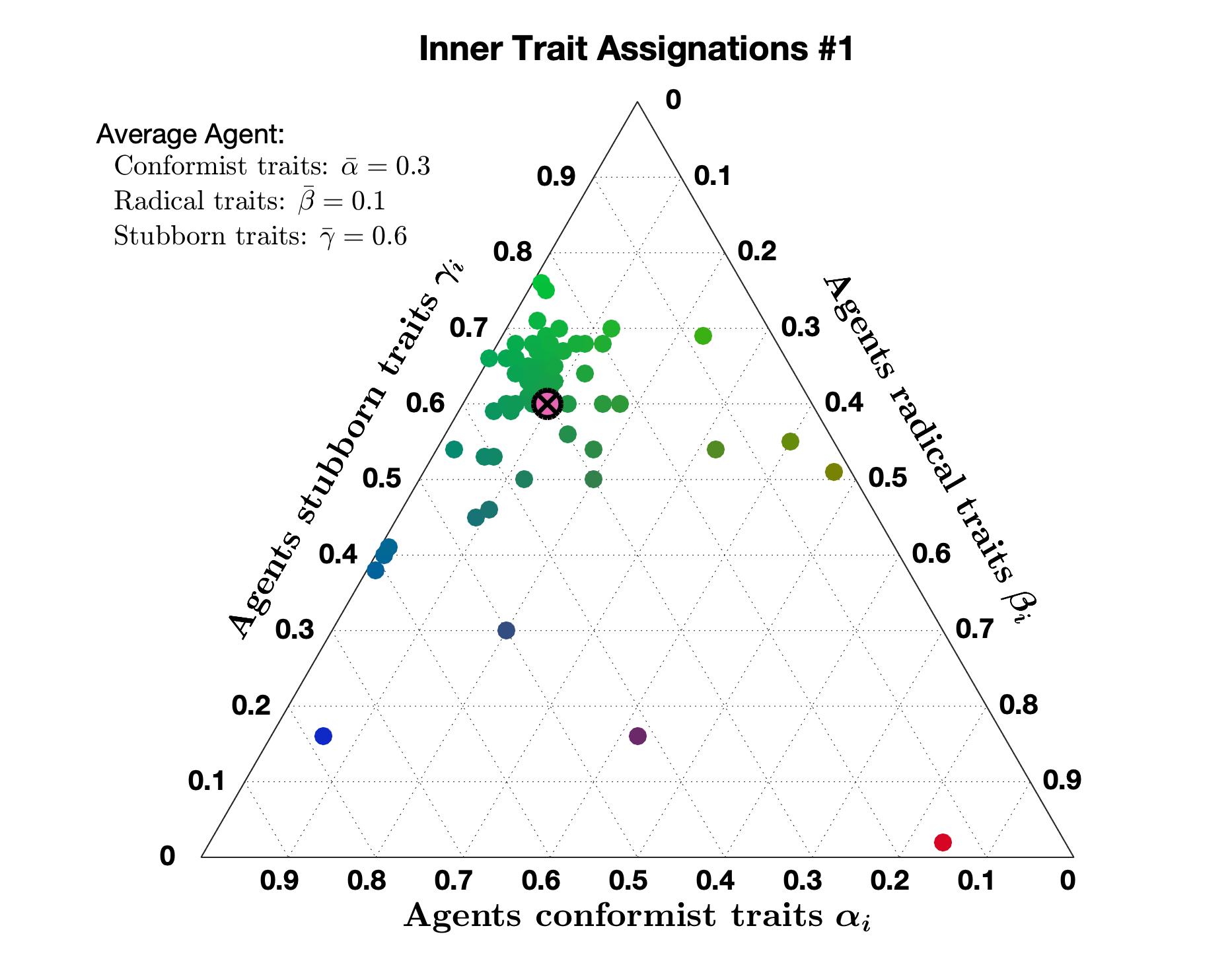}
         \caption{Inner traits assignations 1}
         \label{SugFig:InAs1}
     \end{subfigure}
     \hfill
     \begin{subfigure}[m]{0.42\textwidth}
         \centering
         \includegraphics[width=\textwidth]{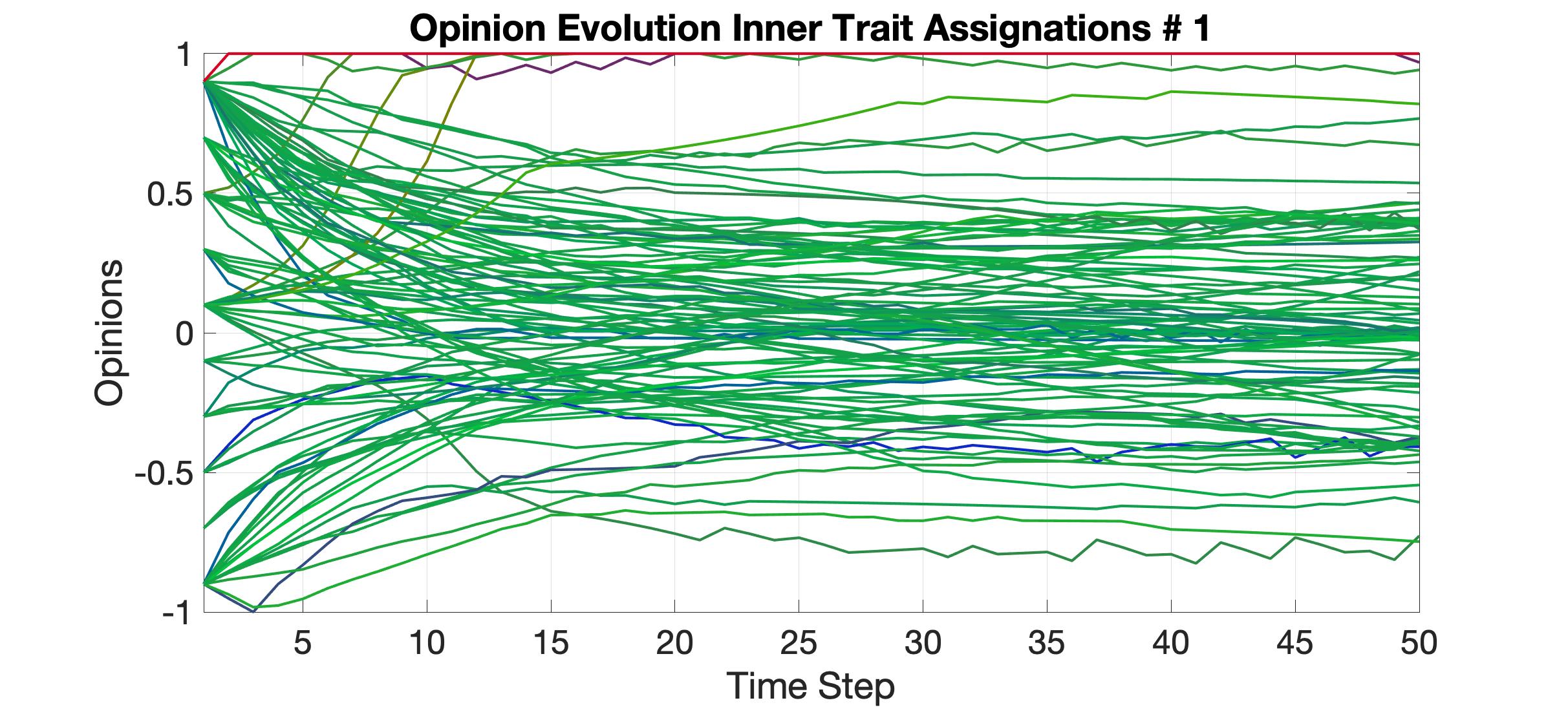}
         \caption{Opinion evolution inner traits assignations 1}
         \label{SugFig:OpEvInAs1}
     \end{subfigure}
     \hfill
     \begin{subfigure}[m]{0.27\textwidth}
         \centering
         \includegraphics[width=\textwidth]{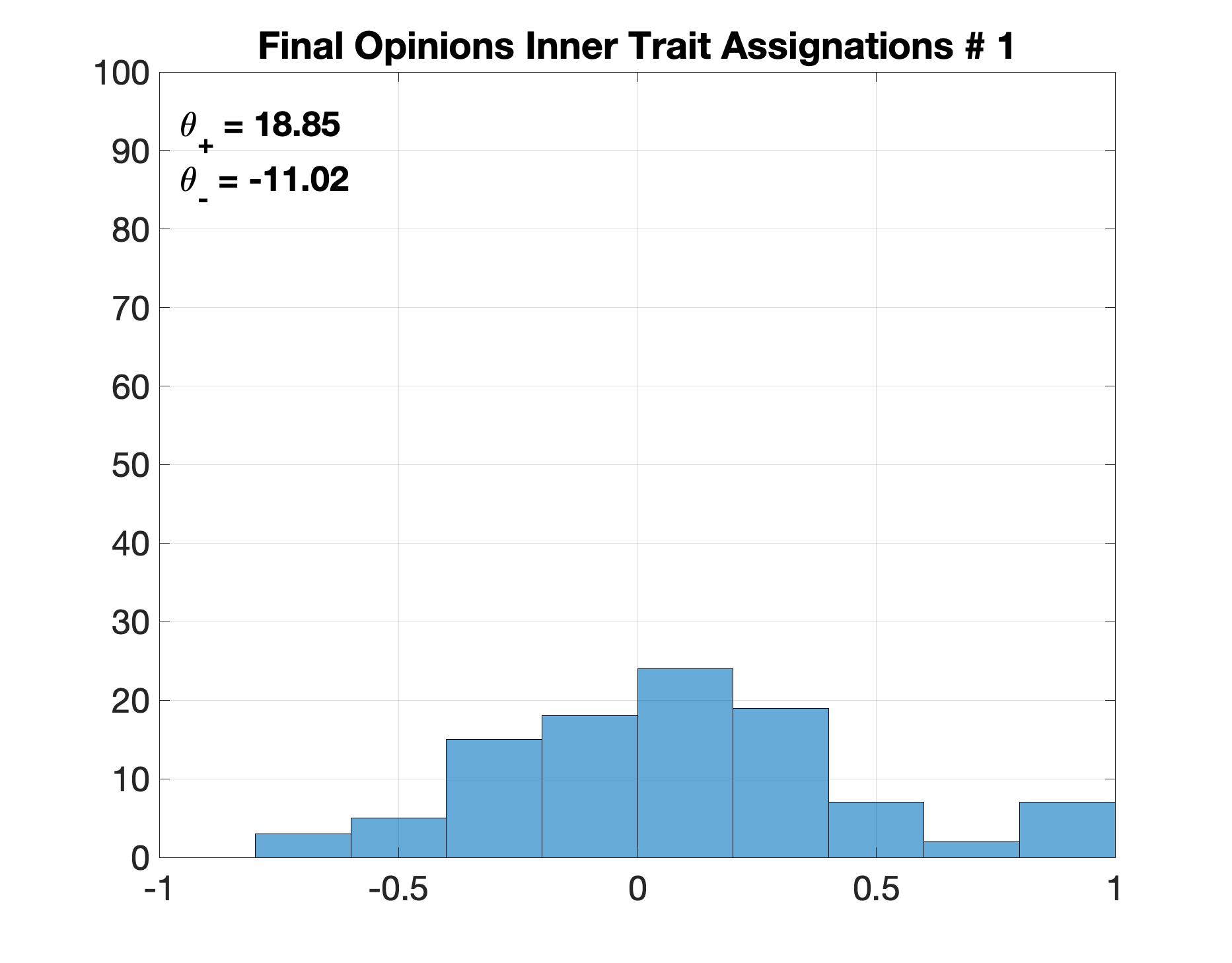}
         \caption{Final opinion histogram}
         \label{SugFig:OpHiInAs1}
     \end{subfigure} \\
     \begin{subfigure}[m]{0.27\textwidth}
         \centering
         \includegraphics[width=\textwidth]{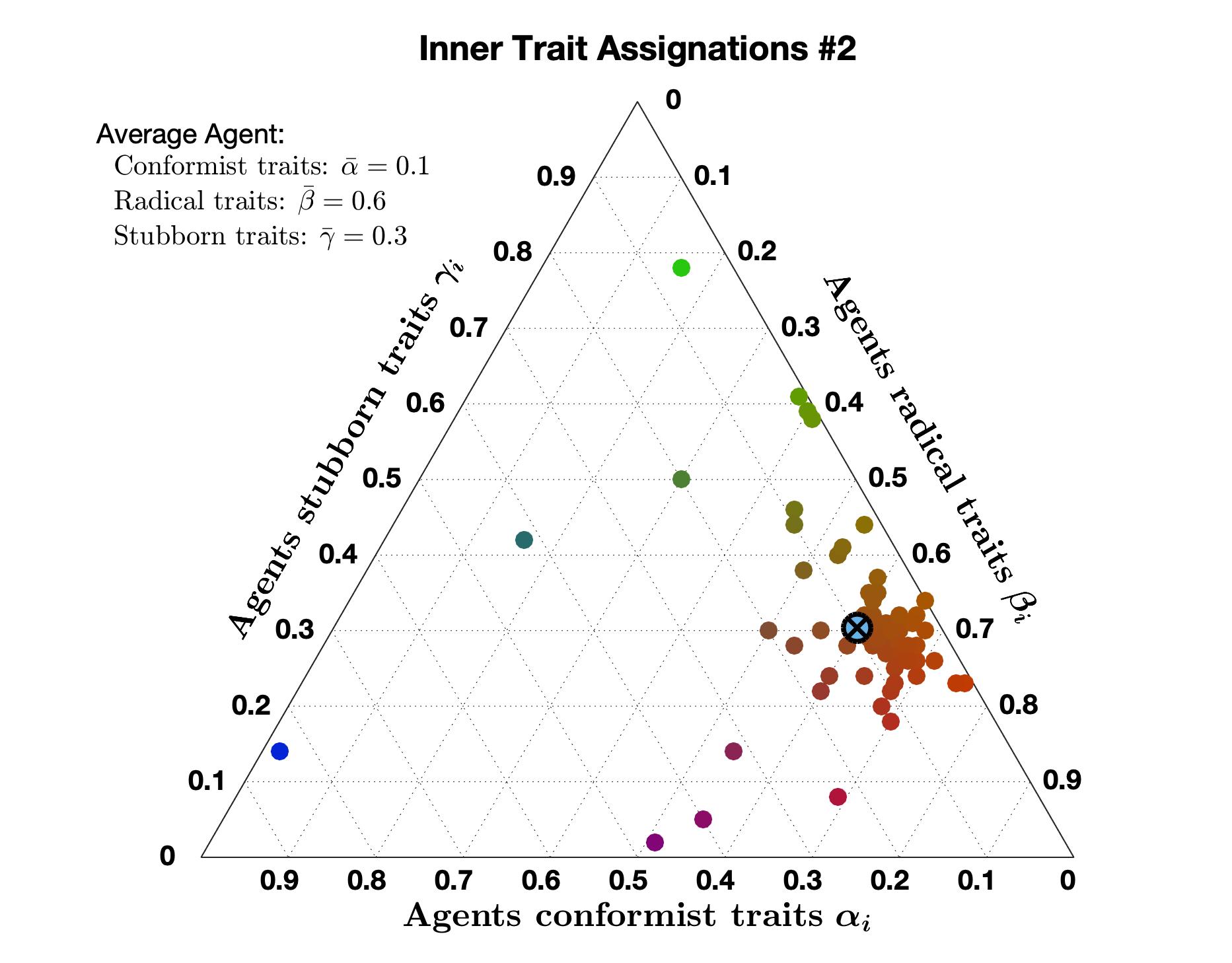}
         \caption{Inner traits assignations 2}
         \label{SugFig:InAs2}
     \end{subfigure}
     \hfill
     \begin{subfigure}[m]{0.42\textwidth}
         \centering
         \includegraphics[width=\textwidth]{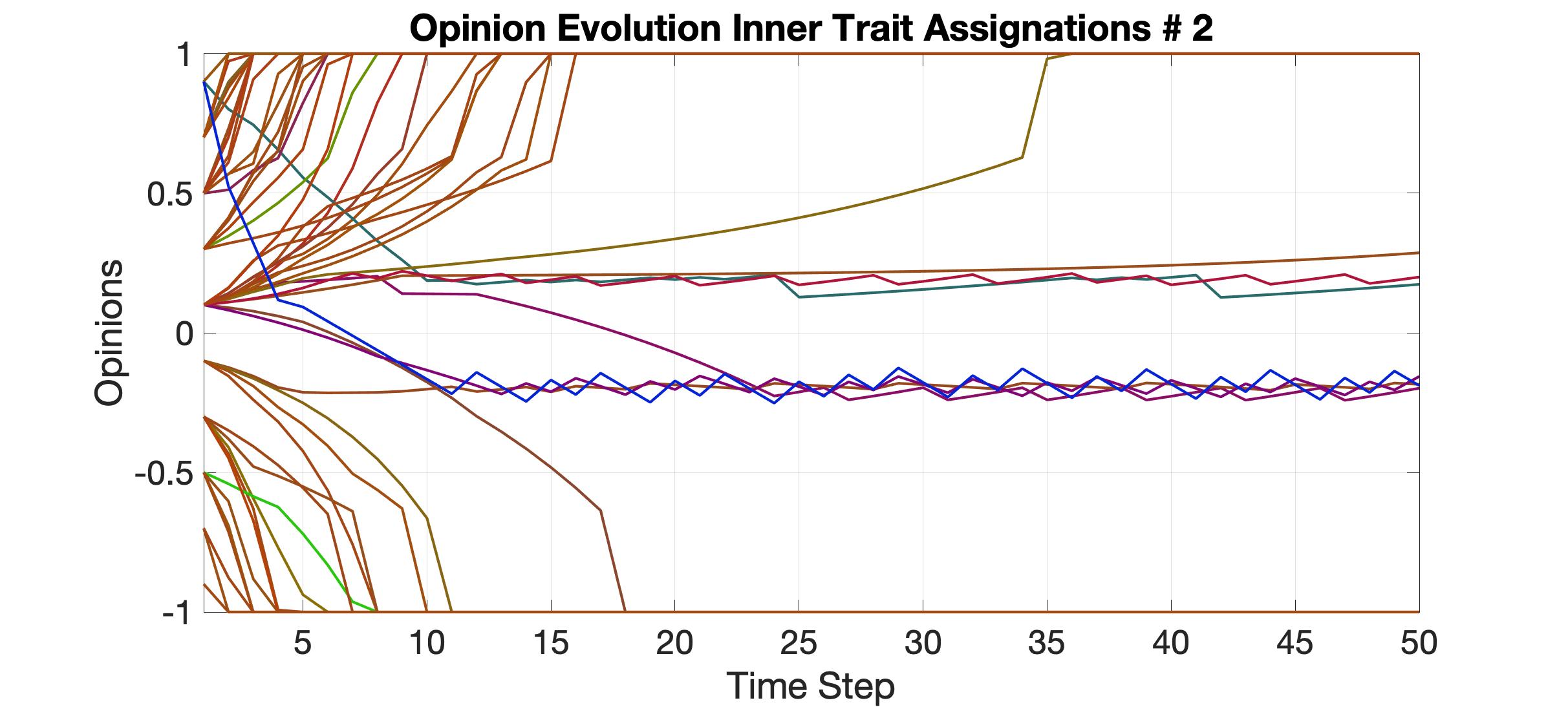}
         \caption{Opinion evolution inner traits assignations 2}
         \label{SugFig:OpEvInAs2}
     \end{subfigure}
     \hfill
     \begin{subfigure}[m]{0.27\textwidth}
         \centering
         \includegraphics[width=\textwidth]{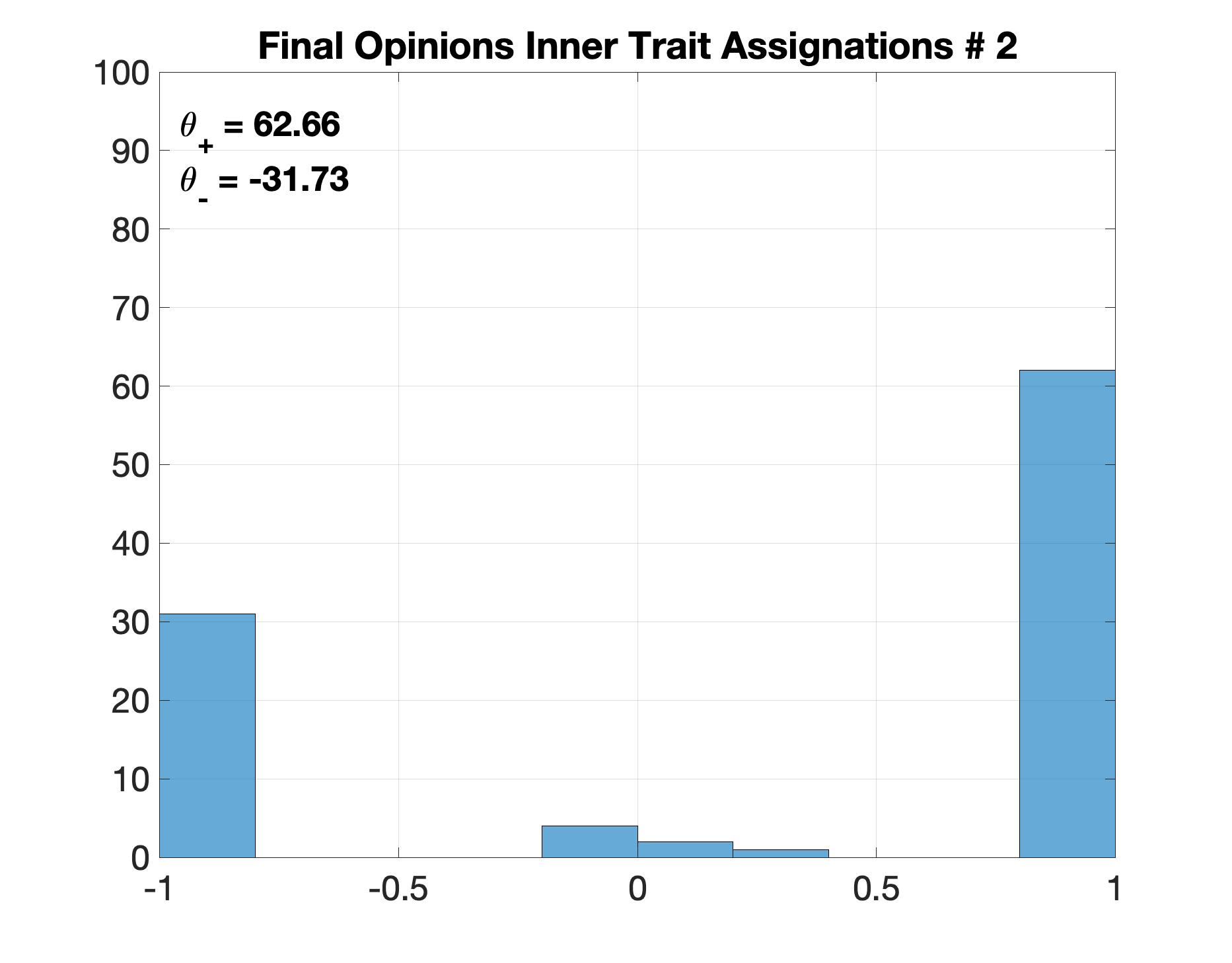}
         \caption{Final opinion histogram}
         \label{SugFig:OpHiInAs2}
     \end{subfigure} 
        \caption{Effect of changing the inner traits assignations  \CAbnew{(evolving 100 agents)}.} 
        \label{Fig:Examples_1_new}
\end{figure}

To evaluate the effect of different inner traits assignations, we change the nominal inner traits assignations of Figure \ref{SubFig:NomInnerAssignations} and simulate the opinion evolution, keeping all the other parameters unchanged. The two new inner traits assignations, shown in Figures \ref{SugFig:InAs1} and \ref{SugFig:InAs2}, are simply rotations of the nominal inner traits assignations. The corresponding opinion evolutions are shown in Figures \ref{SugFig:OpEvInAs1} and \ref{SugFig:OpEvInAs2}, while the final opinion histograms are presented in Figures \ref{SugFig:OpHiInAs1} and \ref{SugFig:OpHiInAs2}.

Comparing the opinion evolutions of Figures \ref{SubFig:NomOpEvo}, \ref{SugFig:OpEvInAs1}, and \ref{SugFig:OpEvInAs2} and the final opinion histograms of Figures \ref{SubFig:NomFOH}, \ref{SugFig:OpHiInAs1}, and \ref{SugFig:OpHiInAs2} reveals the profound effect of different inner traits assignations on the opinion evolution. In the inner traits assignation of Figure \ref{SugFig:InAs1}, the agents are mostly stubborn and conformist. This results in a very slow convergence towards the mean, spurred by conformist traits and slowed down by stubborn traits. Because of the neighbour classification, even completely conformist agents would not reach perfect consensus, but would rather converge to an opinion subinterval where all the agents perceive that the others have a comparable opinion. This tendency towards the mean can be seen in the final opinion histogram of Figure \ref{SugFig:OpHiInAs1}, where both $\theta_- = -11.02$ and $\theta_+ = 18.85$ are much closer to 0.

On the other hand, the inner traits assignation of Figure \ref{SugFig:InAs2} gives agents pronounced radical traits. Both the opinion evolution in Figure \ref{SugFig:OpEvInAs2} and the final opinion histogram in Figure \ref{SugFig:OpHiInAs2} show that agents lean towards extreme opinions. A bunch of agents keeps its opinion closer to zero. The line colours (closer to blue and green) show that these agents do not have very strong radical traits, and instead they are more conformist and stubborn: such traits allow these agents to avoid extreme opinions.

\subsubsection{Varying the Signed Digraph}

\begin{figure}[h!]
     \centering
     \begin{subfigure}[m]{0.27\textwidth}
         \centering
         \includegraphics[width=\textwidth]{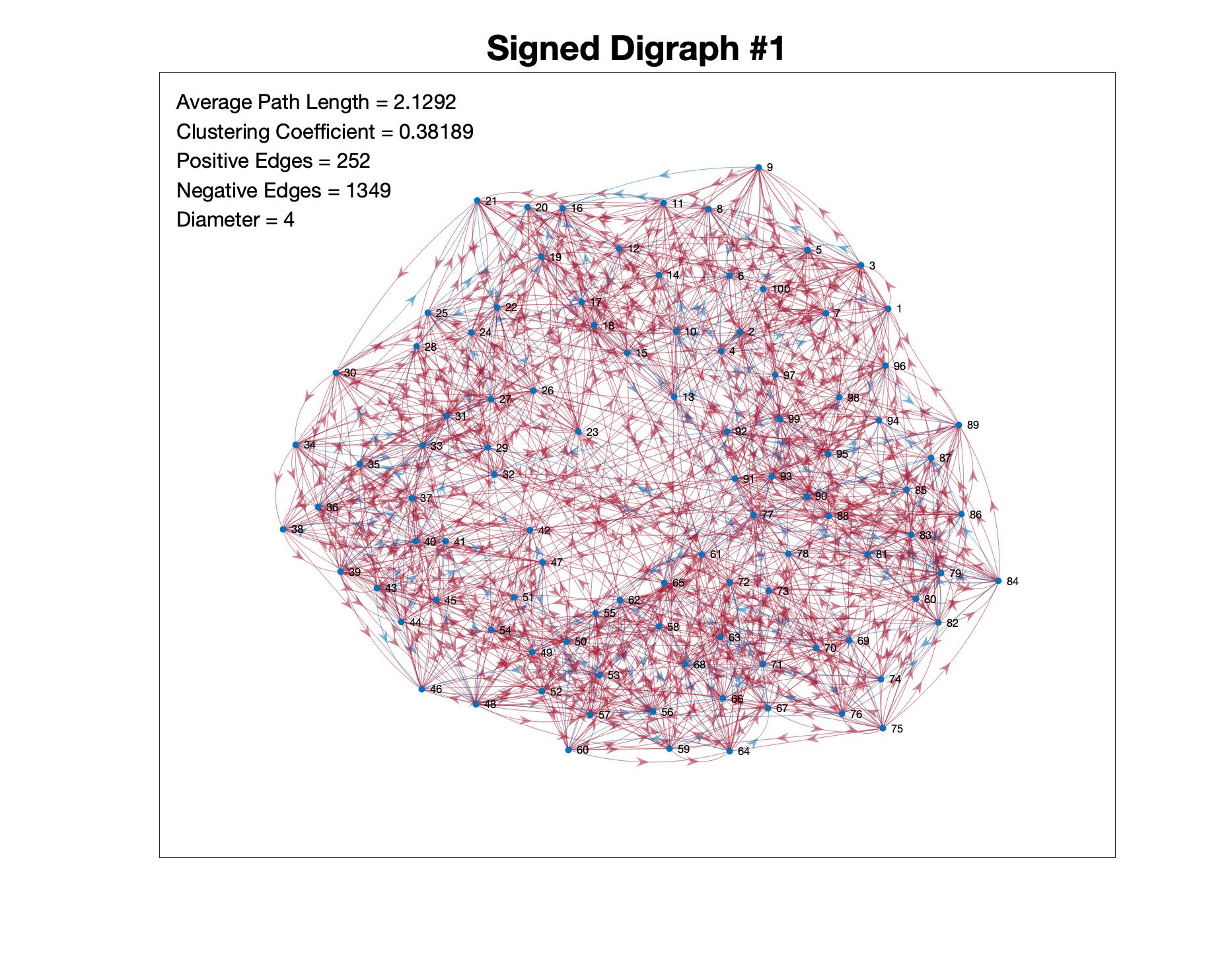}
         \caption{Signed digraph 1}
         \label{SubFig:IntNet1}
     \end{subfigure}
     \hfill
     \begin{subfigure}[m]{0.42\textwidth}
         \centering
         \includegraphics[width=\textwidth]{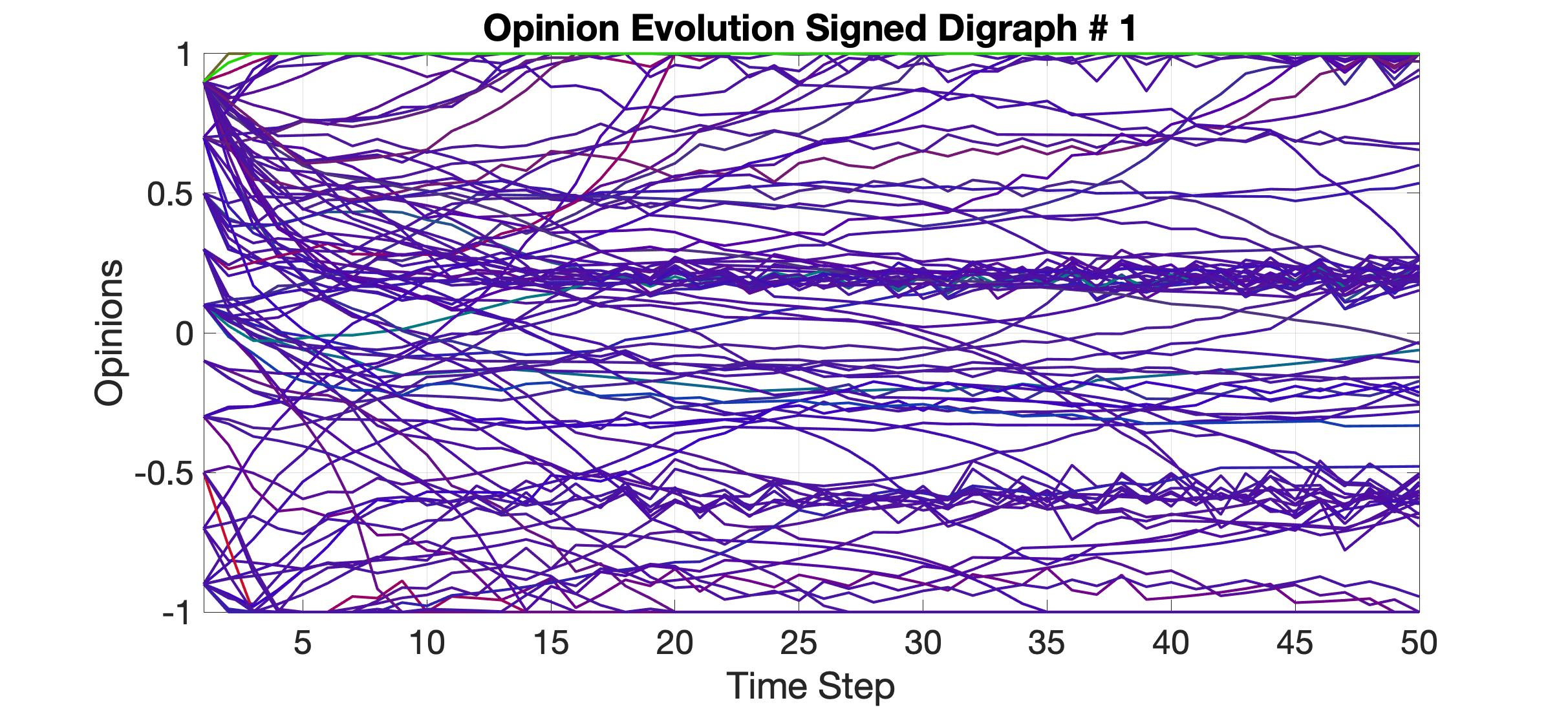}
         \caption{Opinion evolution signed digraph 1}
         \label{SugFig:OpEvInNe1}
     \end{subfigure}
     \hfill
     \begin{subfigure}[m]{0.27\textwidth}
         \centering
         \includegraphics[width=\textwidth]{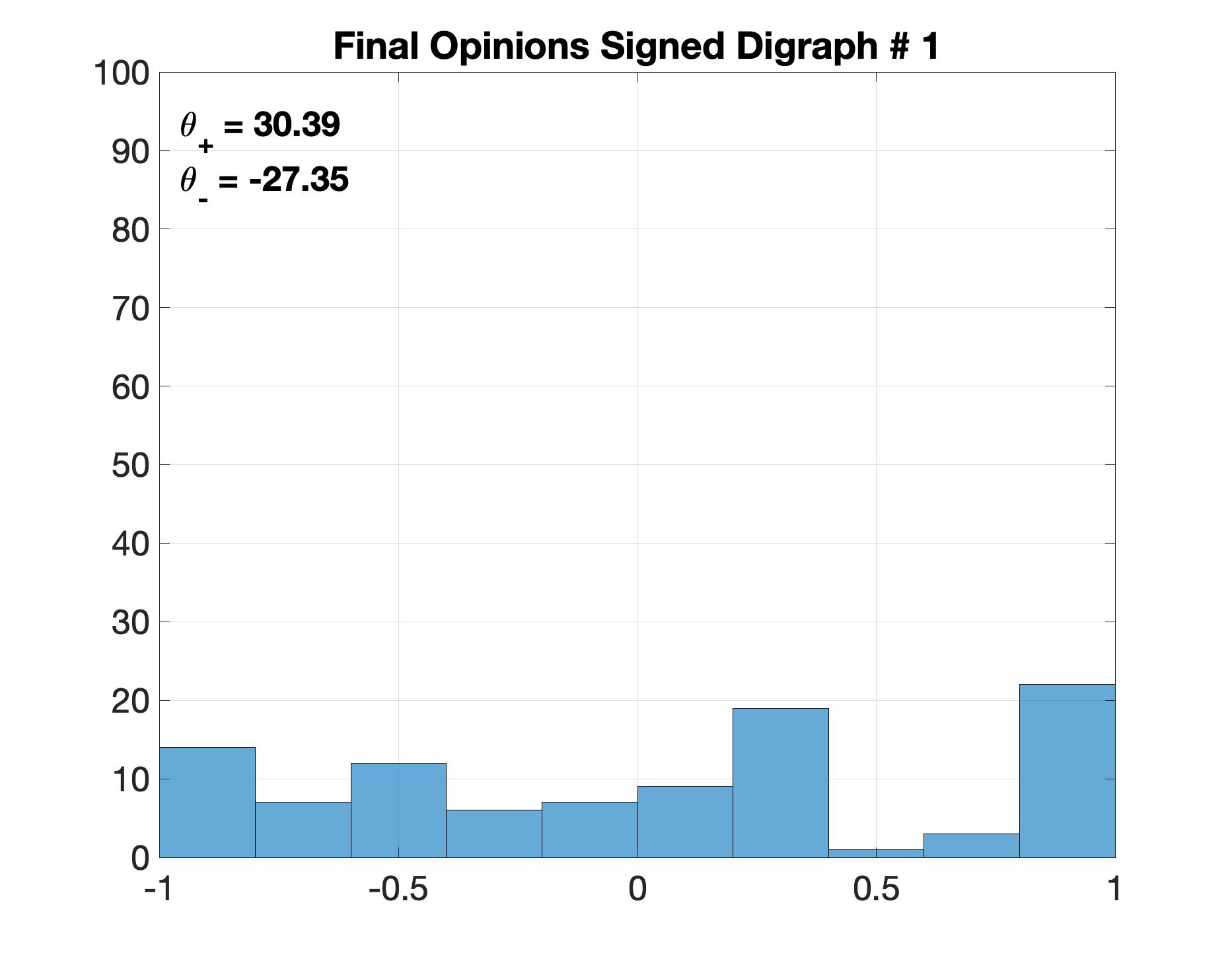}
         \caption{Final opinion histogram}
         \label{SugFig:OpHiInNe1}
     \end{subfigure} \\
     \begin{subfigure}[m]{0.27\textwidth}
         \centering
         \includegraphics[width=\textwidth]{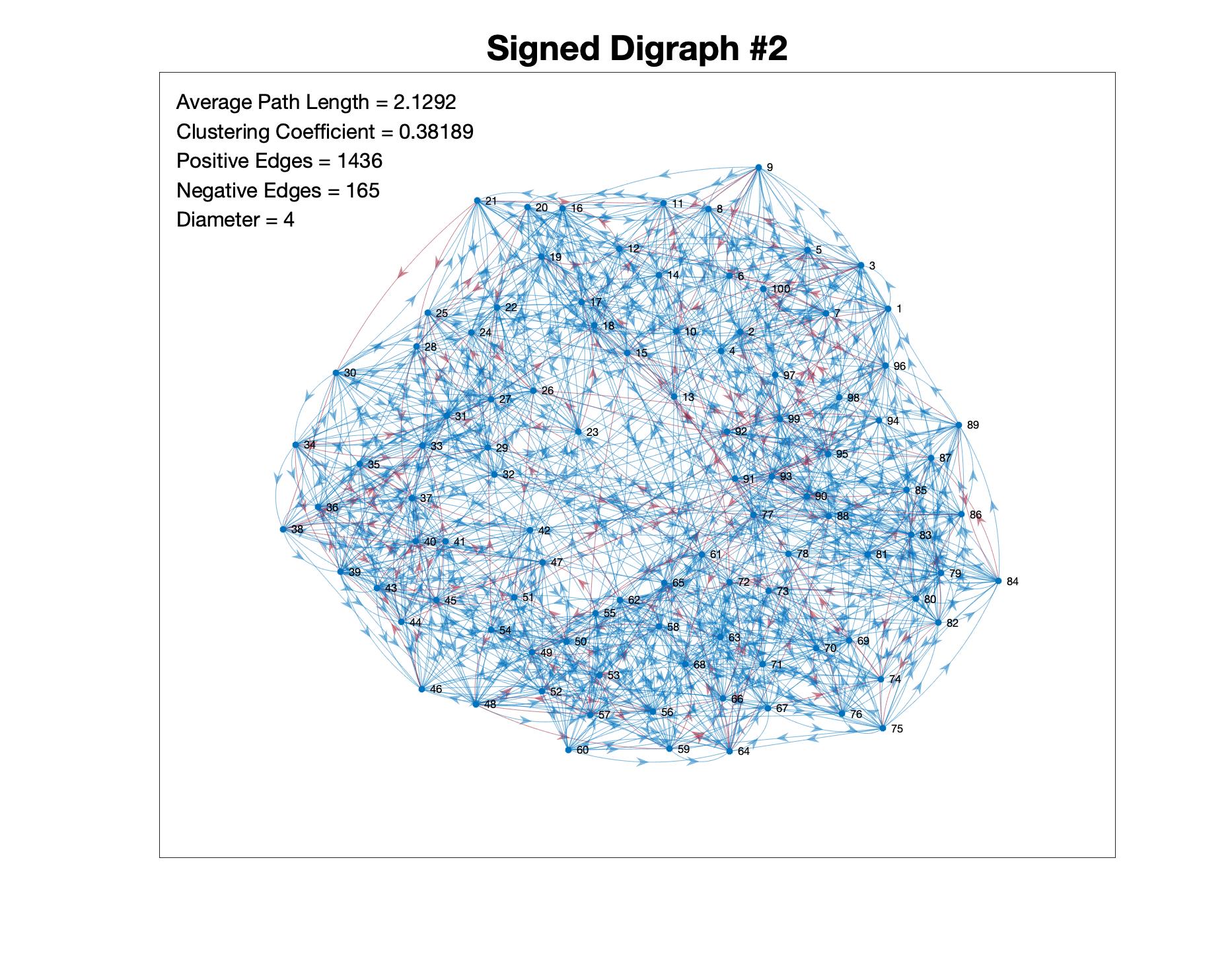}
         \caption{Signed digraph 2}
         \label{SubFig:IntNet2}
     \end{subfigure}
     \hfill
     \begin{subfigure}[m]{0.42\textwidth}
         \centering
         \includegraphics[width=\textwidth]{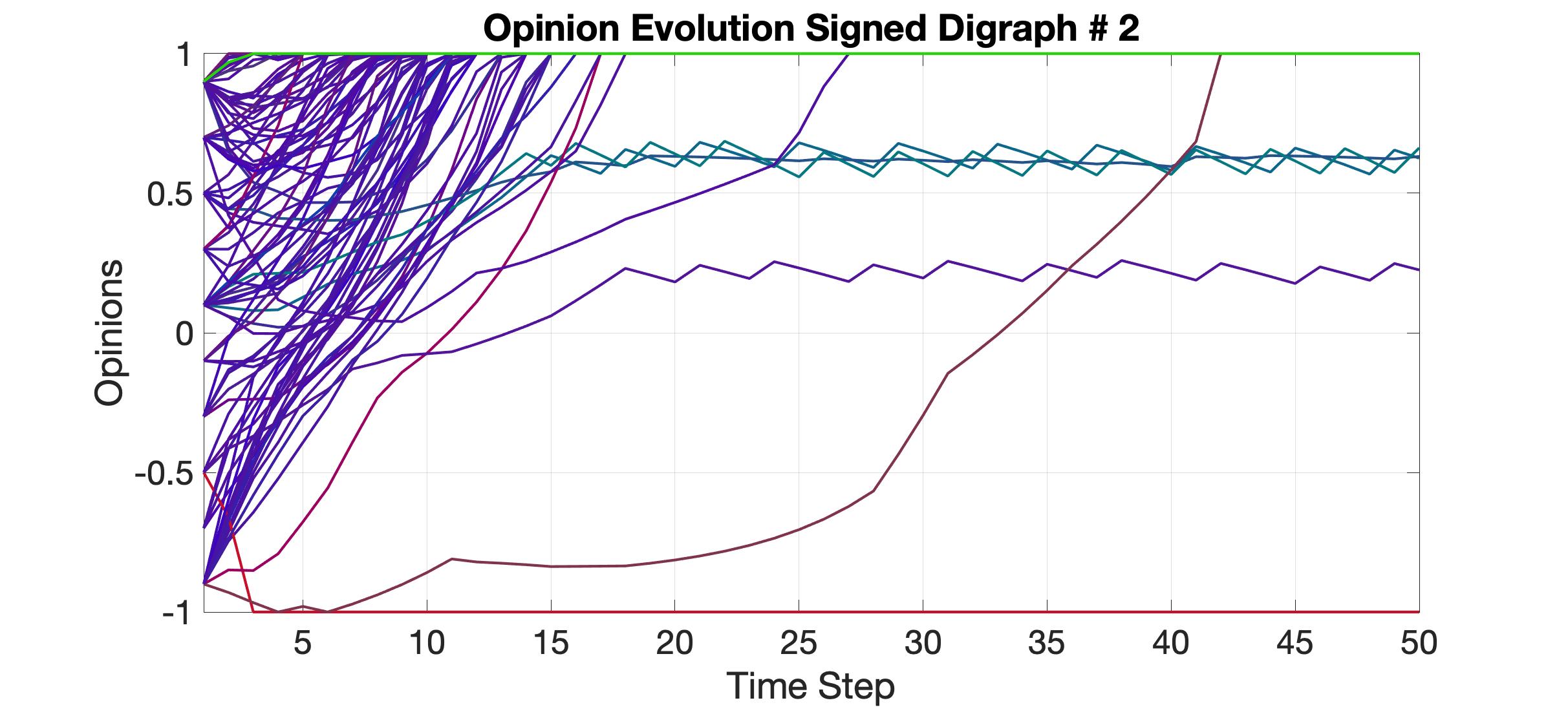}
         \caption{Opinion evolution signed digraph 2}
         \label{SugFig:OpEvInNe2}
     \end{subfigure}
     \hfill
     \begin{subfigure}[m]{0.27\textwidth}
         \centering
         \includegraphics[width=\textwidth]{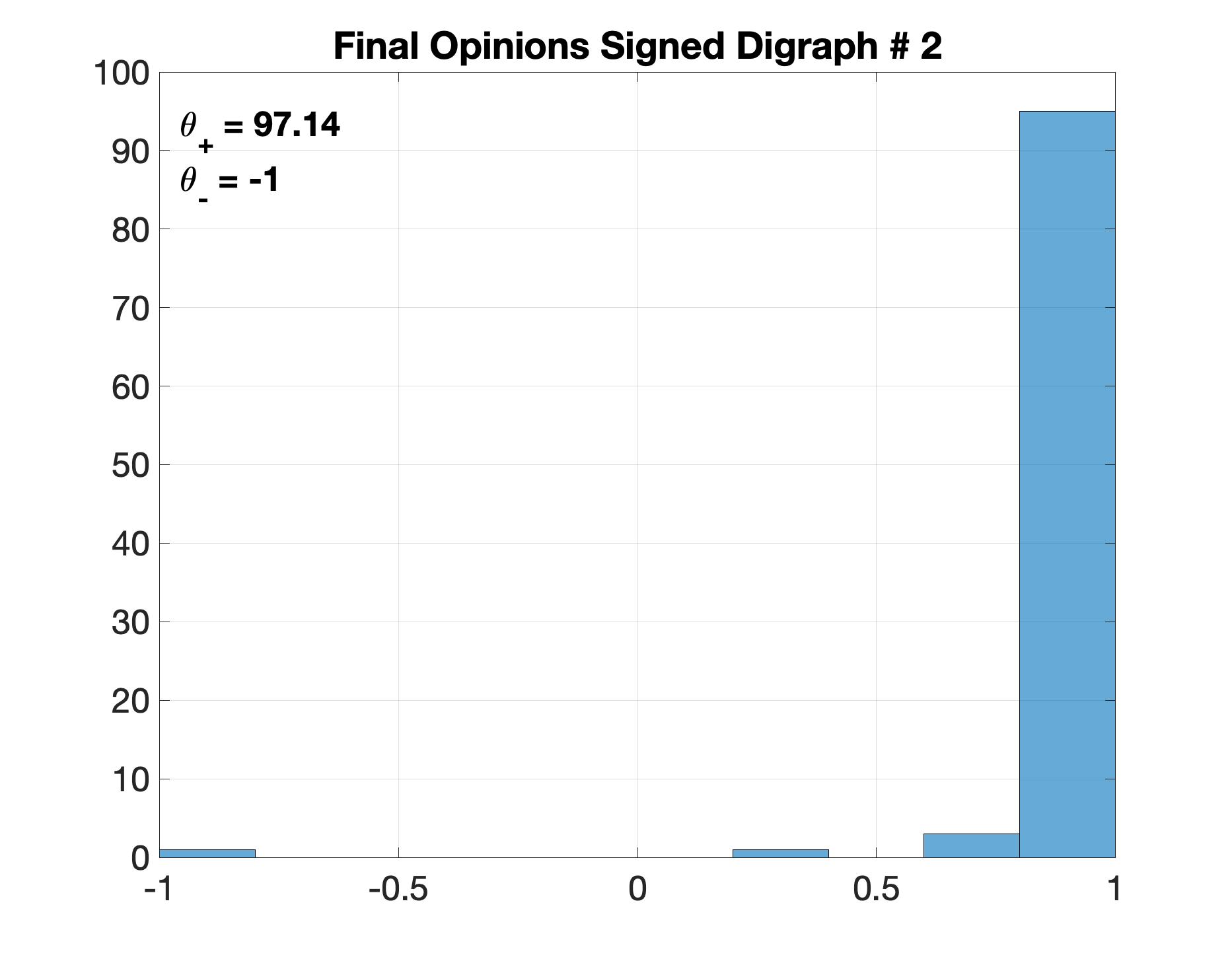}
         \caption{Final opinion histogram}
         \label{SugFig:OpHiInNe2}
     \end{subfigure}
	\caption{Effect of changing the signed digraph  \CAbnew{(evolving 100 agents)}.} 
        \label{Fig:Examples_2}
\end{figure}

To study the effect of changing the signs of the weights of the signed digraph, the nominal signed digraph of Figure \ref{SubFig:NomIntNetw} is modified into the signed digraphs shown in Figures \ref{SubFig:IntNet1} and \ref{SubFig:IntNet2}. The topology is unchanged, but the number of positive and negative edges is changed. The resulting opinion evolution and final opinion histograms are shown in Figures \ref{SugFig:OpEvInNe1} and \ref{SugFig:OpHiInNe1}, and in Figures \ref{SugFig:OpEvInNe2} and \ref{SugFig:OpHiInNe2} respectively.

Compared with the nominal results in Figures \ref{SubFig:NomOpEvo} and \ref{SubFig:NomFOH}, the most different outcome occurs when most edges are positive (digraph in Figure \ref{SubFig:IntNet2}). In this case, the end result is almost perfect consensus for the $+1$ opinion, because the initial opinion, with $\theta_- = -19.3$ and $\theta_+ = 41.5$, is more skewed towards $+1$. The presence of negative edges is crucial to avoid trivial consensus outcomes even when the agents are not completely conformist. The opinion evolution in Figure \ref{SugFig:OpEvInNe2} shows that, initially, conformist traits pull the opinions towards positive values, and then radical traits make them increase in value until they reach $+1$. Purely radical agents would have produced polarisation instead of consensus. 

When increasing the number of negative edges (digraph in Figure \ref{SubFig:IntNet1}), the final opinions in Figure \ref{SugFig:OpHiInNe1} are different from the nominal ones, but the qualitative behaviour is comparable.

\subsubsection{Varying the Opinion Evolution Parameters}

We study the sensitivity with respect to the opinion evolution parameters  $\Omega = (\lambda, \xi, \mu)$, where: $\lambda$ is the overall opinion change magnitude, and can also be thought of as a time scaling parameter; $\xi$ gives more weight to distant opinions for conformist traits; $\mu$ increases the opinion change for radical traits. We change these parameters one at the time, with respect to the nominal parameters, and compare the results with the nominal results in Figure \ref{Fig:Examples_NominalResults}.

\begin{figure}[h!]
     \centering
     \begin{subfigure}[t]{0.42\textwidth}
         \centering
         \includegraphics[width=\textwidth]{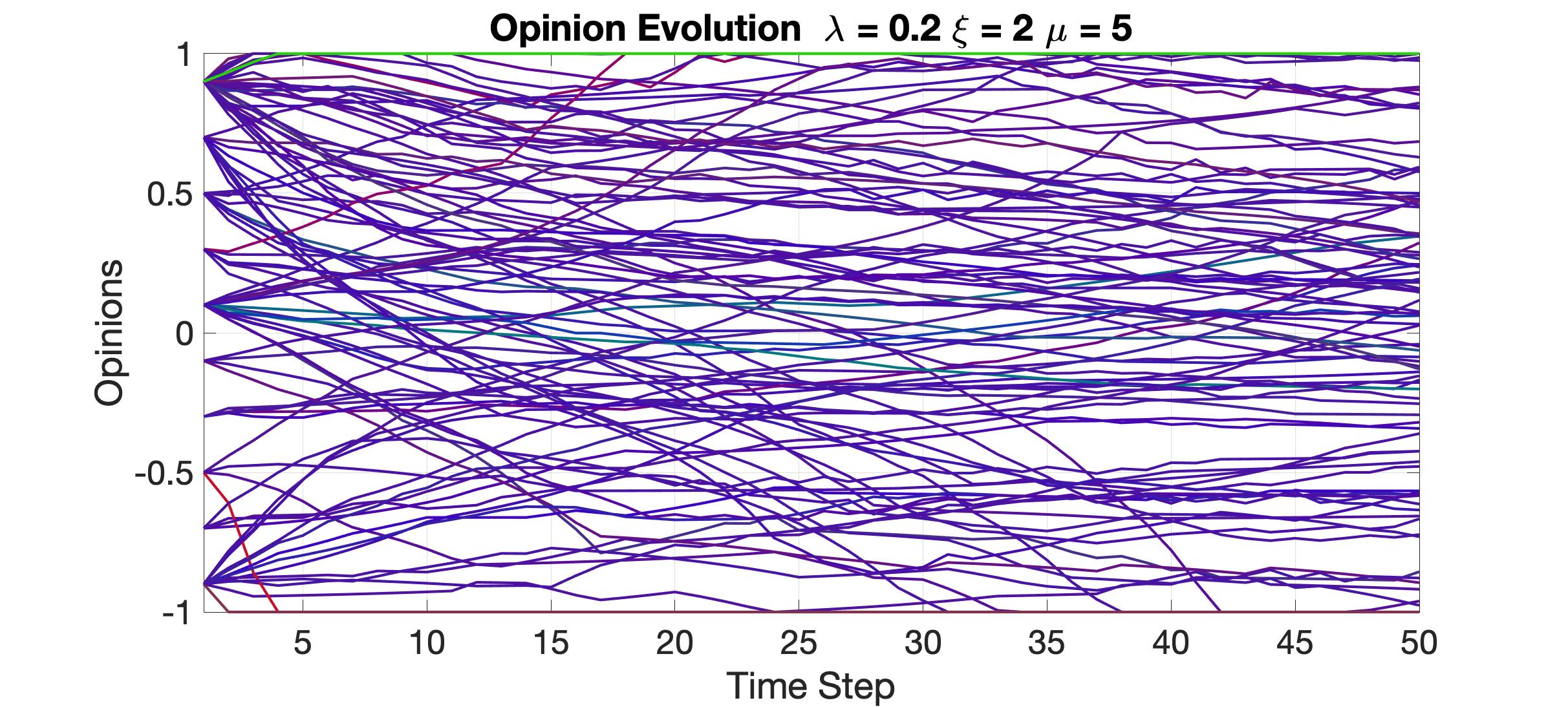}
         \caption{Opinion Evolution}
         \label{SubFig:ChaLam_1}
     \end{subfigure}
     \quad
     \begin{subfigure}[t]{0.27\textwidth}
         \centering
         \includegraphics[width=\textwidth]{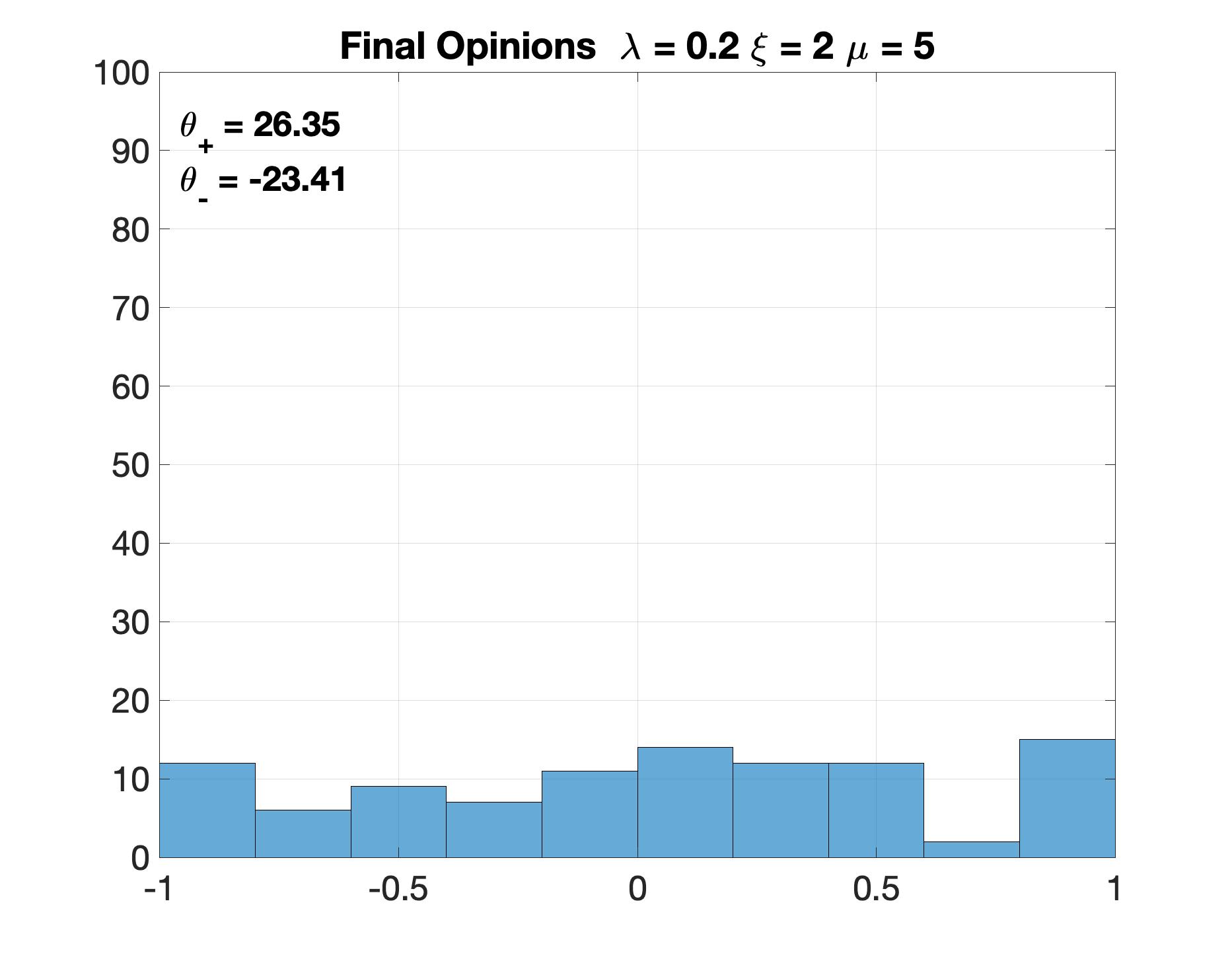}
         \caption{Final Opinion Histogram}
         \label{SubFig:ChaLam_2}
     \end{subfigure} 
     \\
     \begin{subfigure}[t]{0.42\textwidth}
         \centering
         \includegraphics[width=\textwidth]{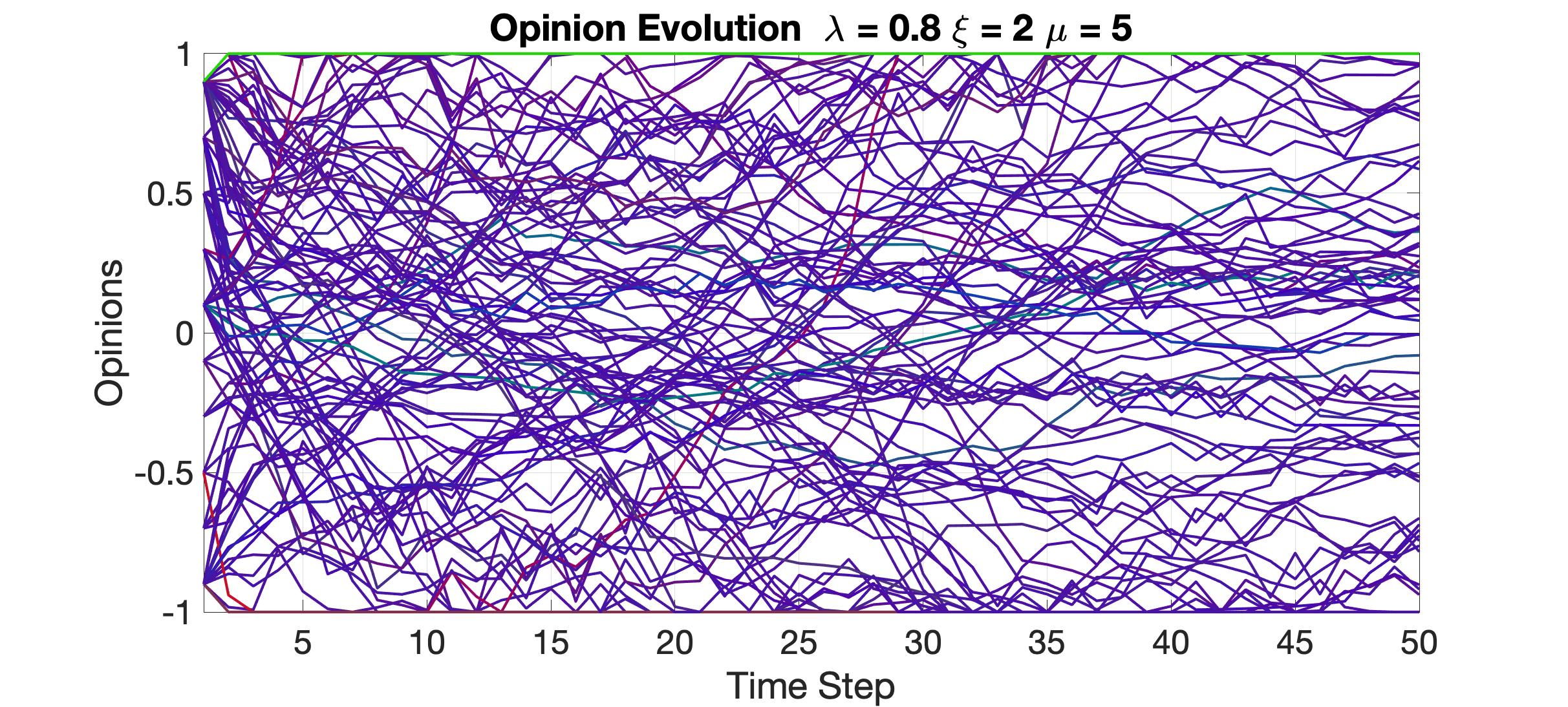}
         \caption{Opinion Evolution}
         \label{SubFig:ChaLam_3}
     \end{subfigure}
     \quad
     \begin{subfigure}[t]{0.27\textwidth}
         \centering
         \includegraphics[width=\textwidth]{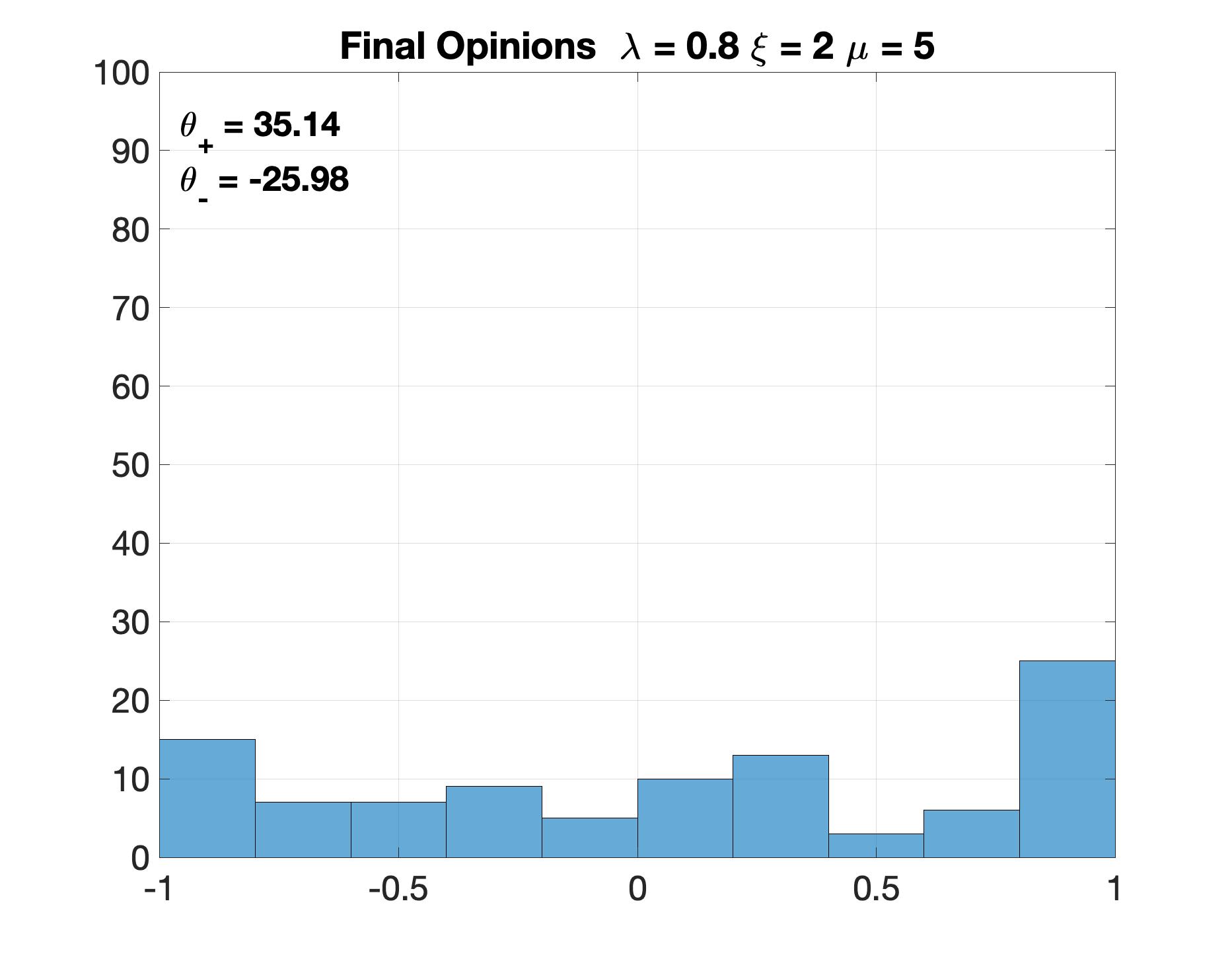}
         \caption{Final Opinion Histogram}
         \label{SubFig:ChaLam_4}
     \end{subfigure}
        \caption{Effect of changing $\lambda$ from the nominal value $\lambda = 0.4$ to $\lambda = 0.2$ (Figures \ref{SubFig:ChaLam_1} and \ref{SubFig:ChaLam_2}) and $\lambda = 0.8$ (Figures \ref{SubFig:ChaLam_3} and \ref{SubFig:ChaLam_4})  \CAbnew{evolving 100 agents.}} 
        \label{Fig:Examples_lambda}
\end{figure}

Figure \ref{Fig:Examples_lambda} shows the opinion evolution and final histogram for $\lambda = 0.2$ and $\lambda = 0.8$. The final histograms in Figures \ref{SubFig:ChaLam_2} and \ref{SubFig:ChaLam_4} do not change much with respect to the nominal. The most significant change can be noticed in Figures \ref{SubFig:ChaLam_1} and \ref{SubFig:ChaLam_3}, showing that indeed a higher value of $\lambda$ produces larger changes in the opinions. Overall, however, the effect of varying $\lambda$ is very limited. 

\begin{figure}[h!]
     \centering
     \begin{subfigure}[t]{0.42\textwidth}
         \centering
         \includegraphics[width=\textwidth]{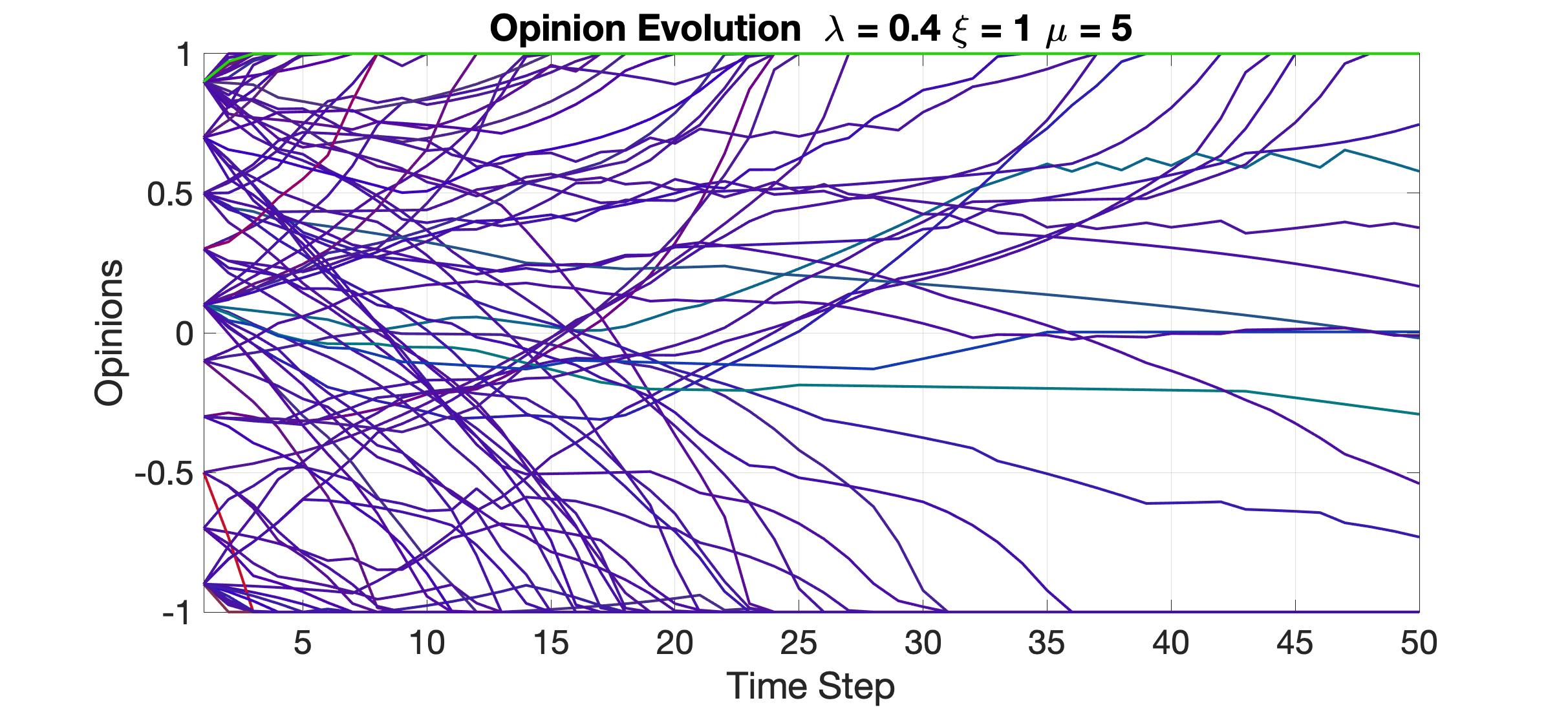}
         \caption{Opinion Evolution}
         \label{SubFig:ChaXi_1}
     \end{subfigure}
     \quad
     \begin{subfigure}[t]{0.27\textwidth}
         \centering
         \includegraphics[width=\textwidth]{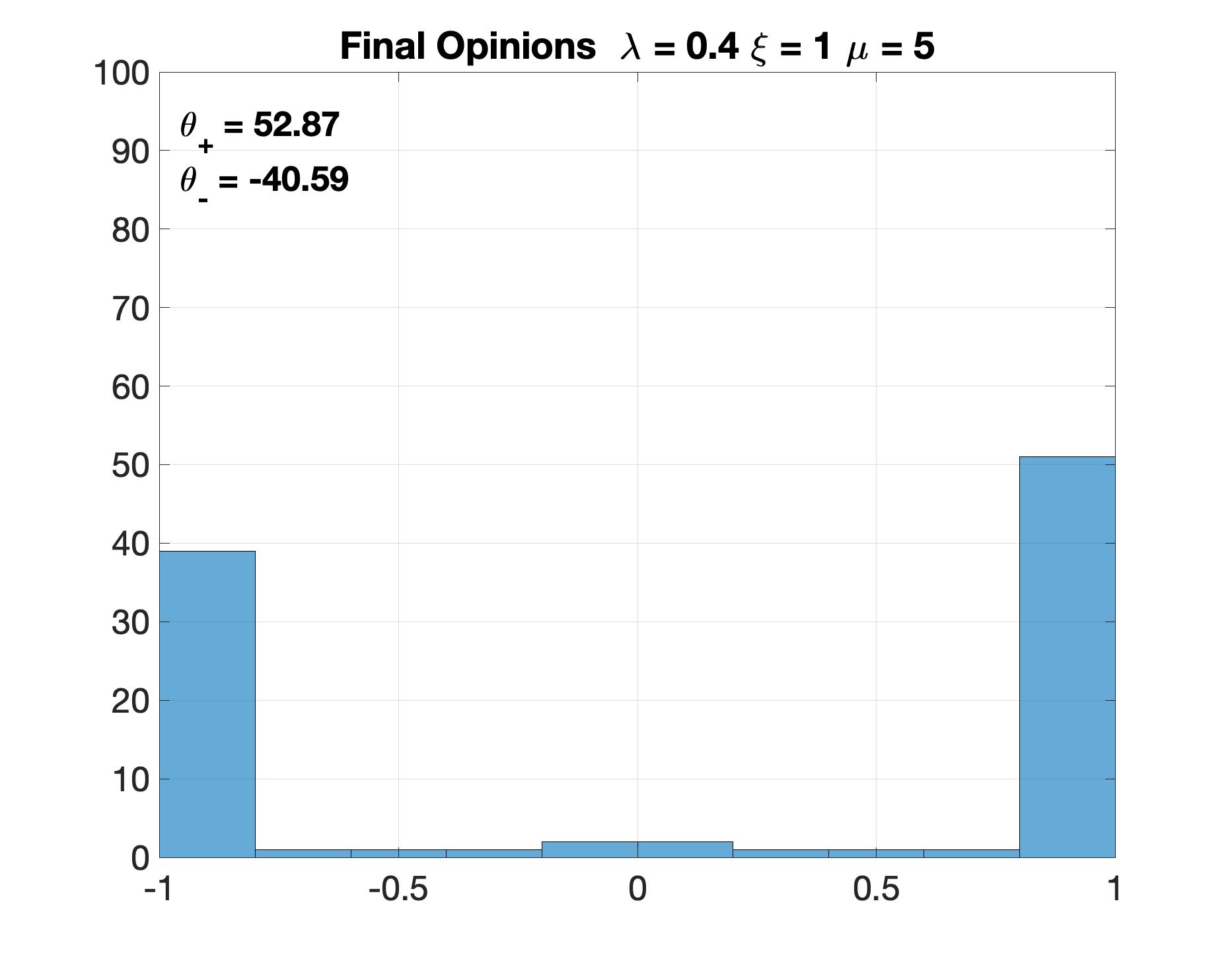}
         \caption{Final Opinion Histogram}
         \label{SubFig:ChaXi_2}
     \end{subfigure} 
          \\
     \begin{subfigure}[t]{0.42\textwidth}
         \centering
         \includegraphics[width=\textwidth]{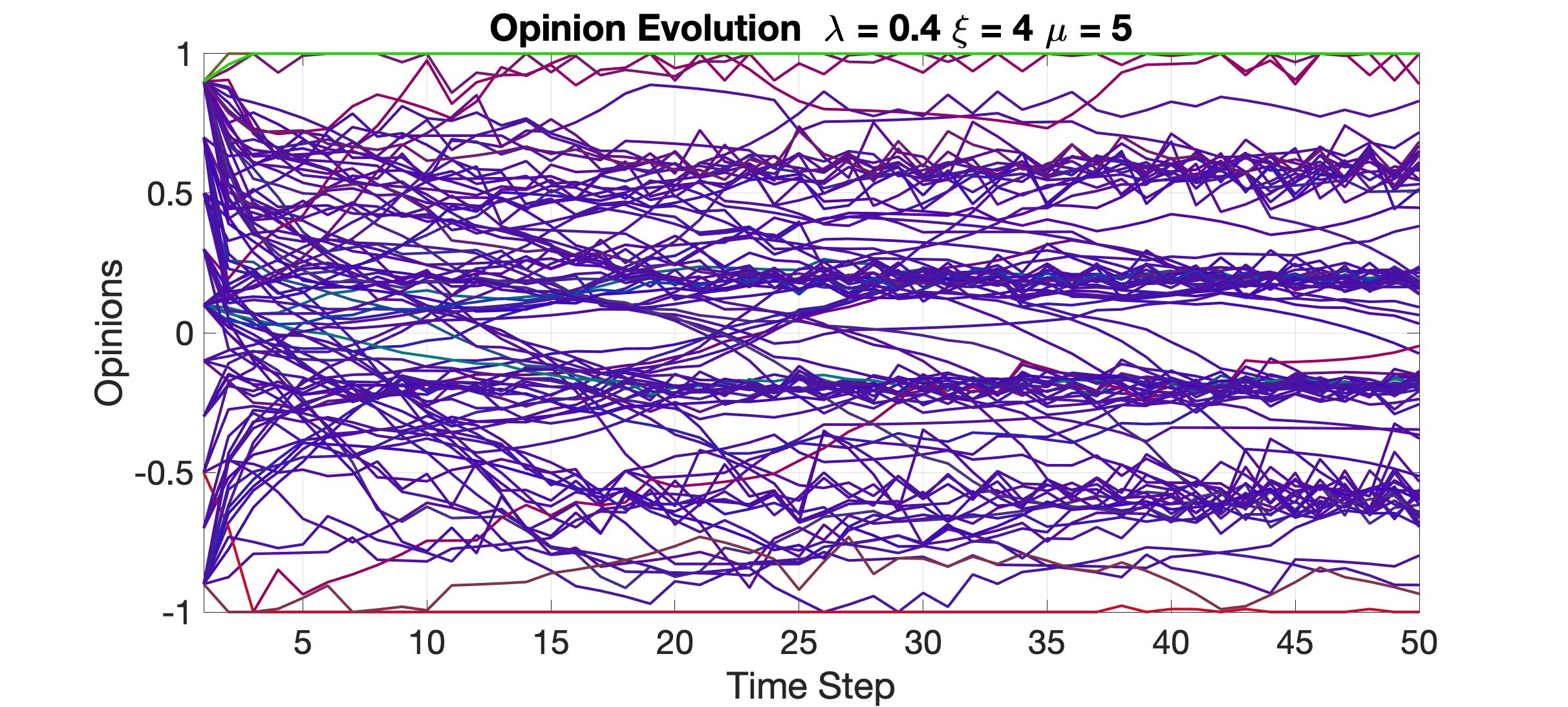}
         \caption{Opinion Evolution}
         \label{SubFig:ChaXi_3}
     \end{subfigure}
     \quad
     \begin{subfigure}[t]{0.27\textwidth}
         \centering
         \includegraphics[width=\textwidth]{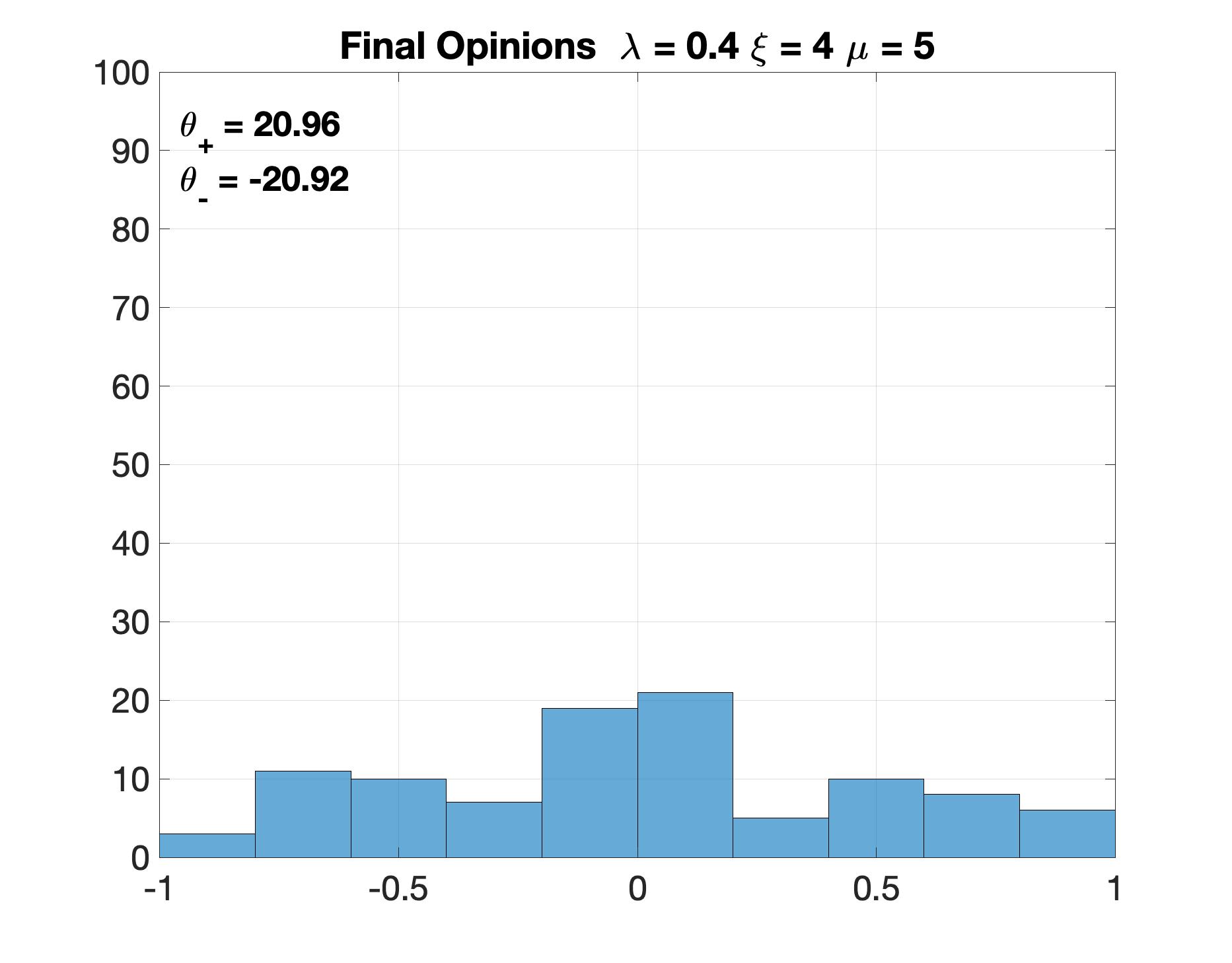}
         \caption{Final Opinion Histogram}
         \label{SubFig:ChaXi_4}
     \end{subfigure}
        \caption{Effect of changing $\xi$ from the nominal value $\xi = 2$ to $\xi = 1$ (Figures \ref{SubFig:ChaXi_1} and \ref{SubFig:ChaXi_2}) and $\xi = 0.8$ (Figures \ref{SubFig:ChaXi_3} and \ref{SubFig:ChaXi_4})  \CAbnew{evolving 100 agents.}} 
        \label{Fig:Examples_xi}
\end{figure}

The effect of varying $\xi$ is shown in Figure \ref{Fig:Examples_xi}. The changes in both the opinion evolution and the final opinion histogram are quite noticeable. A value of $\xi=1$ means that distant opinions have the same attracting power as closer opinions for the conformist traits, hence in general the conformist trait has less influence over the whole opinion change, which is instead dominated by the radical traits. The result is visible in the opinion evolution in Figure \ref{SubFig:ChaXi_1} and the final opinion histogram in Figure \ref{SubFig:ChaXi_2}. On the contrary, increasing the value to $\xi=4$ yields a stronger conformist tendency towards consensus, evident when comparing the nominal final opinions in Figure \ref{SubFig:NomFOH} with the final opinions with $\xi=4$ in Figure \ref{SubFig:ChaXi_4}, and the respective $\theta_-$ and $\theta_+$.

\begin{figure}[h!]
     \centering
     \begin{subfigure}[t]{0.42\textwidth}
         \centering
         \includegraphics[width=\textwidth]{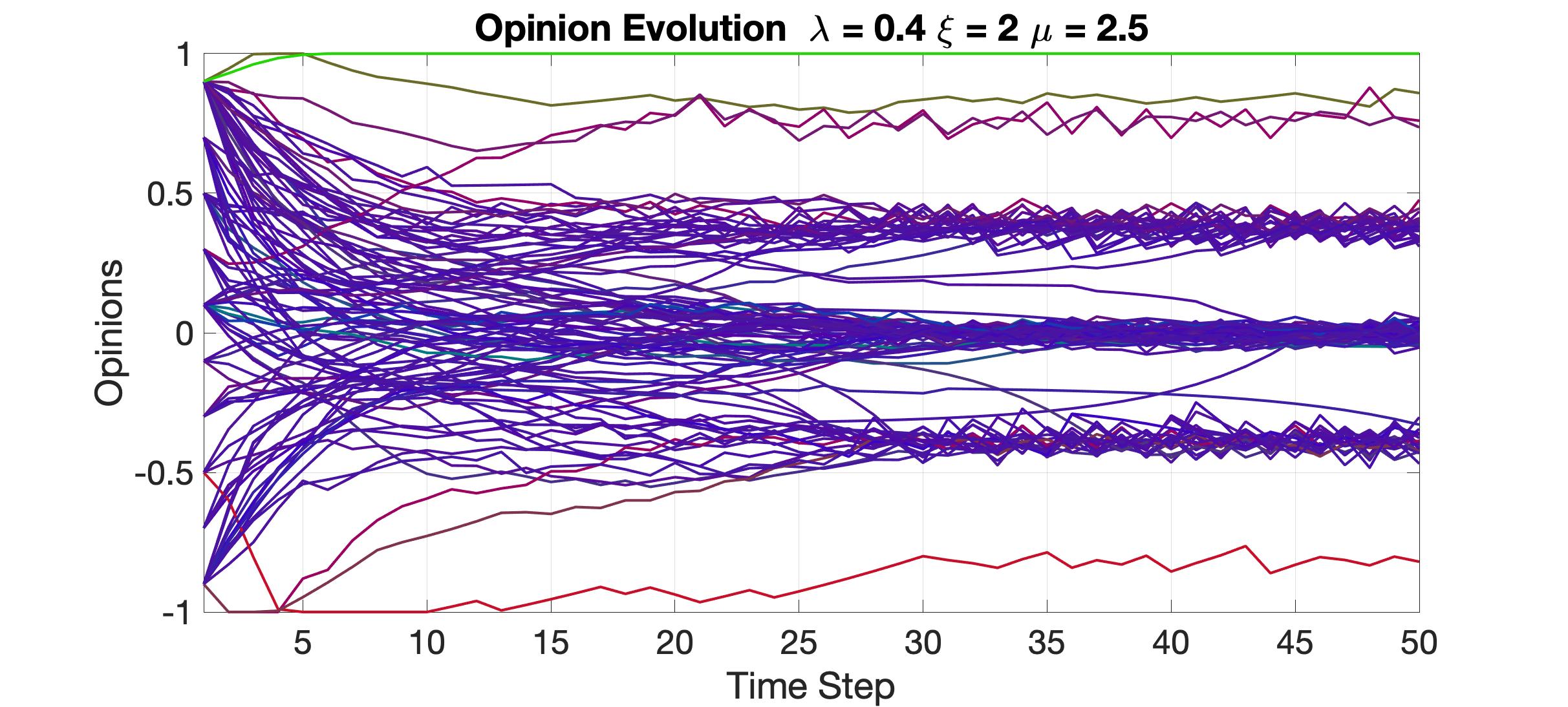}
         \caption{Opinion Evolution}
         \label{SubFig:ChaMu_1}
     \end{subfigure}
     \quad
     \begin{subfigure}[t]{0.27\textwidth}
         \centering
         \includegraphics[width=\textwidth]{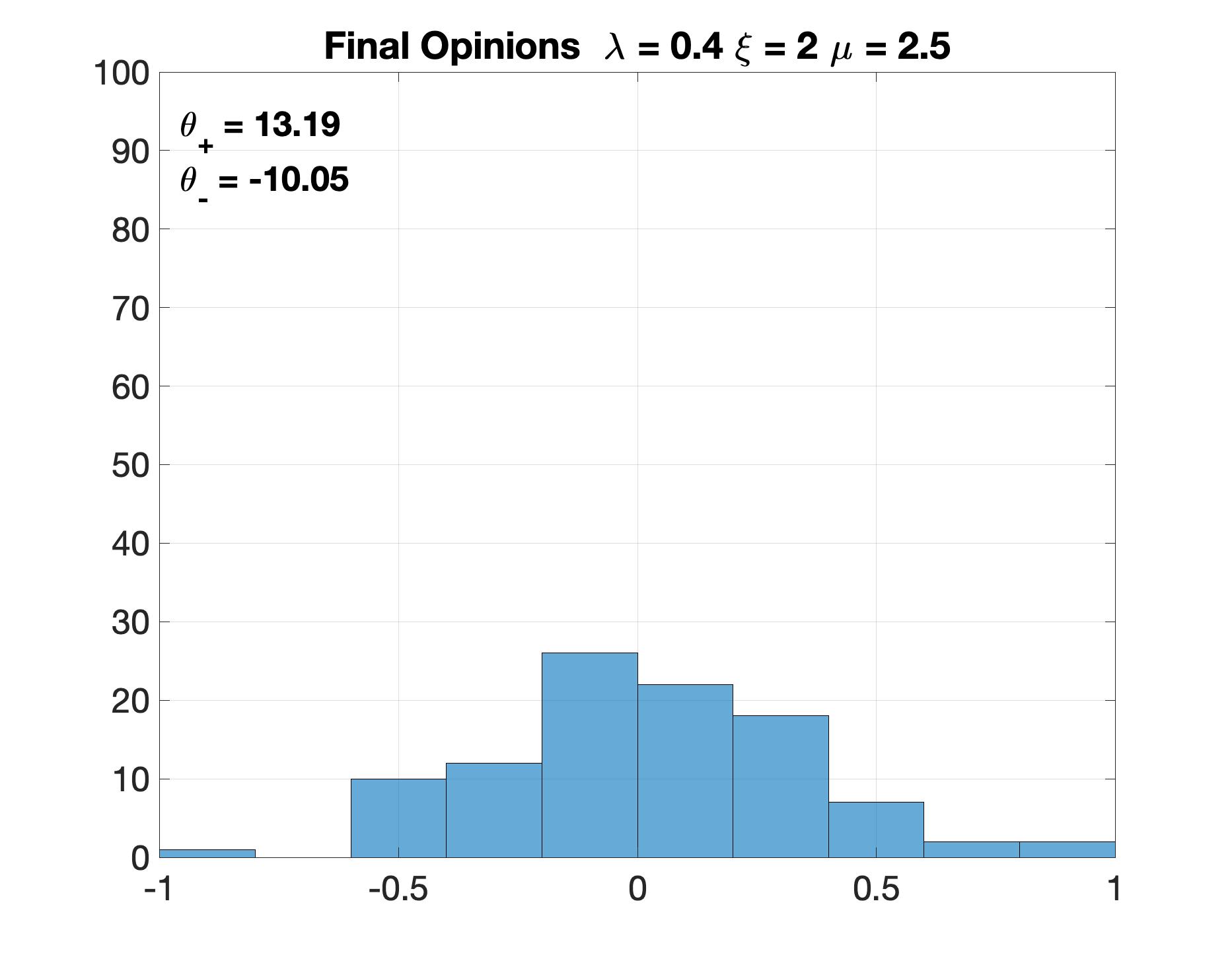}
         \caption{Final Opinion Histogram}
         \label{SubFig:ChaMu_2}
     \end{subfigure} 
          \\
     \begin{subfigure}[t]{0.42\textwidth}
         \centering
         \includegraphics[width=\textwidth]{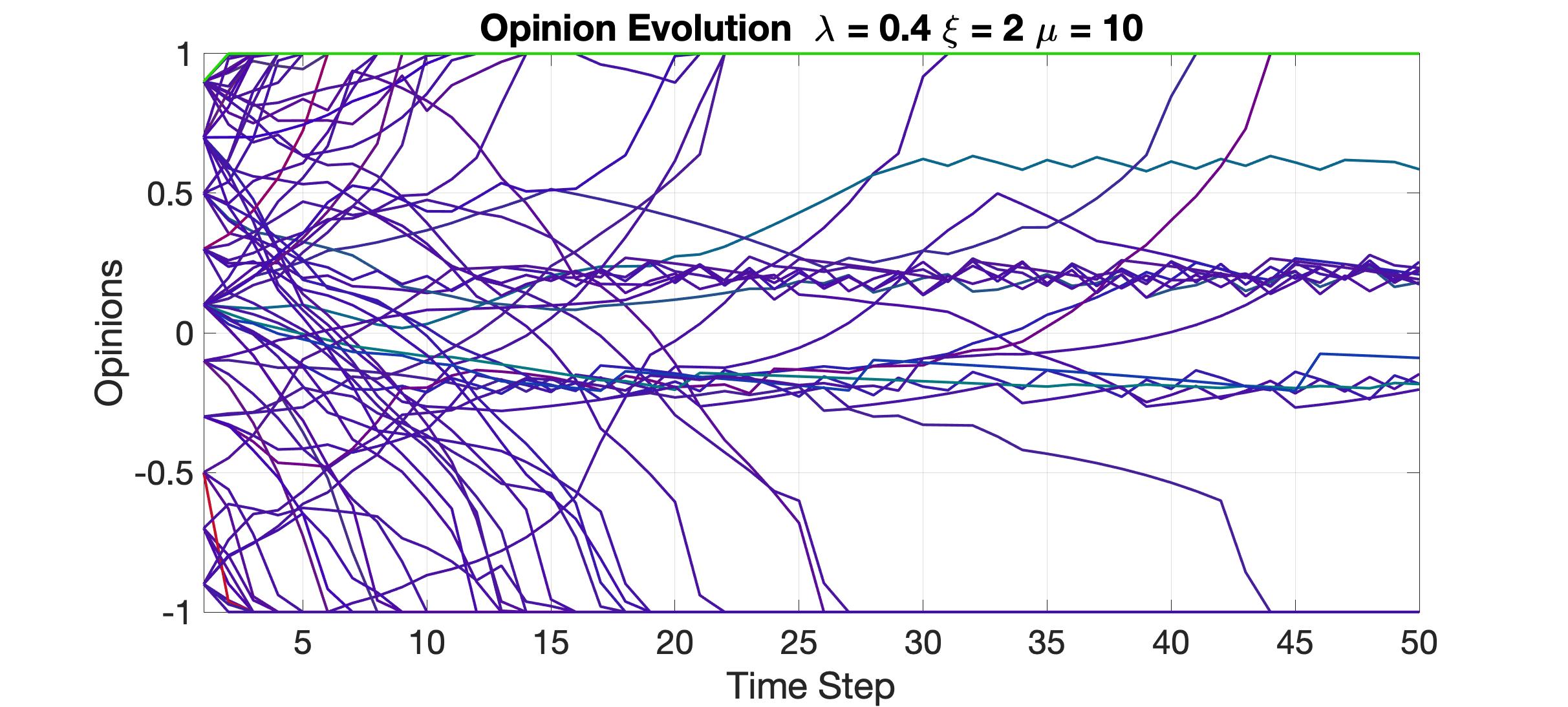}
         \caption{Opinion Evolution}
         \label{SubFig:ChaMu_3}
     \end{subfigure}
     \quad
     \begin{subfigure}[t]{0.27\textwidth}
         \centering
         \includegraphics[width=\textwidth]{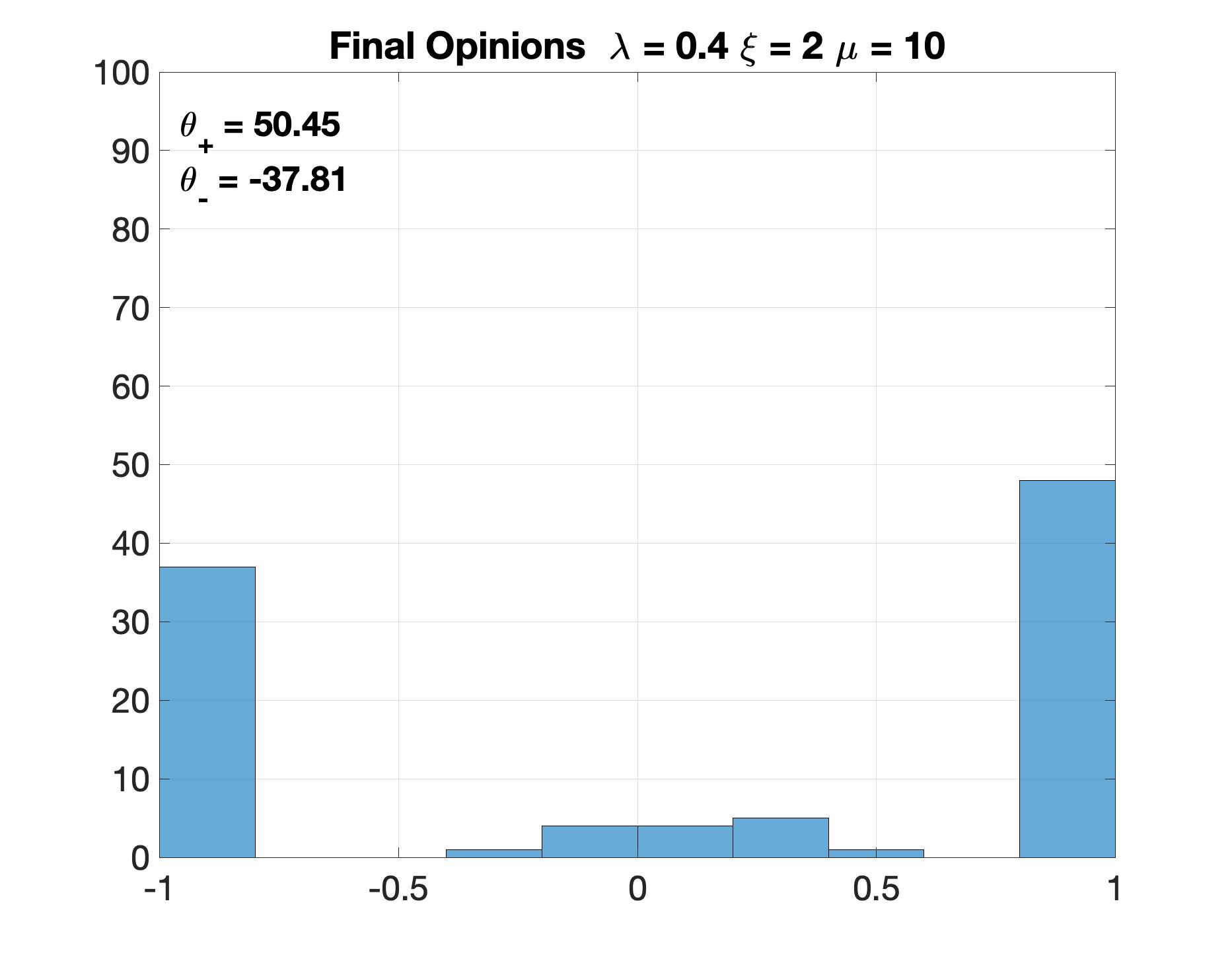}
         \caption{Final Opinion Histogram}
         \label{SubFig:ChaMu_4}
     \end{subfigure} 
        \caption{Effect of changing $\mu$ from the nominal value $\mu = 5$ to $\mu = 2.5$ (Figures \ref{SubFig:ChaMu_1} and \ref{SubFig:ChaMu_2}) and $\mu = 10$ (Figures \ref{SubFig:ChaMu_3} and \ref{SubFig:ChaMu_4})  \CAbnew{evolving 100 agents.}} 
        \label{Fig:Examples_mu}
\end{figure}

Parameter $\mu$ modulates the effect of radical traits on the opinion evolution. Comparing Figure \ref{SubFig:ChaMu_2} with Figure \ref{SubFig:ChaMu_4} shows that a larger $\mu$ increases radicalism in the population, which leads to polarisation for the given initial opinions. A similar effect is achieved by varying $\xi$: in fact, both $\xi$ and $\mu$ affect the balance between the conformist tendency towards consensus and the radical tendency towards polarisation. 
Although both $\xi$ and $\mu$ play a role in the conformist-radical balance, they are not completely complementary: an increase in $\xi$ is not the same as a decrease in $\mu$. This can be seen by comparing Figures \ref{SubFig:ChaXi_4} and \ref{SubFig:ChaMu_2}: increasing $\xi$ produces final opinions that are more evenly distributed than those obtained by decreasing $\mu$. Moreover, increasing radicalism does not always lead to polarisation: this happens only when the opinions have both positive and negative values. If the opinions have only positive values or only negative values, then radicalism will move all of them to a single extreme, resulting in consensus. Therefore, it is not possible to generalise the idea that more radicalism always leads to polarisation, regardless of the initial opinions.

\subsection{Model Validation with Real Data}

Data from the World Values Survey are used to validate the CB model, \CAbnew{namely, show that a suitable choice of the parameters allows the model to produce predicted opinions similar to the real opinions in a society}. The World Values Survey is an international organisation that conducts surveys about ethics and values in different countries around the globe. These surveys are repeated every 5 years. We considered the answers to 30 questions, shown in Table \ref{tab:questions}, in 26 countries, shown in Table \ref{tab:countries}. In each question, the respondents are asked to state the extent to which they agree with a statement in a Likert-scale 10. The answers given in the surveys of 2015 (wave 5) are taken as initial opinions, while the answers of 2020 surveys (wave 6) are taken as final opinions.

Two minimisation problems are stated to find model parameters that produce predicted opinions similar to the ones found in the survey answers. The \textbf{\emph{Free} Optimisation Problem} allows the inner traits assignation to change with questions; in the \textbf{\emph{Constrained} Optimisation Problem}, the inner traits are fixed for all questions. The transition tables for the model with parameters provided by both optimisation problems are also computed.

 \CAbnew{Given real and \CAbnew{model-generated} opinion vectors $r$ and $y$, the cost function $\Cost$ used in the minimisation problems  \eqref{Eq:FirstMP}, \eqref{Eq:LargeOP}, and \eqref{Eq:LargeOP_mod} is defined as}
\begin{equation}
\label{Eq:CostFunction}
\CAbnew{\Cost(r,y) = \sum_{i = 1}^\numag | \tilde{r}_i - \tilde{y}_i |,}
\end{equation}
\CAbnew{where $\tilde{r} = (\tilde{r}_i)_{i=1}^\numag$ is the vector $\hat{r} = (\hat{r}_i)_{i=1}^\numag$ sorted in descending order, and $\hat{\cdot}$ is the quantisation function}
\begin{equation*}
\CAbnew{\hat{r}_i = \argmin_{\zeta \in \XSet}\{|\zeta - r_i|\} \qquad \forall i = 1,\dots , \numag,}
\end{equation*}
\CAbnew{with $\XSet$ defined as $\XSet = \Big\{ \frac{1}{2}(u_k + u_{k+1}) \mid u_k = -1 + k\frac{2}{10} \quad k = 1, \dots , 9 \Big\}$. Quantisation is needed because the World Values Survey answers we consider as real opinions use a Likert scale 10: participants could choose their opinion from 10 different options. These opinions rescaled to be between -1 and 1 produce the set $\XSet$ and, therefore, the predicted opinions also need to be quantified in the same way. Both opinion vectors (real and predicted) are sorted in descending order, so that equal opinions add a zero to the total cost.}

\CAbnew{Even for a relatively small population $\numag = 100$, the size of the sets $\SetN$ (underlying signed digraph structures) and $\SetA$ (inner traits assignations) is enormous.}
 \CAbnew{Given the tremendous size of the parameter space $\SetN\times\SetA$, performing the minimisation over all possible signed digraphs and agent inner traits would be computationally intractable. Therefore, the minimisation occurs over small subsets $\tilde{\SetN}\subset\SetN$, $\tilde{\SetA}\subset\SetA$ of the whole parameter space. As a consequence, there is no guarantee that we are estimating the \emph{real} parameter values or making the absolute best parameter choice: with other parameter choices, not included in $\tilde\SetN\times\tilde\SetA$, the model could reproduce the data with even better accuracy.}
 
 \begin{table}[h]
\centering
\resizebox{\textwidth}{!}{%
\begin{tabular}{||c || c | c | c | c | c | c | c | c | c | c | c | c | c | c | c | c | c | c | c | c | c | c | c | c | c | c | c | c | c | c | c | c | c | c | c ||}
\hline\hline 
ID  & 1 & 2 & 3 & 4 & 5 & 6 & 7 & 8 & 9 & 10 & 11 & 12 \\
\hline 
APL  & 2.13 & 2.13 & 2.13 & 2.13 & 2.13 & 1.95 & 1.95 & 1.95 & 1.95 & 1.95 & 2.04 & 2.04 \\
\hline 
CC  & 0.38 & 0.38 & 0.38 & 0.38 & 0.38 & 0.18 & 0.18 & 0.18 & 0.18 & 0.18 & 0.16 & 0.16 \\
\hline 
PE  & 252 & 558 & 834 & 1115 & 1436 & 258 & 566 & 848 & 1145 & 1438 & 222 & 533 \\
\hline 
NE  & 1349 & 1043 & 767 & 486 & 165 & 1326 & 1018 & 736 & 439 & 146 & 1194 & 883 \\
\hline 
D  & 4 & 4 & 4 & 4 & 4 & 3 & 3 & 3 & 3 & 3 & 3 & 3 \\
\hline 
BI  & 0.00015 & 4.4e-05 & 3.8e-05 & 0.00013 & 0.042 & 0.00023 & 8.1e-05 & 4.8e-05 & 0.00027 & 0.049 & 0.00099 & 0.00025 \\
\hline 
\hline ID  & 13 & 14 & 15 & 16 & 17 & 18 & 19 & 20 & 21 & 22 & 23 & 24 \\
\hline 
APL  & 2.04 & 2.04 & 2.04 & 1.75 & 1.75 & 1.75 & 1.75 & 1.75 & 1.68 & 1.68 & 1.68 & 1.68 \\
\hline 
CC  & 0.16 & 0.16 & 0.16 & 0.25 & 0.25 & 0.25 & 0.25 & 0.25 & 0.35 & 0.35 & 0.35 & 0.35 \\
\hline 
PE  & 746 & 1020 & 1259 & 362 & 864 & 1351 & 1813 & 2344 & 418 & 1079 & 1683 & 2372 \\
\hline 
NE  & 670 & 396 & 157 & 2227 & 1725 & 1238 & 776 & 245 & 2891 & 2230 & 1626 & 937 \\
\hline 
D  & 3 & 3 & 3 & 3 & 3 & 3 & 3 & 3 & 2 & 2 & 2 & 2 \\
\hline 
BI  & 0.00021 & 0.00056 & 0.047 & 2e-08 & 6.1e-09 & 4.1e-09 & 1e-07 & 0.0071 & 3.4e-11 & 5.8e-12 & 7.5e-13 & 6.7e-09 \\
\hline 
\hline ID  & 25 & 26 & 27 & 28 & 29 & 30 & 31 & 32 & 33 & 34 & 35 &  \\
\hline 
APL  & 1.68 & 1.68 & 1.68 & 1.68 & 1.68 & 1.68 & 1.62 & 1.62 & 1.62 & 1.62 & 1.62 &  \\
\hline 
CC  & 0.35 & 0.32 & 0.32 & 0.32 & 0.32 & 0.32 & 0.39 & 0.39 & 0.39 & 0.39 & 0.39 &  \\
\hline 
PE  & 2947 & 456 & 1063 & 1667 & 2329 & 2972 & 457 & 1259 & 1998 & 2717 & 3506 &  \\
\hline 
NE  & 362 & 2823 & 2216 & 1612 & 950 & 307 & 3440 & 2638 & 1899 & 1180 & 391 &  \\
\hline 
D  & 2 & 2 & 2 & 2 & 2 & 2 & 2 & 2 & 2 & 2 & 2 &  \\
\hline 
BI  & 0.00074 & 4.8e-11 & 8.6e-12 & 1.3e-12 & 3.7e-09 & 0.0021 & 4.7e-14 & 3.6e-14 & 3.2e-14 & 4.3e-11 & 0.00033 &  \\
\hline 
\hline 
\end{tabular}}
\caption{Signed Digraph Information: Average Path Length (APL), Clustering Coefficent (CC), Positive Edges (PE), Negative Edges (NE), Diameter (D), and Balance Index (BI)}
\label{tab:network}
\end{table}

\begin{figure}[!t]
\centering
\includegraphics[width=0.4\textwidth]{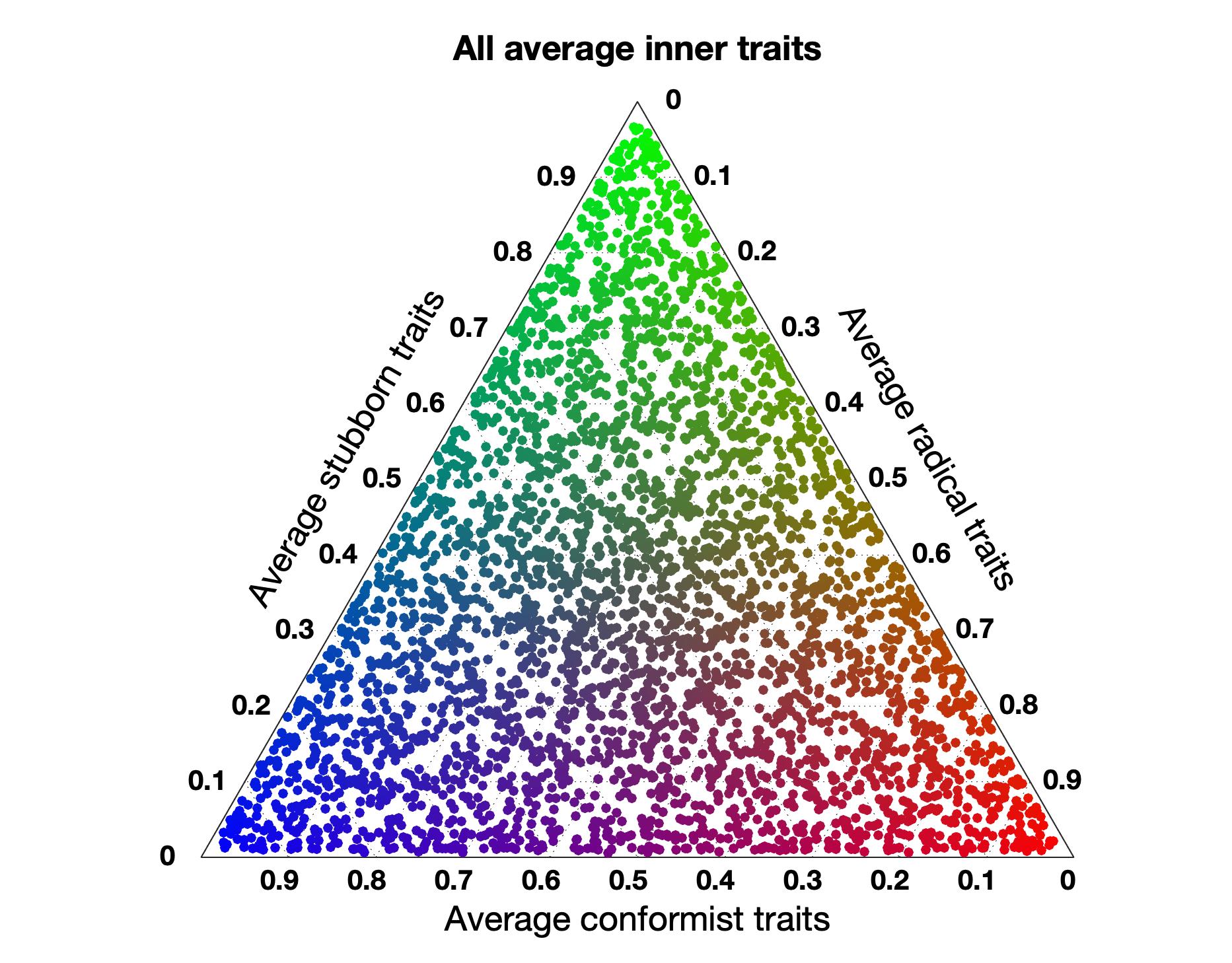}
\caption{All the average inner traits $\bar{\innertraits}$ corresponding to inner traits assignations $\innertraits$ in $\tilde{\SetA}$.}
\label{Fig:PossibleAssignations}
\end{figure}

\CAbnew{The subset $\tilde{\SetN}$ contains 35 different small-world signed strongly connected digraphs. Table \ref{tab:network} shows the main characteristics of the networks. The subset $\tilde{\SetA}$ contains $3528$ randomly generated inner traits assignations $\innertraits = (\innertraits_i)_{i=1}^\numag$. To avoid bias towards societies with average inner traits that are more conformist, radical, or stubborn, the set $\tilde{\SetA}$ satisfies the following property: for every inner traits assignation $\innertraits$, with corresponding average inner trait $\bar{\innertraits} = (\bar{\alpha}, \bar{\beta}, \bar{\gamma}) = (a_1, b_1, c_1)$, there are two inner traits assignations $\innertraits^\prime, \innertraits^{\prime\prime}\in\tilde{\SetA}$ that satisfy $\bar{\innertraits^\prime} = (b_1, c_1, a_1)$, and $\bar{\innertraits^{\prime\prime}}= (c_1,a_1,b_1)$. In other words, the parameter space $\tilde{\SetA}$ is symmetric with respect to permutations of agent traits.  \CAbnew{Besides this property, the elements of this set were chosen at random.} All the average inner traits $\bar{\innertraits}$ corresponding to inner traits assignations $\innertraits$ in $\tilde{\SetA}$ are shown in Figure \ref{Fig:PossibleAssignations}.}

 \CAbnew{Due to the anonymity of the surveys, it is not possible to guarantee that the same people answered the survey in subsequent waves of the WVS. However, if the surveys are done correctly to represent society overall, the results can be anyway assumed to reflect the global opinion distribution of the general population about a given topic at a specific time, and this allows us to use the survey results in different waves in our minimisation problem, \emph{as if} the very same people had answered.}

\subsubsection{\emph{Free} Optimisation Problem}

\begin{figure}[h]
      \centering
          \includegraphics[width=0.7\textwidth]{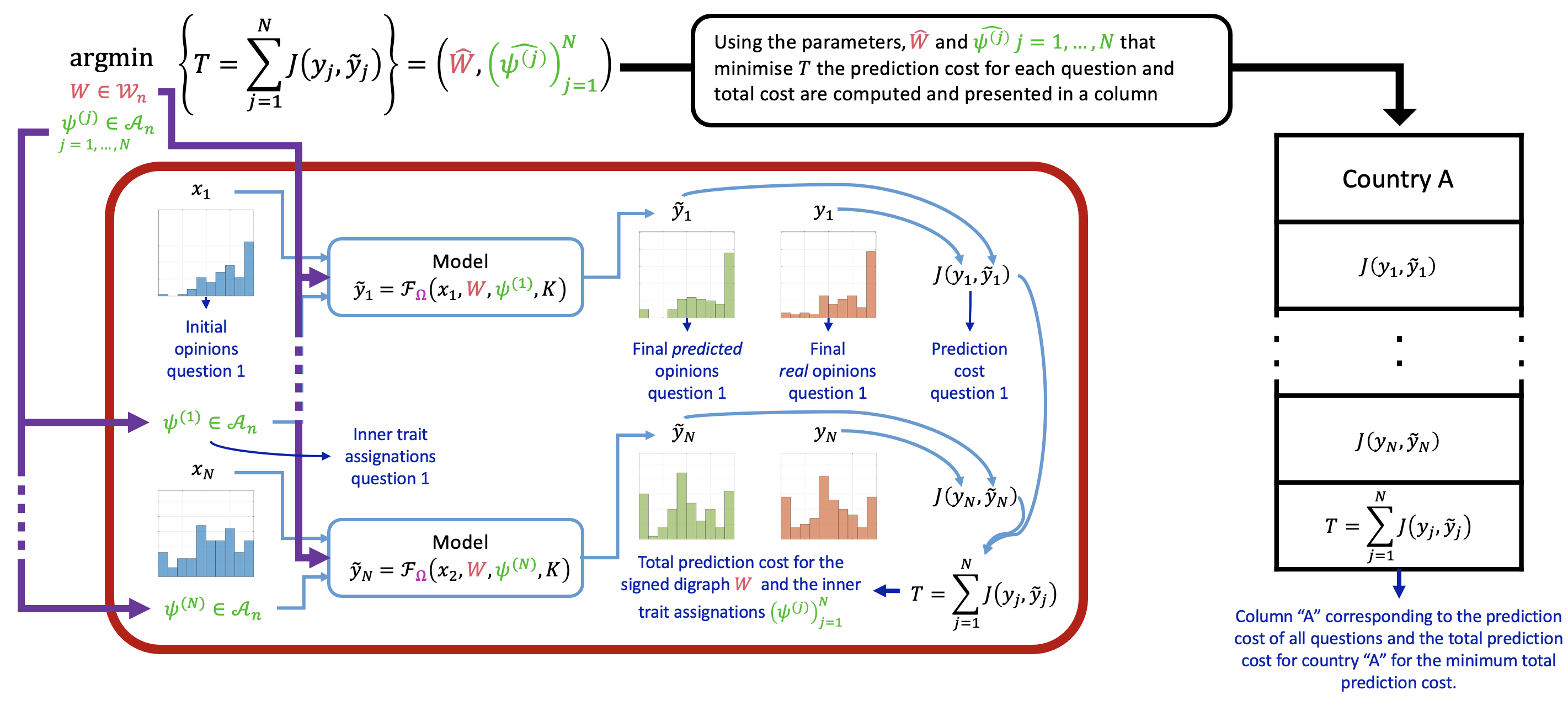}
         \caption{Visualisation of the procedure generating each column in Table \ref{Tab:ResultsCBFree} and of the minimisation problem in Equation \eqref{Eq:LargeOP}. Assume that a survey conducted in two separate occasions  in country A had $\numques$ questions. Given a signed digraph $\Weights$ and $\numques$ inner traits assignations $(\innertraits^{(\newj)})_{\newj=1}^\numques$ for each question, the model predicts a final opinion $\tilde{y}_\newj$. The cost function $\Cost(y_\newj,\tilde{y}_\newj)$ measures how close the predicted final opinion is to the real final opinion $y_\newj$. The sum of all these costs gives the total cost $T$. Minimising the value of $T$ over the signed digraph $\Weights$ and inner traits assignations $(\innertraits^{(\newj)})_{\newj=1}^\numques$ gives the parameters that best reproduce the society, $\widehat{\Weights}$, and $(\widehat{\innertraits^{(\newj)}})_{\newj=1}^\numques$. The cost for each question and the average and total cost obtained using these optimal parameters are reported in the column of Table \ref{Tab:ResultsCBFree} corresponding to the considered country. \CAbnew{All simulations evolved 100 agents.}}
         \label{Fig:GlobalDiagram}
 \end{figure}

\CAg{Assuming that the agents can have different inner traits for each question, Equation \eqref{Eq:LargeOP} was used to  \CAbnew{find model parameters for each country that yield opinions similar to the real ones}}. Once the parameters that solve the minimisation problem \eqref{Eq:LargeOP} were found for each country, the cost associated with the prediction discrepancy for each question-country pair was computed as in Equation \eqref{Eq:CostFunction} (see Figure \ref{Fig:GlobalDiagram}) and is shown in Table \ref{Tab:ResultsCBFree}. Due to its complexity and the huge size of the feasibility set, the minimisation problem is solved approximately: hence, a possibly suboptimal solution is found. By solving the optimisation problem more accurately, over a longer computation time (which we could not afford, due to the very large number of question-country pairs that we consider), even smaller costs could be achieved, and hence even better fits of the real data. 

  \begin{table}
\centering
\resizebox{0.7\textwidth}{!}{%
\begin{tabular}{||c || c | c | c | c | c | c | c | c | c | c | c | c | c | c | c | c | c | c | c | c | c | c | c | c | c | c||} 
\hline\hline 
& C1 &  C2 &  C3 &  C4 &  C5 &  C6 &  C7 &  C8 &  C9 &  C10 &  C11 &  C12 &  C13 &  C14 &  C15 &  C16 &  C17 &  C18 &  C19 &  C20 &  C21 &  C22 &  C23 &  C24 &  C25 &  C26 \\ 
\hline\hline 
 Q1 &  \cellcolor{green!20} 4.2 &  \cellcolor{green!20} 3 &  \cellcolor{green!20} 3 &  \cellcolor{green!20} 3.6 &  \cellcolor{green!20} 3 &  \cellcolor{green!20} 2.8 &  \cellcolor{green!20} 2.6 &  \cellcolor{green!20} 2.6 &  \cellcolor{green!20} 2.4 &  \cellcolor{green!20} 2.4 &  \cellcolor{green!20} 1.6 &  \cellcolor{green!20} 3.2 &  \cellcolor{green!20} 3.2 &  \cellcolor{green!20} 2.2 &  \cellcolor{green!20} 2.8 &  \cellcolor{green!20} 3.6 &  \cellcolor{green!20} 2.2 &  \cellcolor{green!20} 3 &  \cellcolor{green!20} 2.2 &  \cellcolor{green!20} 2.4 &  \cellcolor{green!20} 2.8 &  \cellcolor{green!20} 3.6 &  \cellcolor{green!20} 3.2 &  \cellcolor{green!20} 2 &  \cellcolor{green!20} 3.2 &  \cellcolor{green!20} 3.6 \\ 
 Q2 &  \cellcolor{green!20} 5.2 &  \cellcolor{green!20} 0.8 &  \cellcolor{green!20} 2.4 &  \cellcolor{green!20} 3.6 &  \cellcolor{green!20} 4 &  \cellcolor{green!20} 4.4 &  \cellcolor{green!20} 4 &  \cellcolor{green!20} 3.6 &  \cellcolor{green!20} 2.2 &  \cellcolor{green!20} 2.8 &  \cellcolor{green!20} 2.2 &  \cellcolor{green!20} 2.2 &  \cellcolor{green!20} 4.2 &  \cellcolor{green!20} 2.6 &  \cellcolor{green!20} 3.4 &  \cellcolor{green!20} 3.4 &  \cellcolor{green!20} 3.6 &  \cellcolor{green!20} 4.4 &  \cellcolor{green!20} 2 &  \cellcolor{green!20} 2.2 &  \cellcolor{green!20} 2.4 &  \cellcolor{green!20} 2.4 &  \cellcolor{green!20} 2.6 &  \cellcolor{green!20} 2.8 &  \cellcolor{green!20} 6.2 &  \cellcolor{green!20} 3.4 \\
\hline\hline 
\end{tabular}}
\caption{Results of the \emph{Free} optimisation problem using the Classification-Based model. Each column corresponds to a different country and each row to a different question. The cell values correspond to the optimal cost for all the countries and questions. The average cost along all the countries is 3.2815. The two final rows shows the column average and total. Cells with cost less than 7 are in green, the others are in red. Of the 780 possible question-country pairs, 755 have a cost less than 7 (an accuracy of $97\%$ in total). The average cost of accepted question-country pairs is 2.97.}
\label{Tab:ResultsCBFree}
\end{table}



\begin{figure}[h]
     \centering
     \begin{subfigure}[b]{0.32\textwidth}
         \centering
         \includegraphics[width=\textwidth]{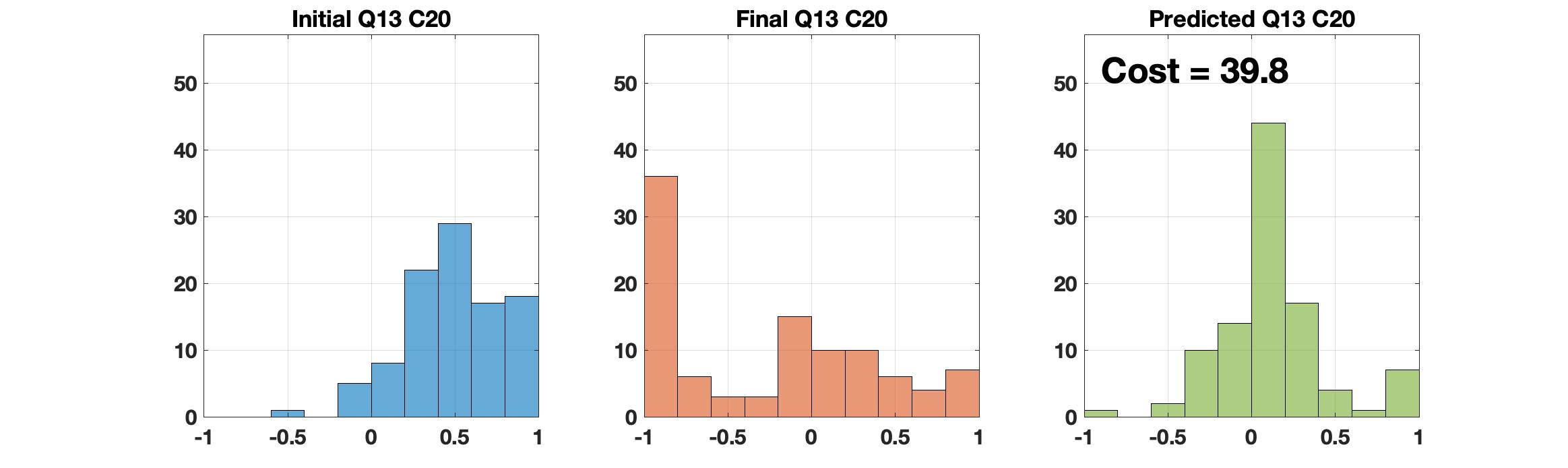}
     \end{subfigure}
     \hfill
     \begin{subfigure}[b]{0.32\textwidth}
         \centering
         \includegraphics[width=\textwidth]{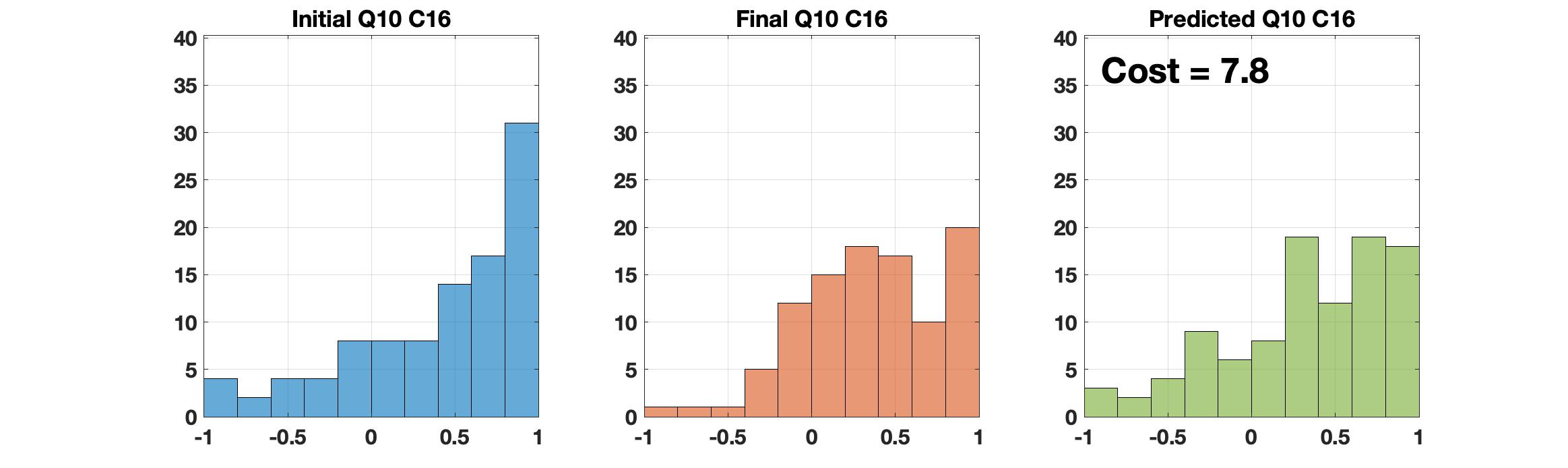}
     \end{subfigure}
     \hfill
     \begin{subfigure}[b]{0.32\textwidth}
         \centering
         \includegraphics[width=\textwidth]{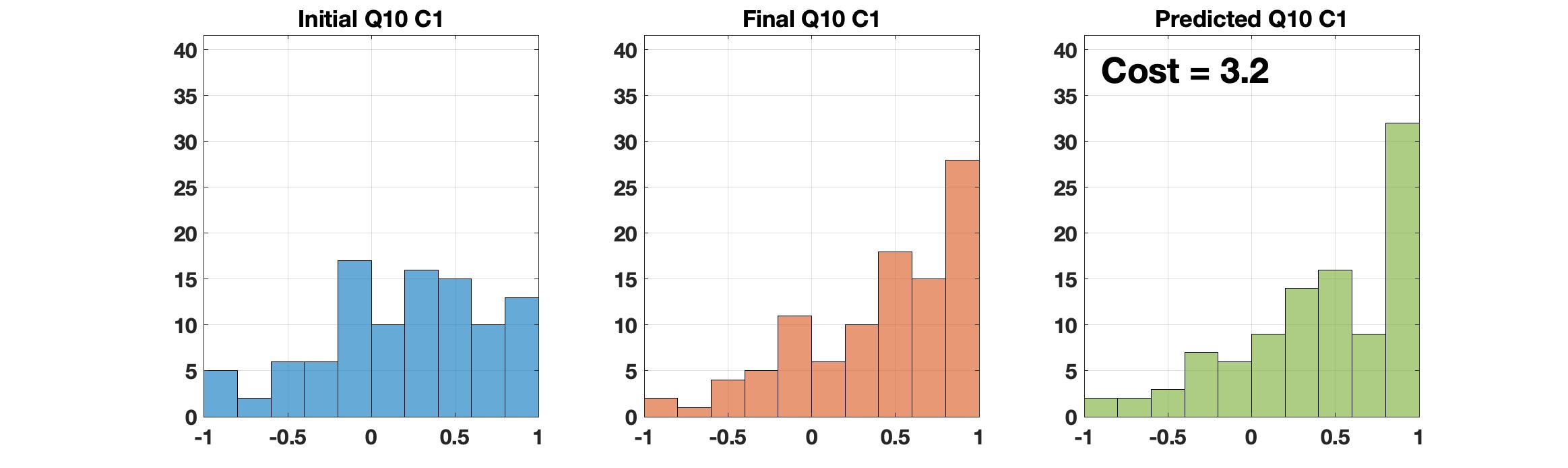} 
     \end{subfigure} \\
     \begin{subfigure}[b]{0.32\textwidth}
         \centering
         \includegraphics[width=\textwidth]{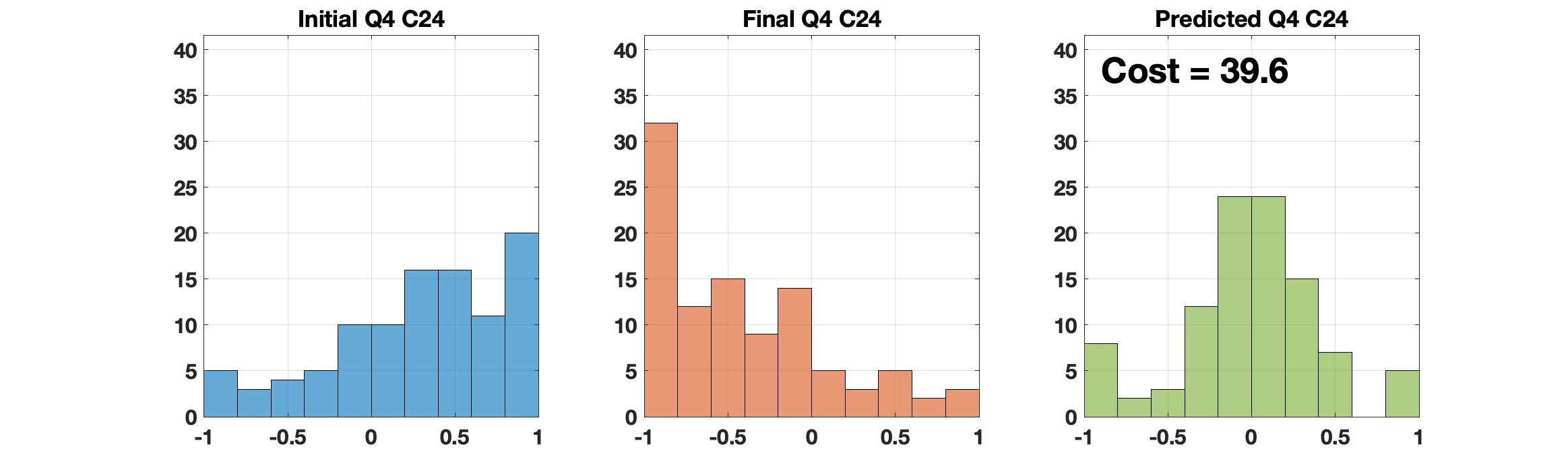}
     \end{subfigure}
     \hfill
     \begin{subfigure}[b]{0.32\textwidth}
         \centering
         \includegraphics[width=\textwidth]{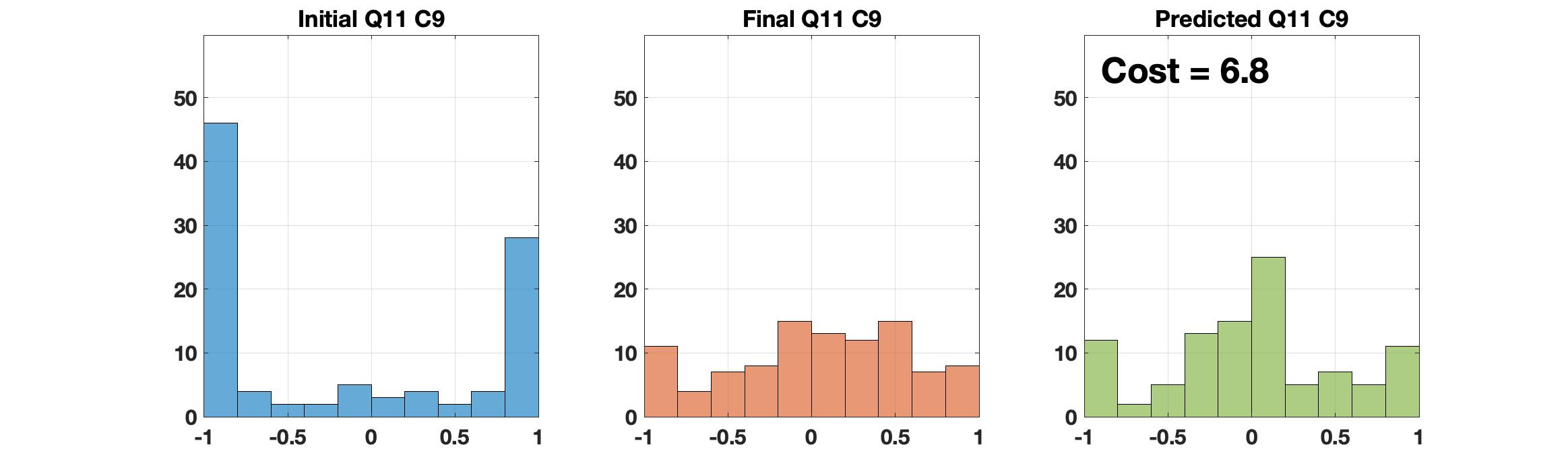}
     \end{subfigure}
     \hfill
     \begin{subfigure}[b]{0.32\textwidth}
         \centering
         \includegraphics[width=\textwidth]{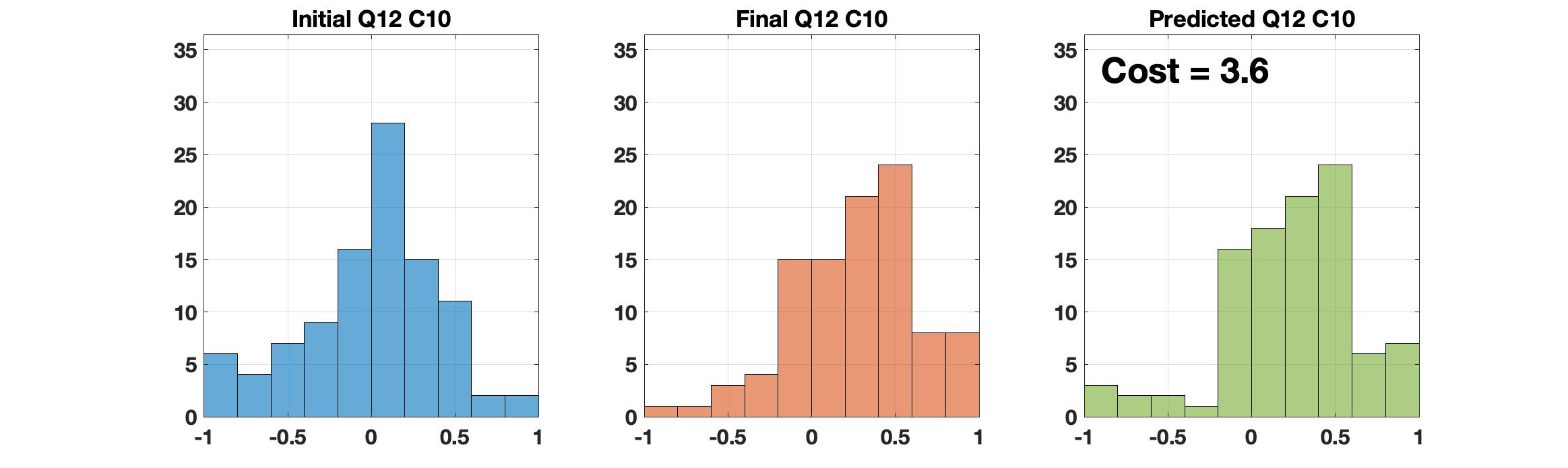}
     \end{subfigure} 
        \caption{Examples of the opinion predictions achieved with the CB model, with varying resulting costs. For each trio of histograms, the initial opinions are in blue, the real final opinions in orange, and the predicted final opinions in green. The value of the cost $J$ (in Equation \eqref{Eq:CostFunction}) is shown in the predicted histogram: cost values within 7 are shown to correspond to an accurate reproduction of the real opinion distribution.} 
        \label{Fig:Examples}
\end{figure}

Figure \ref{Fig:Examples} shows the model predictions for some question-country pairs. The original opinion is shown in blue, the real final opinion in orange, and the predicted final opinion in green; the corresponding cost (discrepancy) is reported. For costs less than 7, the model produces predicted final opinions that accurately represent the real final opinions.
These cases correspond to green cells in Table \ref{Tab:ResultsCBFree}, while cells with a cost higher than or equal to 7 are highlighted in red and constitute a small minority.

\CAh{To carry out a thorough comparison with standard models of opinion formation, an analysis equivalent to the one reported in Table \ref{Tab:ResultsCBFree} is performed also for the Null model (the model that assumes that the opinions do not change over time) and the French-DeGroot (FG) model \cite{French1956,Harary1959,Harary1965,DeGroot1974}. The results are reported in Tables \ref{Tab:ResultsNull} and \ref{Tab:ResultsFG}, respectively.}

 \CAbnew{To make Tables \ref{Tab:ResultsCBFree} and \ref{Tab:ResultsFG} comparable, for the FG model the digraphs used in each country are selected following the same minimisation problem as the one solved for the CB model. Since the FG model does not involve agent parameters, we only minimise over the set of digraphs $\SetN_\text{FG}$. Both the set of digraphs for the CB model, $\SetN$, and for the FG model, $\SetN_\text{FG}$, have the same number of elements and there is a one-to-one topology correspondence; the digraphs in $\SetN$ are signed and unweighted, while those in $\SetN_\text{FG}$ are unsigned and row-stochastic (as required by the different nature of the two models).}

 \CAh{Comparing Table \ref{Tab:ResultsCBFree} with Tables \ref{Tab:ResultsFG} and \ref{Tab:ResultsNull} shows that the CB model performs remarkably well, yielding a 97\% accuracy in contrast to the 43\% accuracy of the Null model and the 2\% accuracy of the French-DeGroot model.
In fact, from Table \ref{Tab:ResultsNull} it is clear that, although there is a strong tendency towards stubbornness and opinion distribution tend to change only slightly over time, keeping the opinions exactly constant does not lead to good predictions. As shown in Table \ref{Tab:ResultsFG}, the predictions of the French-DeGroot model are also not accurate, consistently with the evidence that perfect consensus is uncommon in real life.}

\begin{table}[!h]
\centering
\resizebox{0.7\textwidth}{!}{%
\begin{tabular}{||c || c | c | c | c | c | c | c | c | c | c | c | c | c | c | c | c | c | c | c | c | c | c | c | c | c | c||} 
\hline\hline 
& C1 &  C2 &  C3 &  C4 &  C5 &  C6 &  C7 &  C8 &  C9 &  C10 &  C11 &  C12 &  C13 &  C14 &  C15 &  C16 &  C17 &  C18 &  C19 &  C20 &  C21 &  C22 &  C23 &  C24 &  C25 &  C26 \\ 
\hline\hline 
 Q1 &  \cellcolor{green!20} 2 &  \cellcolor{red!20} 8 &  \cellcolor{green!20} 5.4 &  \cellcolor{red!20} 7.4 &  \cellcolor{green!20} 3.4 &  \cellcolor{green!20} 6 &  \cellcolor{red!20} 8.4 &  \cellcolor{red!20} 13.4 &  \cellcolor{red!20} 12.8 &  \cellcolor{red!20} 7.6 &  \cellcolor{green!20} 3 &  \cellcolor{green!20} 1.2 &  \cellcolor{green!20} 3.4 &  \cellcolor{green!20} 6 &  \cellcolor{red!20} 9 &  \cellcolor{red!20} 12.8 &  \cellcolor{green!20} 4 &  \cellcolor{green!20} 3.8 &  \cellcolor{green!20} 2.8 &  \cellcolor{red!20} 11.6 &  \cellcolor{red!20} 9.6 &  \cellcolor{green!20} 2.2 &  \cellcolor{green!20} 6.2 &  \cellcolor{red!20} 7.8 &  \cellcolor{green!20} 3.4 &  \cellcolor{green!20} 5.6 \\ 
 Q2 &  \cellcolor{green!20} 6 &  \cellcolor{green!20} 4.6 &  \cellcolor{green!20} 5.2 &  \cellcolor{red!20} 7.8 &  \cellcolor{red!20} 10.4 &  \cellcolor{red!20} 9.8 &  \cellcolor{red!20} 7 &  \cellcolor{red!20} 16.8 &  \cellcolor{red!20} 16.6 &  \cellcolor{green!20} 3.8 &  \cellcolor{green!20} 5.2 &  \cellcolor{green!20} 5 &  \cellcolor{green!20} 3.4 &  \cellcolor{red!20} 16 &  \cellcolor{green!20} 4.4 &  \cellcolor{red!20} 12.4 &  \cellcolor{red!20} 11 &  \cellcolor{green!20} 2 &  \cellcolor{green!20} 4 &  \cellcolor{green!20} 6.6 &  \cellcolor{green!20} 4 &  \cellcolor{green!20} 5.2 &  \cellcolor{green!20} 5.2 &  \cellcolor{green!20} 3.2 &  \cellcolor{green!20} 4.4 &  \cellcolor{green!20} 4.8 \\
\hline\hline 
\end{tabular}}
\caption{Results using the Null model, for comparison with Table \ref{Tab:ResultsCBFree}.  Out of 780 possible question-country pairs, 332 have a cost less than 7 (an accuracy of $43\%$ in total). The average cost of accepted question-country pairs is 4.17. }
\label{Tab:ResultsNull}
\end{table}

\begin{table}[!h]
\centering
\resizebox{0.7\textwidth}{!}{%
\begin{tabular}{||c || c | c | c | c | c | c | c | c | c | c | c | c | c | c | c | c | c | c | c | c | c | c | c | c | c | c||} 
\hline\hline 
& C1 &  C2 &  C3 &  C4 &  C5 &  C6 &  C7 &  C8 &  C9 &  C10 &  C11 &  C12 &  C13 &  C14 &  C15 &  C16 &  C17 &  C18 &  C19 &  C20 &  C21 &  C22 &  C23 &  C24 &  C25 &  C26 \\ 
\hline\hline 
 Q1 &  \cellcolor{red!20} 28.6 &  \cellcolor{red!20} 41.2 &  \cellcolor{red!20} 32.4 &  \cellcolor{red!20} 32 &  \cellcolor{red!20} 35 &  \cellcolor{red!20} 37.6 &  \cellcolor{red!20} 32.6 &  \cellcolor{red!20} 33.6 &  \cellcolor{red!20} 34.6 &  \cellcolor{red!20} 31.6 &  \cellcolor{red!20} 26.8 &  \cellcolor{red!20} 32 &  \cellcolor{red!20} 33.6 &  \cellcolor{red!20} 36 &  \cellcolor{red!20} 36.2 &  \cellcolor{red!20} 33 &  \cellcolor{red!20} 28.8 &  \cellcolor{red!20} 27.4 &  \cellcolor{red!20} 30.6 &  \cellcolor{red!20} 36.8 &  \cellcolor{red!20} 32 &  \cellcolor{red!20} 31.8 &  \cellcolor{red!20} 33 &  \cellcolor{red!20} 39.2 &  \cellcolor{red!20} 25.8 &  \cellcolor{red!20} 34.6 \\ 
 Q2 &  \cellcolor{red!20} 32.2 &  \cellcolor{red!20} 34.8 &  \cellcolor{red!20} 28.6 &  \cellcolor{red!20} 30.6 &  \cellcolor{red!20} 34 &  \cellcolor{red!20} 37.6 &  \cellcolor{red!20} 39.2 &  \cellcolor{red!20} 38.4 &  \cellcolor{red!20} 33.2 &  \cellcolor{red!20} 31.2 &  \cellcolor{red!20} 28 &  \cellcolor{red!20} 32.4 &  \cellcolor{red!20} 31.8 &  \cellcolor{red!20} 41.2 &  \cellcolor{red!20} 31.8 &  \cellcolor{red!20} 36.6 &  \cellcolor{red!20} 26.6 &  \cellcolor{red!20} 23.4 &  \cellcolor{red!20} 32.8 &  \cellcolor{red!20} 32.8 &  \cellcolor{red!20} 35 &  \cellcolor{red!20} 36.4 &  \cellcolor{red!20} 32.4 &  \cellcolor{red!20} 40.4 &  \cellcolor{red!20} 30.2 &  \cellcolor{red!20} 28.8 \\
\hline\hline 
\end{tabular}}
\caption{Results of the optimisation problem using the French-DeGroot model, for comparison with Table \ref{Tab:ResultsCBFree}. The average cost along all the countries is 38.4323. Of the 780 possible question-country pairs, 13 have a cost less than 7 (an accuracy of $2\%$ in total). The average cost of accepted question-country pairs is 5.43.}
\label{Tab:ResultsFG}
\end{table}



%

\begin{figure}[h]
     \centering
          \begin{subfigure}[t]{0.49\textwidth}
         \centering
         \includegraphics[width=0.5\textwidth]{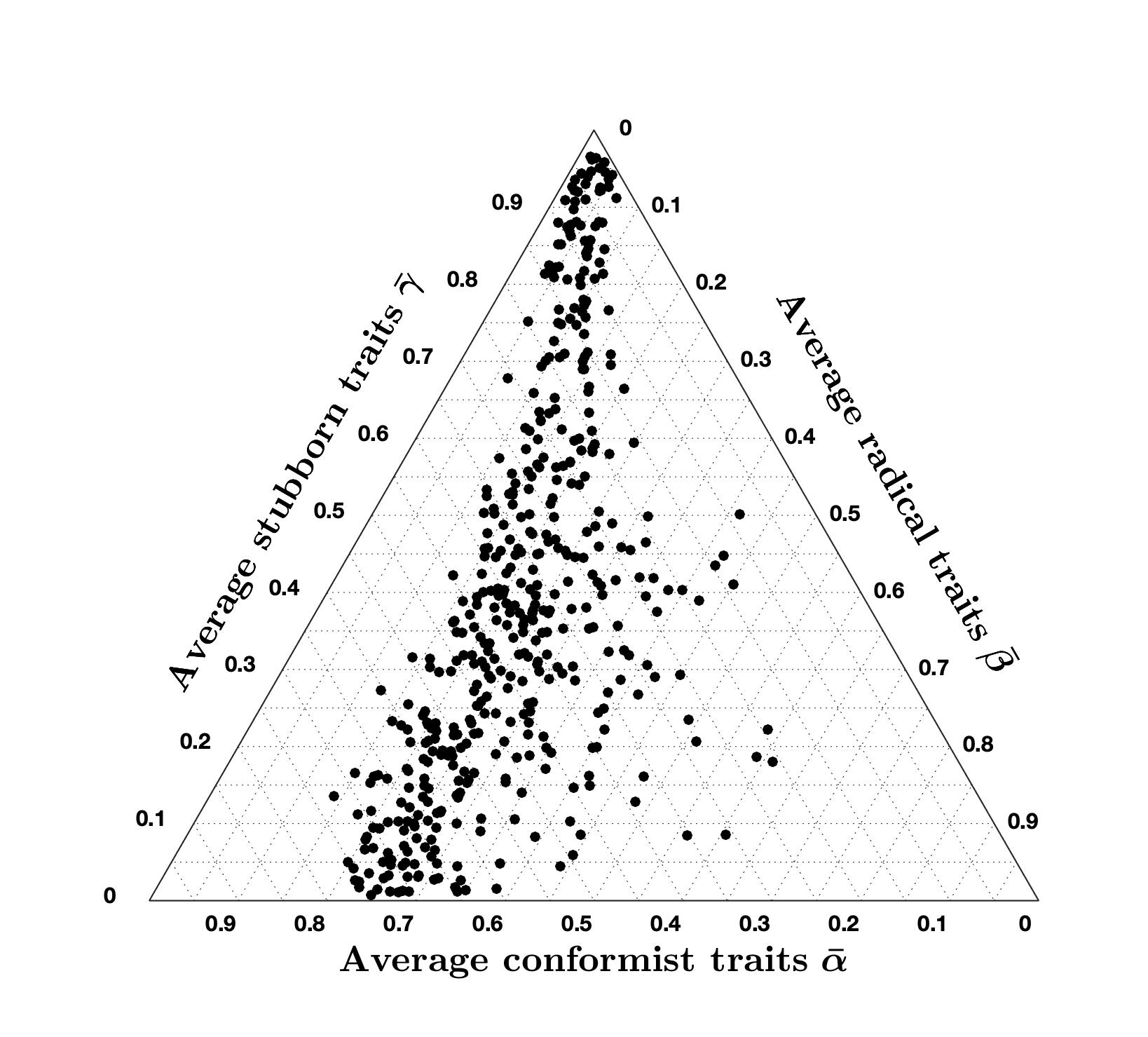}
         \caption{Ternary diagram plotting the average inner traits for the question-country pairs with cost $< 7$ according to Table \ref{Tab:ResultsCBFree}.}
         \label{Fig:TD1a}
     \end{subfigure}
     \hfill
     \begin{subfigure}[t]{0.49\textwidth}
         \centering
         \includegraphics[width=0.5\textwidth]{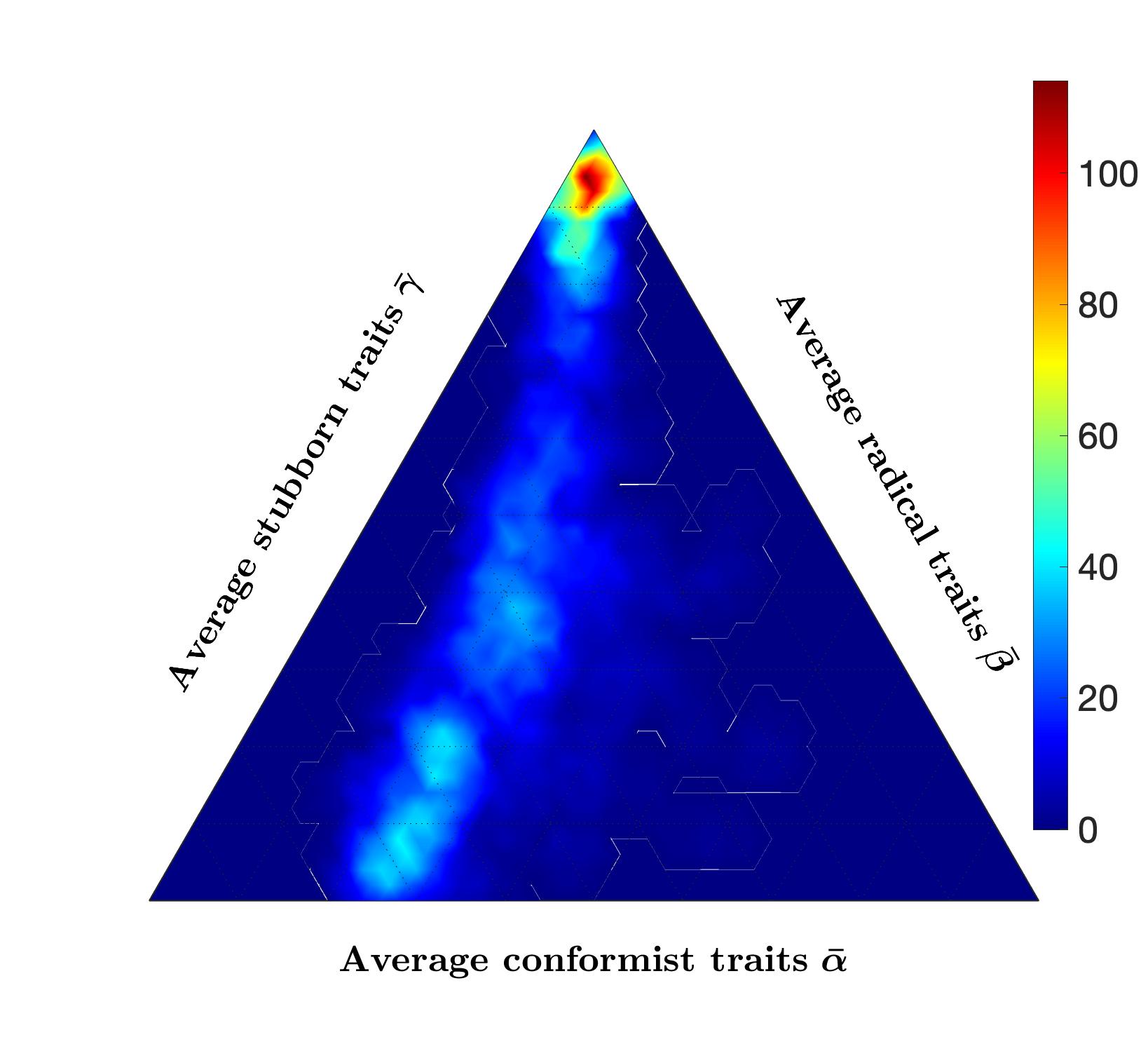}
         \caption{Density plot of the ternary diagram: most question-country pairs have an average agent with more than $95\%$ stubborn traits.}
         \label{Fig:TD1b}
     \end{subfigure}
        \caption{Analysis of the location of the average agents for all the question-country pairs with cost less than 7.}
        \label{Fig:TD1}
\end{figure}

Plotting the average inner traits $\bar{\innertraits}$ for all question-country pairs for which the cost is less than 7  \CAbnew{provides possible hints on how these societies could potentially be formed. However, because of the large parameter space and relatively small data set, we cannot make conclusive statements on actual societies just based on the optimisation results, as very different inner traits assignations may produce similarly low costs: we just propose a \emph{possible} explanation}. The resulting ternary diagram is presented in Figure \ref{Fig:TD1}. Figure  \ref{Fig:TD1a} shows the position of each question-country pair. Figure \ref{Fig:TD1b} shows a density plot over the ternary diagram indicating the regions where most question-country pairs are found.

 \CAbnew{Despite the small data set and possible multiple local minima with similar low cost, fitting real data to gain an insight into the composition of actual societies reveals a clear trend: most average inner traits include a strong stubborn component, as shown by the high density in the stubbornness corner in Figure \ref{Fig:TD1b}. Also, the non-stubborn part can be roughly divided into $70\%$ conformist and $30\%$ radical, as shown by the trend in Figure \ref{Fig:TD1a}. This distribution is almost constant across all question-country pairs. Again, this is a possible explanation, and more data and more thorough explorations of the parameter space (extremely challenging from a computational standpoint) would be needed to make more conclusive statements.}
\CAf{Hence, this is not conclusive evidence that most people are stubborn. There may be other explanations, for instance that not too many opinion exchange events take place in an average person's life. Graph-theoretically speaking, \emph{isolation} due to the lack of outgoing edges from a node (i.e., lack of interactions) is associated with the concept of \emph{stubborness}. However, from a mathematical model it is impossible to draw conclusions on whether the opinion of an agent remains unchanged because the agent refuses to consider the different opinions it is exposed to, or because the agent intentionally avoids exposure to different opinions, or because the agent simply lacks the opportunity to come into contact with different opinions.}  \CAbnew{Furthermore, the traits themselves can be interpreted in different ways: for instance, a lower value of stubbornness can be regarded as a greater openness to change.}

\textbf{Parameter Variation:}
The results presented in Table \ref{Tab:ResultsCBFree} and Figures \ref{Fig:Examples} and \ref{Fig:TD1} are obtained by solving the minimisation problem \eqref{Eq:LargeOP} with nominal opinion evolution parameters $\lambda = 0.4$, $\xi = 2$, and $\mu = 5$. We now analyse the results of the minimisation problem when these parameters are changed. Tables \ref{table:var_lambda} to \ref{table:var_mu} present how this variation affects the percentage of accurate question-country pairs (namely, those associated with a cost smaller than 7), the average cost of accurate question-country pairs, and the ternary diagram plot.


\begin{table}[h!]
\centering
\begin{tabular}{||p{5cm} || p{1.2in} | p{1.2in} | p{1.2in} ||} 
	\hline\hline
 & $\lambda = 0.2$ & $\lambda = 0.4$ & $\lambda = 0.8$ \\ \hline\hline
\% of accepted country-question pairs & 93.7 & 96.8 & 97.8 \\ \hline
Average cost of accepted country-question pairs & 2.79 & 2.97 & 3.02 \\ \hline
Ternary Diagram Plot &  \parbox[l]{0.05em}{\includegraphics[width=1.2in]{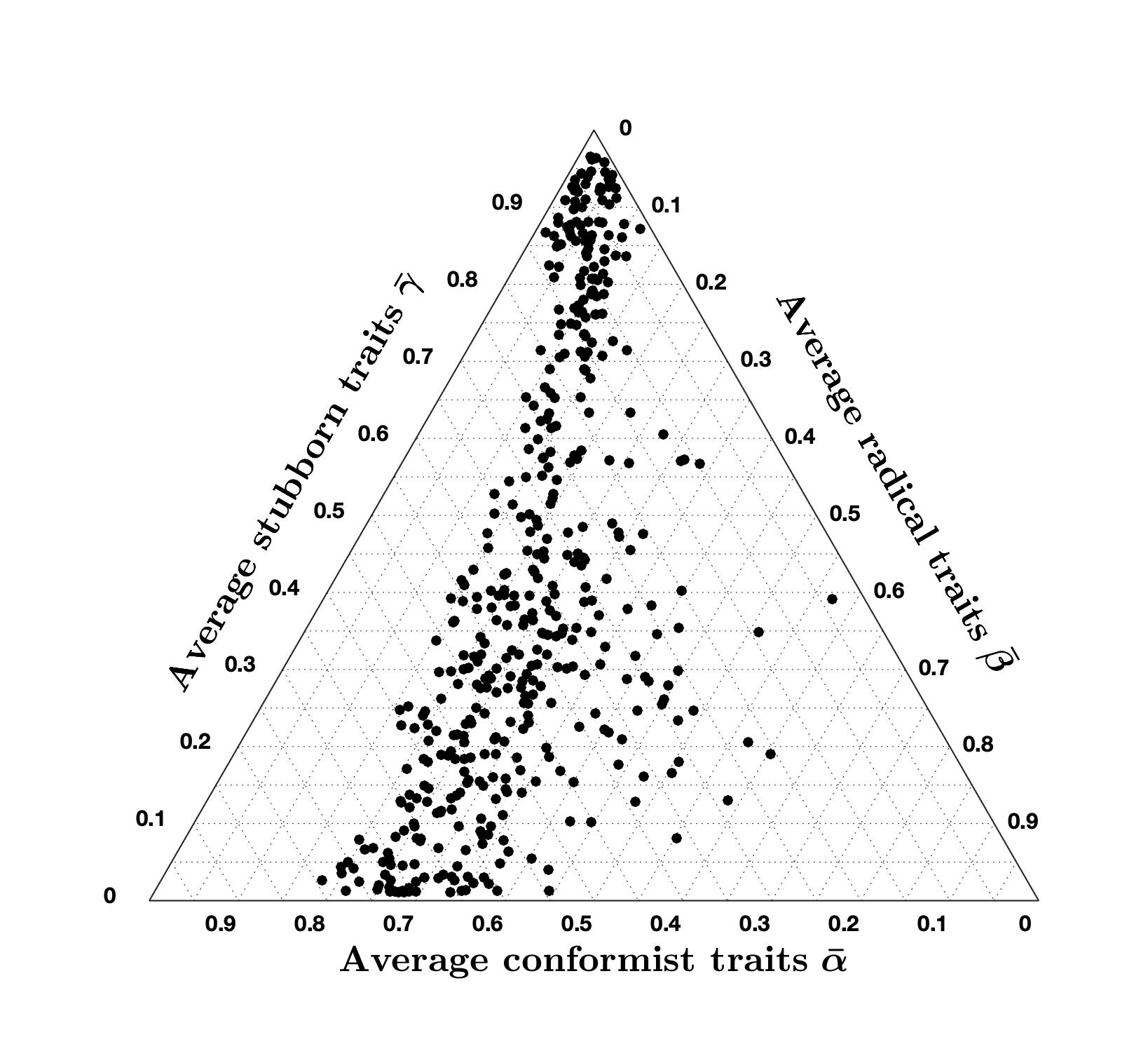}}  &  \parbox[l]{0.05em}{\includegraphics[width=1.2in]{Ternary1_lambda_40_xi_200_mu_500.jpg}}  &  \parbox[l]{0.05em}{\includegraphics[width=1.2in]{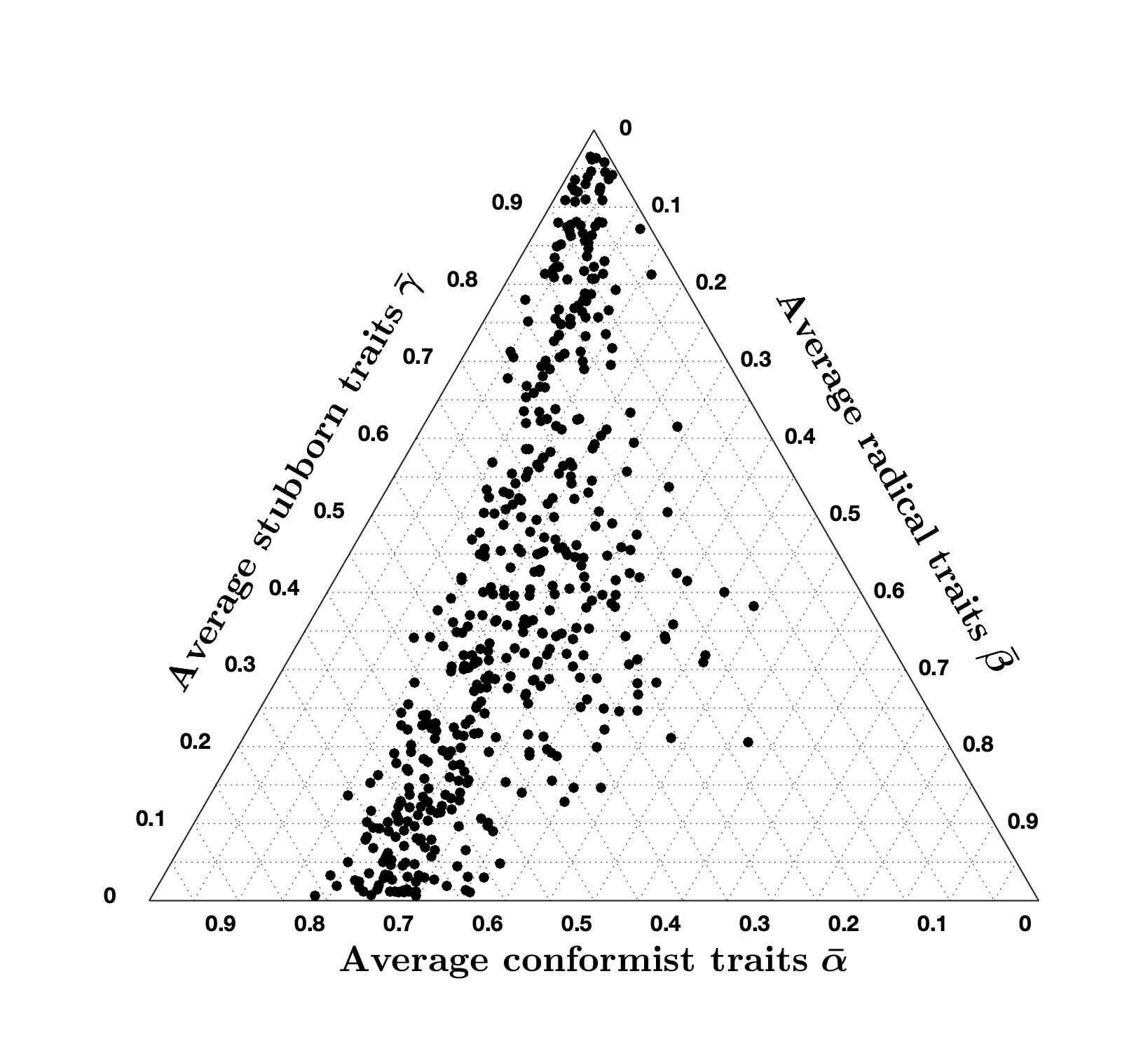}}  \\ 
	\hline\hline
\end{tabular} 
\caption{Effects of varying $\lambda$ while keeping the nominal values $\xi = 2$, and $\mu=5$.} 
\label{table:var_lambda}
	\end{table}
\begin{table}[h!]
\centering
\begin{tabular}{||p{5cm} || p{1.2in} | p{1.2in} | p{1.2in} ||} 
	\hline\hline
 & $\xi = 1$ & $\xi = 2$ & $\xi = 4$ \\ \hline\hline
\% of accepted country-question pairs & 96 & 96.8 & 95.8 \\ \hline
Average cost of accepted country-question pairs & 2.84 & 2.97 & 3.45 \\ \hline
Ternary Diagram Plot &  \parbox[l]{0.05em}{\includegraphics[width=1.2in]{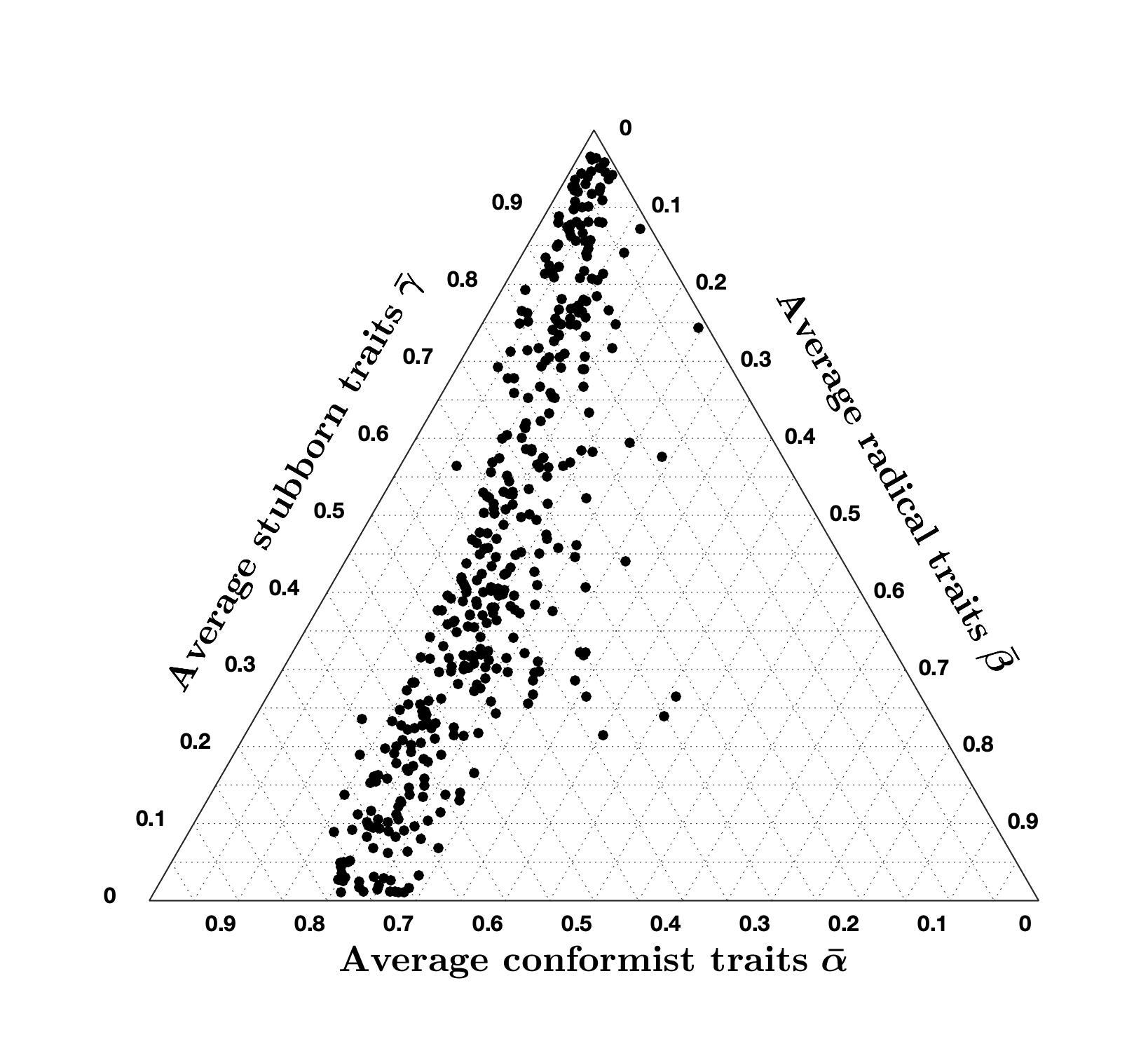}}  &  \parbox[l]{0.05em}{\includegraphics[width=1.2in]{Ternary1_lambda_40_xi_200_mu_500.jpg}}  &  \parbox[l]{0.05em}{\includegraphics[width=1.2in]{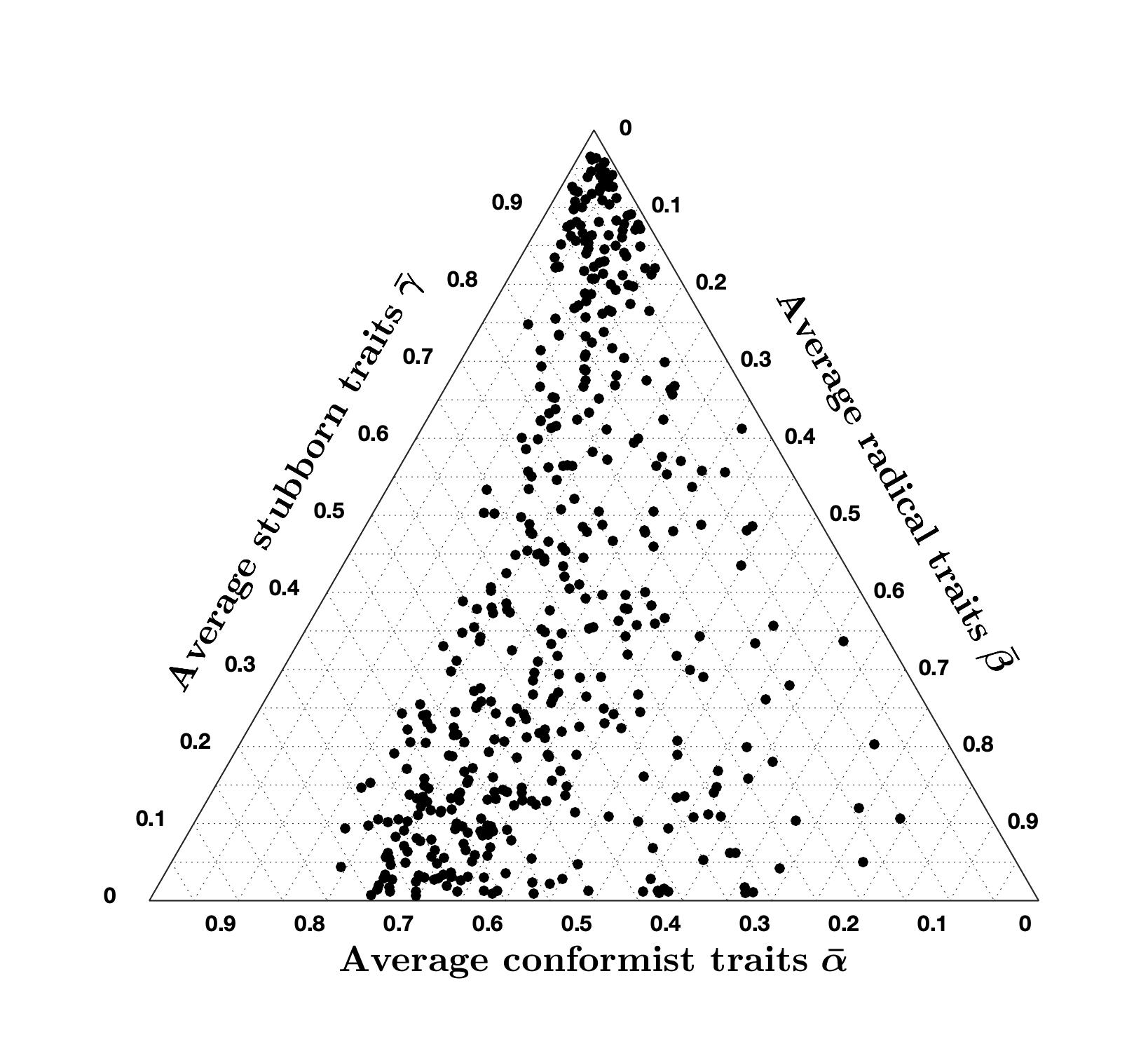}}  \\ 
	\hline\hline
\end{tabular} 
\caption{Effects of varying $\xi$ while keeping the nominal values $\mu = 5$, and $\lambda=0.4$.} 
\label{table:var_xi}
	\end{table}
\begin{table}[h!]
\centering
\begin{tabular}{||p{5cm} || p{1.2in} | p{1.2in} | p{1.2in} ||} 
	\hline\hline
 & $\mu = 2.5$ & $\mu = 5$ & $\mu = 10$ \\ \hline\hline
\% of accepted country-question pairs & 96.8 & 96.8 & 97.2 \\ \hline
Average cost of accepted country-question pairs & 2.84 & 2.97 & 3.15 \\ \hline
Ternary Diagram Plot &  \parbox[l]{0.05em}{\includegraphics[width=1.2in]{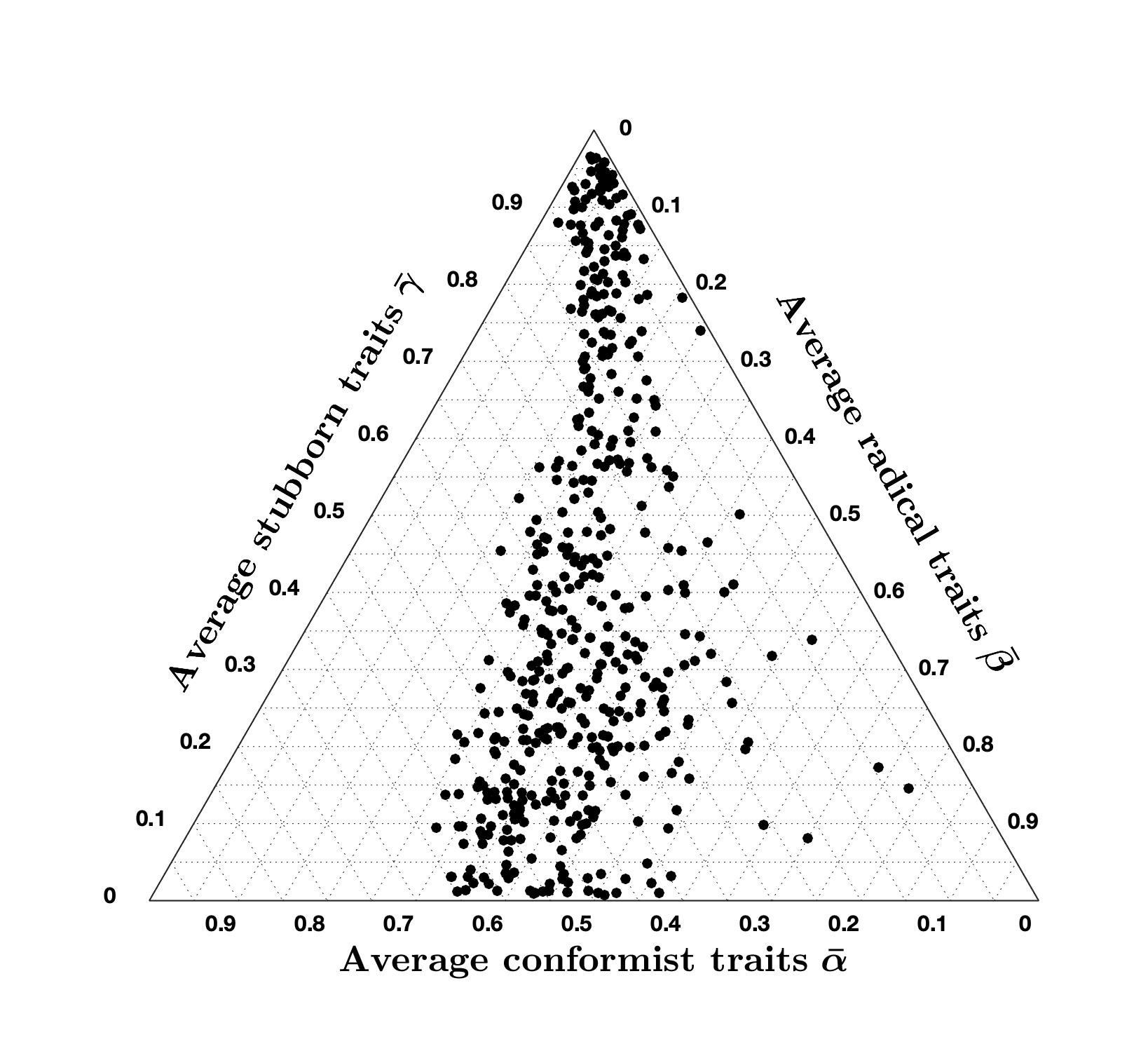}}  &  \parbox[l]{0.05em}{\includegraphics[width=1.2in]{Ternary1_lambda_40_xi_200_mu_500.jpg}}  &  \parbox[l]{0.05em}{\includegraphics[width=1.2in]{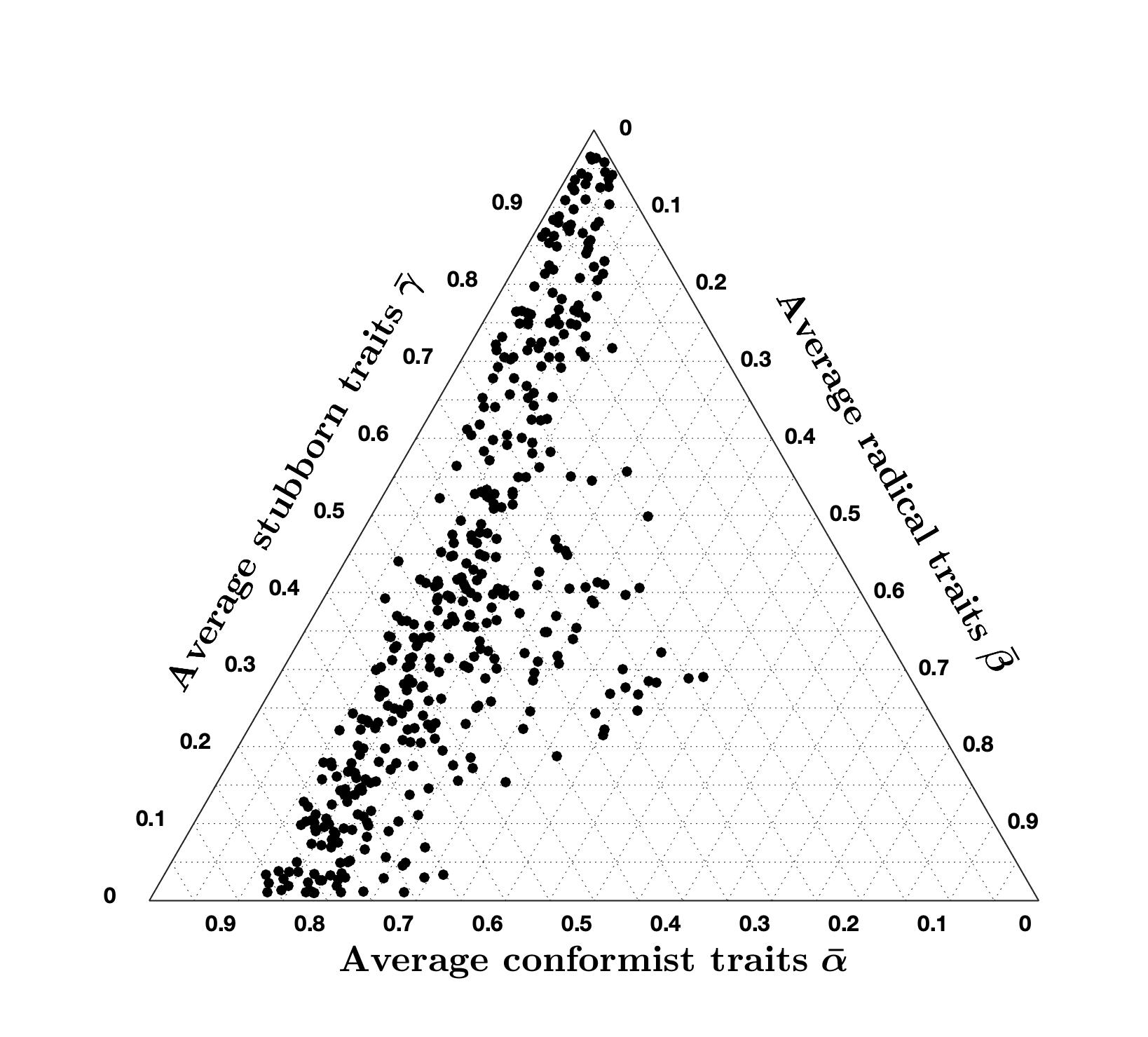}}  \\ 
	\hline\hline
\end{tabular} 
\caption{Effects of varying $\mu$ while keeping the nominal values $\lambda = 0.4$, and $\xi=2$.} 
\label{table:var_mu}
	\end{table}

Tables \ref{table:var_lambda} to \ref{table:var_mu} show that, even after varying the values of $\lambda$, $\xi$, and $\mu$, the percentage of accurate question-country pairs remains around $96\%$, and the average cost of accurate question-country pairs is between $2.79$ and $3.45$ which is quite remarkable since it means that the high accuracy achieved with the CB model is very robust to parameter variations. 

Comparing the ternary diagrams shows the persistent tendency of question-country pairs to lie along a line where the proportion between conformist and radical traits is constant. For most simulation results, this proportion is still $70\%$ conformist and $30\%$ radical, as in the nominal case (Figure \ref{Fig:TD1a}). The proportion only changes when varying $\mu$: for $\mu = 2.5$, we have $60\%$ conformist and $40\%$ radical agents, while for $\mu = 10$ we have $80\%$ conformist and $20\%$ radical agents. Therefore, it appears that $\mu$ can be tuned to regulate this proportion.

\subsubsection{\emph{Constrained} Optimisation Problem}

\CAg{If the agents are assumed to have the same inner traits for every question, then the model parameters can be found using the \emph{constrained} optimisation problem in Equation \eqref{Eq:LargeOP_mod}. One advantage of using this approach is that, since each country has the same topology and inner traits assignation for all the questions, these parameters can be identified by solving the \emph{constrained} optimisation problem \eqref{Eq:LargeOP_mod} for a subset of all available questions (training dataset), and then tested on the remaining questions (test dataset). This was not possible previously, when assuming a different inner traits assignation associated with each question.}\\
\CAg{This procedure is commonly known as cross-validation. Generally, a subset of available data is used to train an algorithm (in this case, to identify the model parameters $\widehat{\Weights}$ and $\widehat{\innertraits}$) and the remaining data is used to test the trained algorithm (in this case, the model with identified parameters $\widehat{\Weights}$ and $\widehat{\innertraits}$). To eliminate result biases due to the selected training datasets and test datasets, cross-validation is performed multiple times for different partitions of the data. A common approach is to divide the data in $K$ subsets and validate the model $K$ times so that, at each iteration, only one subset is taken as the test dataset. This is known as $K$-fold cross-validation.}\\
\CAg{Table \ref{tab:Rest_Opt} shows the result of sixfold cross-validation on the available data (the questions are divided in six subsets of five questions each: $\{1, \dots, 5\}$, $\{6, \dots, 10\}$, $\dots$, $\{26, \dots, 30\}$ ). The first six rows show the mean cost for the five questions in the test dataset for each country for each cross-validation (CV1 to CV6). The last row shows the mean of the first six rows.}\\
\CAg{The simulation results summarised in Table \ref{tab:Rest_Opt} show that the model is able to accurately reproduce the final opinions for the tested data. Although the values are higher than 7, it is important to note that these predictions are done based on the assumption that the inner traits are the same for every question, while in reality the inner traits of the agents may change when considering their attitude towards different types of questions (which is taken into account by the free optimisation approach).}


\begin{table}[h!]
\centering
\resizebox{\textwidth}{!}{%
\begin{tabular}{||c || c | c | c | c | c | c | c | c | c | c | c | c | c | c | c | c | c | c | c | c | c | c | c | c | c | c||} 
\hline\hline 
 & C1 & C2 & C3 & C4 & C5 & C6 & C7 & C8 & C9 & C10 & C11 & C12 & C13 & C14 & C15 & C16 & C17 & C18 & C19 & C20 & C21 & C22 & C23 & C24 & C25 & C26 \\ 
\hline 
CV1 & 7.4 & 8.5 & 12.6 & 7.4 & 13.8 & 21.3 & 18.5 & 13.9 & 24.7 & 9.6 & 10.4 & 8.2 & 6.3 & 10.4 & 11.4 & 11 & 7.8 & 9.4 & 9.1 & 13.6 & 12 & 7.3 & 11.2 & 9.9 & 5.4 & 9.4 \\ 
CV2 & 6.3 & 8.4 & 9.3 & 11.4 & 6.6 & 16.8 & 15.6 & 19.6 & 14.3 & 12.6 & 9.7 & 7.6 & 8.6 & 15.5 & 5.7 & 10.7 & 7.8 & 7.5 & 6 & 18.5 & 16.9 & 8.6 & 7.5 & 7.1 & 6.3 & 8.6 \\ 
CV3 & 8.5 & 10.7 & 10.5 & 10.6 & 7.7 & 12.5 & 20.1 & 34.6 & 14.7 & 10.4 & 14.5 & 10.3 & 8.8 & 8.1 & 11.9 & 12.6 & 10.9 & 8.3 & 10.2 & 16.3 & 10 & 8.8 & 12.9 & 6.8 & 10.4 & 12.7 \\ 
CV4 & 10.1 & 10.6 & 11.6 & 5.7 & 10.9 & 19.1 & 10.6 & 19.7 & 20.4 & 13.6 & 13.9 & 9 & 12.8 & 9.4 & 11.5 & 14 & 12.1 & 7.7 & 7.7 & 26.8 & 14.6 & 11.7 & 9.8 & 18.1 & 7.6 & 10.9 \\ 
CV5 & 9.7 & 6.8 & 9.9 & 6.6 & 20.5 & 7.7 & 10.1 & 13.6 & 7.9 & 8.8 & 8.4 & 8.2 & 8.9 & 7.5 & 11.4 & 13 & 8.1 & 15.2 & 7.5 & 15.2 & 8.2 & 5.3 & 10.7 & 7.8 & 5.8 & 15.2 \\ 
CV6 & 11.2 & 9 & 14.9 & 13.7 & 20.3 & 11.4 & 11.4 & 22.5 & 14.9 & 16.6 & 18.2 & 8.7 & 11.9 & 21.1 & 11.2 & 9 & 16.2 & 16.2 & 6.8 & 19.8 & 8.6 & 9 & 12.4 & 20.3 & 10.9 & 9.4 \\ 
\hline
 Mean & 8.9 & 9 & 11.5 & 9.2 & 13.3 & 14.8 & 14.4 & 20.7 & 16.1 & 11.9 & 12.5 & 8.7 & 9.5 & 12 & 10.5 & 11.7 & 10.5 & 10.7 & 7.9 & 18.4 & 11.7 & 8.4 & 10.8 & 11.7 & 7.7 & 11 \\ 
\hline\hline  
\end{tabular}} 
\caption{\CAg{Results of the sixfold cross validation. Each column corresponds to a country, and each row to one of the six cross validations.  The value in cell $(i,j)$ is the average cost of the test data in cross validation $i$ for country $j$.  The last row represents the mean over all the rows.}} 
\label{tab:Rest_Opt} 
\end{table}

\CAnewcomment{Table \ref{Tab:ResultsCBRest} is analogous to Table \ref{Tab:ResultsCBFree}, but now the model parameters are obtained with the \emph{Constrained} optimisation problem \eqref{Eq:LargeOP_mod}, which yields a higher cost, as expected, since the optimised inner traits assignations can be very different when unconstrained, see Figure \ref{Fig:TD1a}.}

\begin{table}
\centering
\resizebox{0.7\textwidth}{!}{%
\begin{tabular}{||c || c | c | c | c | c | c | c | c | c | c | c | c | c | c | c | c | c | c | c | c | c | c | c | c | c | c||} 
\hline\hline 
& C1 &  C2 &  C3 &  C4 &  C5 &  C6 &  C7 &  C8 &  C9 &  C10 &  C11 &  C12 &  C13 &  C14 &  C15 &  C16 &  C17 &  C18 &  C19 &  C20 &  C21 &  C22 &  C23 &  C24 &  C25 &  C26 \\ 
\hline\hline 
 Q1 &  \cellcolor{red!20} 9.4 &  \cellcolor{red!20} 11.8 &  \cellcolor{red!20} 18.6 &  \cellcolor{green!20} 5.2 &  \cellcolor{red!20} 10.2 &  \cellcolor{red!20} 12.2 &  \cellcolor{red!20} 9.2 &  \cellcolor{red!20} 22 &  \cellcolor{green!20} 6 &  \cellcolor{red!20} 15.8 &  \cellcolor{red!20} 26.8 &  \cellcolor{green!20} 5.8 &  \cellcolor{red!20} 8.4 &  \cellcolor{red!20} 7.2 &  \cellcolor{red!20} 7 &  \cellcolor{green!20} 4.8 &  \cellcolor{red!20} 12.2 &  \cellcolor{red!20} 15.6 &  \cellcolor{green!20} 6.6 &  \cellcolor{green!20} 4.6 &  \cellcolor{green!20} 5.4 &  \cellcolor{green!20} 5.8 &  \cellcolor{red!20} 12.6 &  \cellcolor{red!20} 8 &  \cellcolor{red!20} 7.6 &  \cellcolor{red!20} 7 \\ 
 Q2 &  \cellcolor{red!20} 11.6 &  \cellcolor{red!20} 10.4 &  \cellcolor{red!20} 13 &  \cellcolor{red!20} 8 &  \cellcolor{red!20} 22.2 &  \cellcolor{red!20} 14.2 &  \cellcolor{red!20} 8 &  \cellcolor{red!20} 13.6 &  \cellcolor{red!20} 8.8 &  \cellcolor{red!20} 14 &  \cellcolor{red!20} 20.4 &  \cellcolor{green!20} 3.4 &  \cellcolor{red!20} 8.4 &  \cellcolor{red!20} 25 &  \cellcolor{green!20} 6.4 &  \cellcolor{green!20} 6.2 &  \cellcolor{red!20} 26.4 &  \cellcolor{red!20} 15.4 &  \cellcolor{green!20} 6 &  \cellcolor{red!20} 12.2 &  \cellcolor{red!20} 8 &  \cellcolor{green!20} 6.8 &  \cellcolor{red!20} 12.4 &  \cellcolor{green!20} 6.8 &  \cellcolor{red!20} 7.6 &  \cellcolor{red!20} 8.2 \\
\hline\hline 
\end{tabular}}
\caption{Results of the \emph{Constrained} optimisation problem using the Classification-Based model. The average cost along all the countries is 11.6746. Out of 780 possible question-country pairs, 220 have a cost less than 7 (an accuracy of $28\%$ in total). The average cost of accepted question-country pairs is 5.16. }
\label{Tab:ResultsCBRest}
\end{table}



\subsubsection{ Transition Tables}

\begin{figure}[h]
      \centering
          \includegraphics[width=0.9\textwidth]{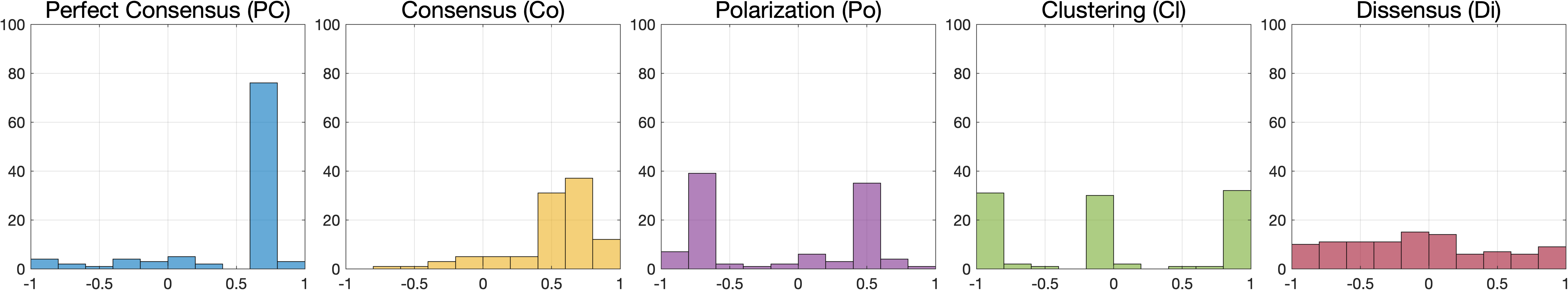}
         \caption{  \CAbnew{Histograms of opinion vectors that are representative of the five possible qualitative opinion distribution categories: Perfect Consensus, Consensus, Polarization, Clustering, and Dissensus.}}
         \label{Fig:ClassifiedHistograms}
 \end{figure}

  \CAbnew{Let $x_o$ be an initial opinion vector, and $x_f$ the final opinion vector predicted by the model. Both $x_o$ and $x_f$ can be sorted into one of five possible opinion distribution categories shown in Figure \ref{Fig:ClassifiedHistograms}: 1) \emph{perfect consensus}, PC; 2) \emph{consensus}, Co; 3) \emph{polarization}, Po; 4) \emph{clustering}, Cl; 5) \emph{dissensus}, Di. The reader is referred to \cite{Devia2022A} for more details.}

  \CAbnew{Now, let $X_o$ be a collection of initial opinions $X_o=\{x_o\}$ and $X_f$ the corresponding collection of predicted opinions $X_f= \{x_f\}$. A \emph{transition table} $T$, with 5 rows and 5 columns corresponding to the five possible opinion categories, is computed so that the coefficient in cell $(a,b)$ is the number of initial opinion vectors  $x_o\in X_o$ belonging to category $a$ for which the corresponding predicted opinion $x_f$ belongs to the category $b$. The table shows whether the model can evolve initial opinions belonging to any category into predicted opinions belonging to any other category.}
  
\begin{figure}[!h]
\centering
\includegraphics[width=\textwidth]{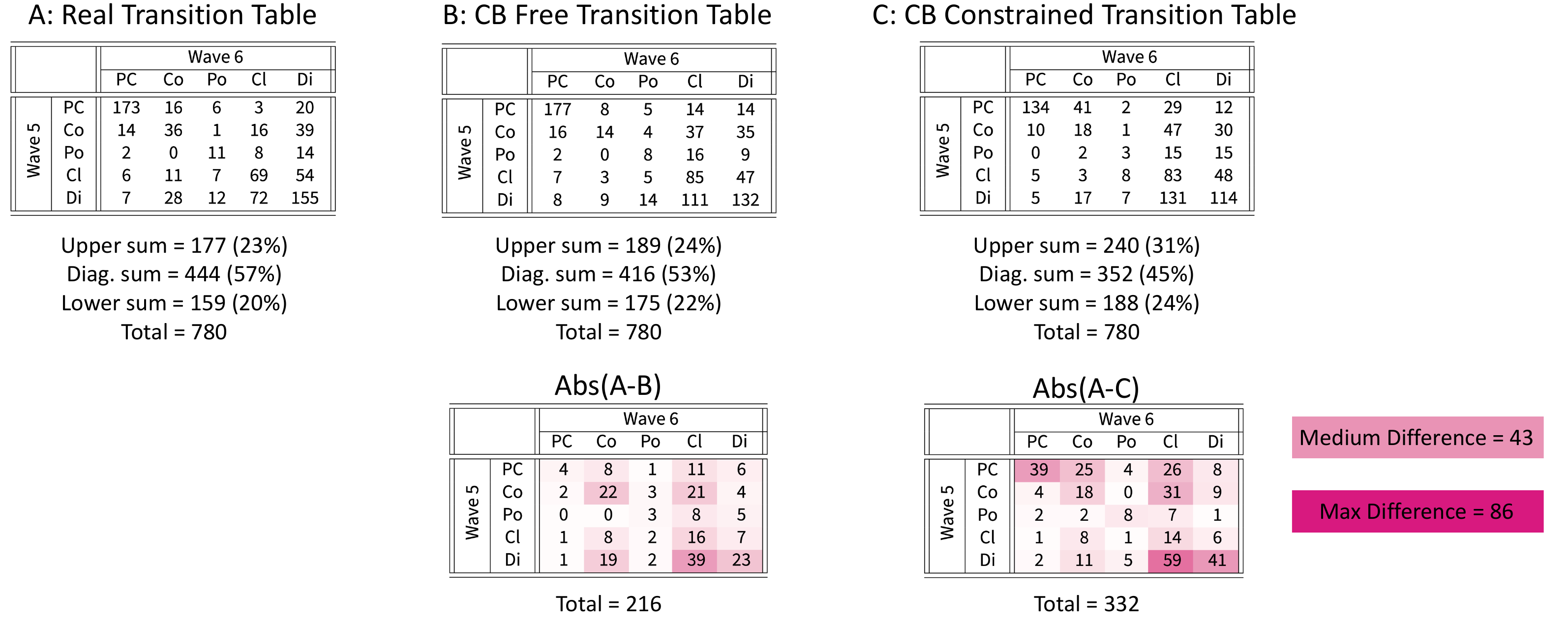}
\caption{ \CAbnew{Real transition table and transition tables produced by the CB model using the digraphs and inner traits assignations given by the \emph{Free} and \emph{Constrained} optimisation problems, along with the sum and percentage of cells above / on / below the diagonal. The two tables below show the absolute value of the difference between predicted and real transition tables.}}
\label{Fig:TT_CB}
\end{figure}

  \CAbnew{Figure \ref{Fig:TT_CB} shows three transition tables, where the set $X_o$ represents the set of all World Values Survey answers to wave 5 for all questions and countries. For the real transition table A, the set $X_f$ represents all the corresponding survey answers in wave 6, which are the true final opinions. For the transition table B (respectively, C) the set $X_f$ contains all the corresponding predicted opinions produced by the CB model with the parameters obtained through the \emph{Free} (respectively, \emph{Constrained}) optimisation, namely, all the predicted final opinions $\tilde{y}$ used to compute Table \ref{Tab:ResultsCBFree} (respectively, Table \ref{Tab:ResultsCBRest}). }
 
 \CAbnew{Interestingly, with the suitable choice of parameters, the CB model is capable of evolving opinions from any initial category into any other final category, as it happens with real opinion distributions. As expected, better results are achieved with the \emph{Free} optimisation, in line with the results from Tables \ref{Tab:ResultsCBFree} and \ref{Tab:ResultsCBRest}; still, both the free and the constrained transition tables show the versatility of the CB model, which can yield all transitions between opinion distribution categories that are seen in real life.}
 
%
%



\subsection{Comparison with the Friedkin-Johnsen Model}

\CAe{The classification-based model can be seen as an extension of the Friedkin-Johnsen (FJ) model \cite{Friedkin1986,Friedkin1999},  \CAbnew{in the sense that} both models include in the agents' behaviour \CAbnew{inner traits described by tuples. In the FJ model, each agent $i$ is characterised by two parameters: \emph{susceptibility} $a_i\in[0,1]$, determining how strongly the agent is affected by its neighbours' opinions and forgets the initial opinion \cite{Friedkin1999}, and \emph{prejudice} $b_i = 1-a_i\in[0,1]$}. A value of $a_i = 1$ means that the agent has complete susceptibility to interpersonal influence (similar to complete conformism), while a value of $a_i=0$ means that the opinion remains the same for all times (similar to complete stubbornness). \CAnewcomment{When all agents are completely susceptible, the FJ model becomes the classic French-DeGroot (FG) model \cite{French1956,Harary1959,Harary1965,DeGroot1974}; when all agents are completely prejudgemental, the FJ model becomes the Null model (opinions do not change over time).}
\CAbnew{In the CB model, each agent $i$ is associated with three parameters: \emph{conformism} $\alpha_i\in[0,1]$, \emph{radicalism} $\beta_i\in[0,1]$, and \emph{stubbornness} $\gamma_i\in[0,1]$, such that $\alpha_i+\beta_i+\gamma_i = 1$ for all $i\in\Vertices$}. Therefore, a FJ model where all agents have a susceptibility of $a$ is similar to a CB model where all agents have inner traits $\alpha = a$ and $\gamma = 1-a$ (hence $\beta=0$).} \CAbnew{Still, the interpretations of stubbornness and prejudice are slightly different: prejudice in the FJ model means that agents tend to remain with their \emph{initial opinion}, while stubbornness in the CB model means that agents tend to remain with their \emph{current opinion}, which leads to the same outcome only when all the agents are completely stubborn.
Apart from the outlined similarity, the FJ and CB models are different: crucially, the FJ model is linear, while the CB model is highly non-linear, which severely limits the applicability of closed-form analysis tools.}

\begin{figure}[h]
      \centering
          \includegraphics[width=0.9\textwidth]{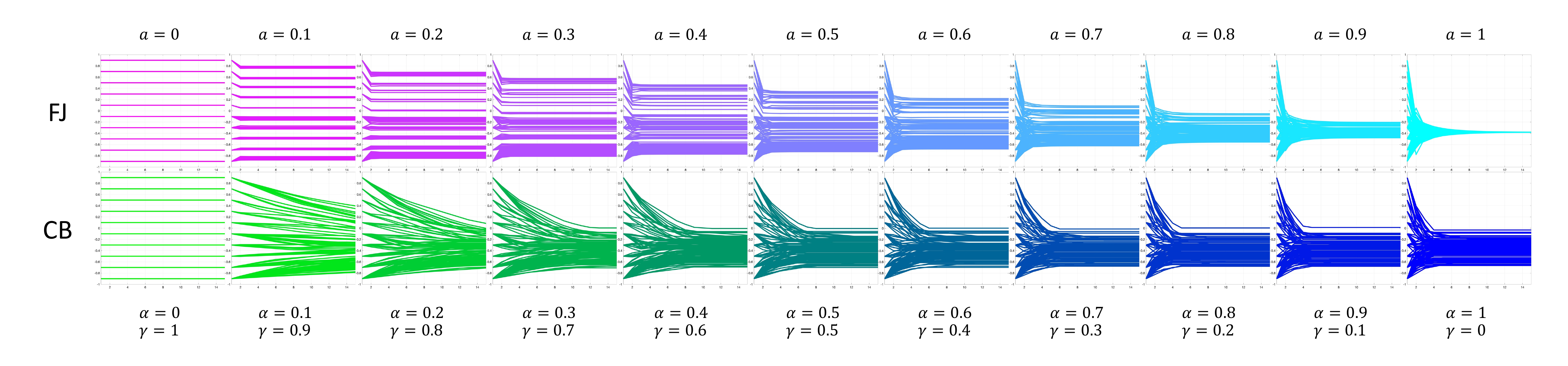}
         \caption{\CAe{Comparison between the Friedkin-Johnsen (FJ) model, for different values of susceptibility $a$, and the Classification-Based (CB) model, for corresponding values of conformist ($\alpha$) and stubborn ($\gamma$) weights. All the 100 agents have the same values of $a$ (FJ) and of $\alpha$ and $\gamma$ (CB). The simulations start from the same initial opinions and evolve over digraphs with the same topology. The degree of susceptibility, prejudice, conformism, and stubbornness is represented by the colours cyan, magenta, blue, and green respectively.}}
         \label{Fig:Comp_FJ}
 \end{figure}

\CAe{Figure \ref{Fig:Comp_FJ} shows the evolution of the same initial opinions according to the two models, for different values of $a$, $\alpha$, and $\gamma$. The digraphs had the same topology; randomly generated weights are considered for the FJ model, while for the CB model all the edge signs are taken positive to match the absence of antagonism in the FJ model.}
\CAe{The FJ model exhibits a slower change as $a$ increases, while with the CB model, as soon as conformism is introduced, the opinions converge to an interval where all the agents perceive that their neighbours' opinion is similar enough to theirs. \CAbnew{This difference is caused by the two different interpretations and implementations of stubbornness and prejudice}. Another important difference is that, as the susceptibility value increases, the final opinions of the FJ model tend to converge to a single opinion, and yield perfect consensus when $a=1$. Conversely, the final opinions of the CB model never converge to perfect consensus, even when $\alpha=1$ and $\gamma=0$, as a consequence of the classification-based approach.}

\CAc{The differences between CB and FJ model help visualise the strong implications of the classification-based mechanism for assessing the opinion of others, which captures the fact that opinions cannot be perceived with perfect resolution and accuracy, and hence changes the model behaviour significantly: it grants the model new properties, such as the existence of multiple equilibria that can span the complete spectrum of opinions. For instance, in Figure \ref{Fig:Comp_FJ}, the CB model with $\alpha = 1$ and $\gamma = 0$ generates equilibrium opinions that span almost $40\%$ of the opinion interval $[-1,1]$ (a wider span can be achieved with different topologies), while the FJ model with $a = 1$ leads to identical equilibrium opinions.
The non-linearity introduced by the classification-based assessment of the opinion of others can completely change the resulting dynamics and lead to the emergence of peculiar features, which would not emerge from models where the agents have perfect access to the opinion of others. This is highlighted, for instance, by the comparison with the Friedkin-Johnsen model and with the French-DeGroot model (corresponding to the FJ model with $a=1$).
The formal analysis of the distinctive mathematical properties of the CB model is the subject of ongoing research.}


\CAnewcomment{The results of an analysis for the FJ model, equivalent to the one reported in Tables \ref{Tab:ResultsCBFree} and \ref{Tab:ResultsCBRest} for the CB model, are reported in Tables \ref{Tab:ResultsFJFree} and \ref{Tab:ResultsFJRest}. To make the results comparable, when solving the optimisation problems \eqref{Eq:LargeOP} and \eqref{Eq:LargeOP_mod} for the FJ model the sets $\tilde{\SetA}$ and $\tilde{\SetN}$ are modified as follows: the set of digraphs is the same used for the French-DeGroot model, since both models require row-stochastic adjacency matrices;
the inner traits assignations in $\tilde{\SetA}$ are transformed into parameters of the FJ model using the mapping $a_i = \alpha_i/(\alpha_i + \gamma_i)$; if $\alpha_i + \gamma_i = 0$, then $a_i = 0.5$.}

\begin{table}[h!]
\centering
\resizebox{0.7\textwidth}{!}{%
\begin{tabular}{||c || c | c | c | c | c | c | c | c | c | c | c | c | c | c | c | c | c | c | c | c | c | c | c | c | c | c||} 
\hline\hline 
& C1 &  C2 &  C3 &  C4 &  C5 &  C6 &  C7 &  C8 &  C9 &  C10 &  C11 &  C12 &  C13 &  C14 &  C15 &  C16 &  C17 &  C18 &  C19 &  C20 &  C21 &  C22 &  C23 &  C24 &  C25 &  C26 \\ 
\hline\hline 
 Q1 &  \cellcolor{green!20} 1.8 &  \cellcolor{red!20} 8 &  \cellcolor{green!20} 2.4 &  \cellcolor{green!20} 1.2 &  \cellcolor{green!20} 1.8 &  \cellcolor{green!20} 2.2 &  \cellcolor{green!20} 2.4 &  \cellcolor{green!20} 6 &  \cellcolor{red!20} 7.2 &  \cellcolor{green!20} 6.2 &  \cellcolor{green!20} 2.6 &  \cellcolor{green!20} 1 &  \cellcolor{green!20} 1.4 &  \cellcolor{green!20} 4.8 &  \cellcolor{green!20} 5.4 &  \cellcolor{red!20} 9.4 &  \cellcolor{green!20} 4 &  \cellcolor{green!20} 2.6 &  \cellcolor{green!20} 1.6 &  \cellcolor{red!20} 10.6 &  \cellcolor{green!20} 4.2 &  \cellcolor{green!20} 1.8 &  \cellcolor{green!20} 1.2 &  \cellcolor{green!20} 5.6 &  \cellcolor{green!20} 2.4 &  \cellcolor{green!20} 5.4 \\ 
 Q2 &  \cellcolor{green!20} 5 &  \cellcolor{green!20} 4.2 &  \cellcolor{green!20} 1 &  \cellcolor{green!20} 1.4 &  \cellcolor{red!20} 8.8 &  \cellcolor{red!20} 7.4 &  \cellcolor{green!20} 2.6 &  \cellcolor{red!20} 14 &  \cellcolor{green!20} 4 &  \cellcolor{green!20} 3.6 &  \cellcolor{green!20} 3.8 &  \cellcolor{green!20} 3.6 &  \cellcolor{green!20} 1 &  \cellcolor{red!20} 13.8 &  \cellcolor{green!20} 2.4 &  \cellcolor{red!20} 8.6 &  \cellcolor{red!20} 9 &  \cellcolor{green!20} 1.4 &  \cellcolor{green!20} 2.8 &  \cellcolor{green!20} 6.4 &  \cellcolor{green!20} 3.2 &  \cellcolor{green!20} 3 &  \cellcolor{green!20} 1.6 &  \cellcolor{green!20} 3 &  \cellcolor{green!20} 4 &  \cellcolor{green!20} 3.8 \\
\hline\hline 
\end{tabular}}
\caption{Results of the \emph{Free} optimisation problem using the Friedkin-Johnsen model. The average cost along all the countries is 7.4897. Out of 780 possible question-country pairs, 460 have a cost less than 7 (an accuracy of $59\%$ in total). The average cost of accepted question-country pairs is 3.3. }
\label{Tab:ResultsFJFree}
\end{table} 

\begin{table}[h!]
\centering
\resizebox{0.7\textwidth}{!}{%
\begin{tabular}{||c || c | c | c | c | c | c | c | c | c | c | c | c | c | c | c | c | c | c | c | c | c | c | c | c | c | c||} 
\hline\hline 
& C1 &  C2 &  C3 &  C4 &  C5 &  C6 &  C7 &  C8 &  C9 &  C10 &  C11 &  C12 &  C13 &  C14 &  C15 &  C16 &  C17 &  C18 &  C19 &  C20 &  C21 &  C22 &  C23 &  C24 &  C25 &  C26 \\ 
\hline\hline 
 Q1 &  \cellcolor{green!20} 2.6 &  \cellcolor{red!20} 8.2 &  \cellcolor{green!20} 6.6 &  \cellcolor{green!20} 3 &  \cellcolor{green!20} 3.6 &  \cellcolor{green!20} 3.6 &  \cellcolor{red!20} 7 &  \cellcolor{red!20} 9.2 &  \cellcolor{red!20} 9 &  \cellcolor{red!20} 7.6 &  \cellcolor{green!20} 3.2 &  \cellcolor{green!20} 1.6 &  \cellcolor{green!20} 2.8 &  \cellcolor{green!20} 6 &  \cellcolor{red!20} 7.4 &  \cellcolor{red!20} 10.8 &  \cellcolor{green!20} 4.6 &  \cellcolor{green!20} 4 &  \cellcolor{green!20} 2.8 &  \cellcolor{red!20} 11.8 &  \cellcolor{red!20} 9.4 &  \cellcolor{green!20} 2.2 &  \cellcolor{green!20} 5.2 &  \cellcolor{red!20} 7.6 &  \cellcolor{green!20} 4 &  \cellcolor{green!20} 5.6 \\ 
 Q2 &  \cellcolor{green!20} 6.2 &  \cellcolor{green!20} 4.8 &  \cellcolor{green!20} 3.2 &  \cellcolor{green!20} 5.6 &  \cellcolor{red!20} 10.4 &  \cellcolor{red!20} 10.2 &  \cellcolor{red!20} 7.2 &  \cellcolor{red!20} 17.6 &  \cellcolor{red!20} 7.6 &  \cellcolor{green!20} 4.4 &  \cellcolor{green!20} 5.2 &  \cellcolor{green!20} 4.6 &  \cellcolor{green!20} 3.2 &  \cellcolor{red!20} 16 &  \cellcolor{green!20} 4.2 &  \cellcolor{red!20} 11.4 &  \cellcolor{red!20} 10.8 &  \cellcolor{green!20} 2 &  \cellcolor{green!20} 4.6 &  \cellcolor{red!20} 7.2 &  \cellcolor{green!20} 4.4 &  \cellcolor{green!20} 5 &  \cellcolor{green!20} 4.8 &  \cellcolor{green!20} 3.2 &  \cellcolor{green!20} 4.6 &  \cellcolor{green!20} 5.2 \\
\hline\hline 
\end{tabular}}
\caption{Results of the \emph{Constrained} optimisation problem using the Friedkin-Johnsen model. The average cost along all the countries is 10.3918. Out of 780 possible question-country pairs, 330 have a cost less than 7 (an accuracy of $42\%$ in total). The average cost of accepted question-country pairs is 4.1. }
\label{Tab:ResultsFJRest}
\end{table}

\CAnewcomment{Comparing Tables \ref{Tab:ResultsCBFree} and \ref{Tab:ResultsFJFree} shows that the CB model outperforms the FJ model, yielding a 97\% accuracy in contrast to 59\%. Also, the average cost of accepted country-question pairs is lower for the CB model (2.97) compared with the one produced by the FJ model (3.3), indicating that not only more question-country pairs are predicted satisfactorily, but also the predictions are more accurate.}

Since the French-DeGroot (FG) and the Null model can be seen as extreme cases of the FJ model, as expected, the FJ model produces better results than the FG model and the Null model, as can be seen by comparing Tables \ref{Tab:ResultsFJFree} and \ref{Tab:ResultsFJRest} with Tables \ref{Tab:ResultsFG} and \ref{Tab:ResultsNull} (note that the optimisation was done for the total cost, not the number of accepted question-country pairs). The fact that the Null model yields better predictions than the FG model suggests that opinions do not change much from one wave to the other, an observation that is also confirmed by the transition tables in Figure \ref{Fig:TT_CB}, where the sum (and percentage) of the values on the diagonal cells (associated with cases where initial and final opinion distributions both belong to the same qualitative category) is always the largest.

\begin{figure}[!h]
\centering
\includegraphics[width=\textwidth]{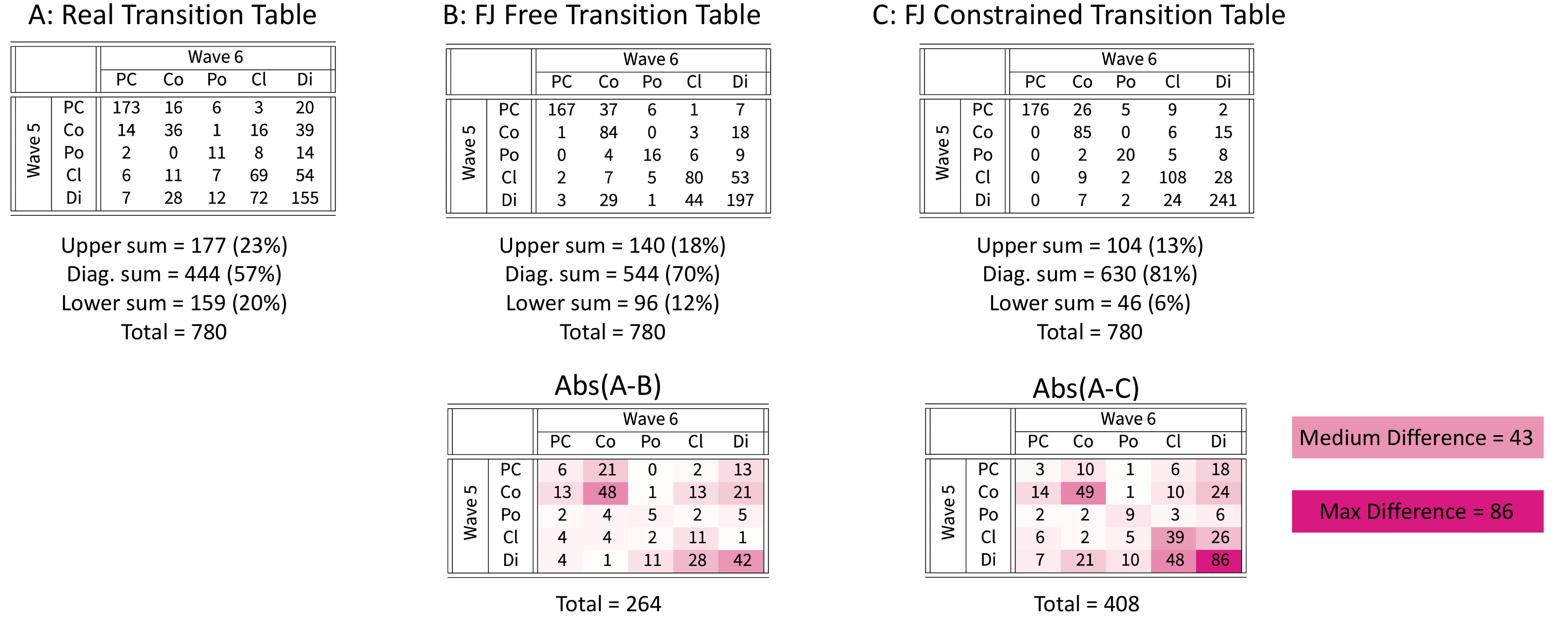}
\caption{ \CAbnew{Real transition table and transition tables produced by the FJ model using the digraphs and inner traits assignations from the \emph{Free} and \emph{Constrained} optimisation problems. The sum and percentage of cells above, on, and below the diagonal is shown. The two bottom tables correspond to the absolute value of the difference between the predicted transition tables and the real one. }}
\label{Fig:TT_FJ}
\end{figure}

\CAnewcomment{Figure \ref{Fig:TT_FJ} shows the transition tables for the FJ model (computed as for the CB model in Figure \ref{Fig:TT_CB}).} 
\CAnewcomment{In the CB transition tables, the diagonal sums ($416$ and $352$) are always smaller than the real diagonal sum ($444$), while in the FJ transition tables the diagonal sums ($544$ and $630$) are always larger, probably due to prejudice having a stronger effect than stubbornness in preserving the initial opinion. The lack of a ``radical'' trait makes the FJ model less versatile, which is reflected in the lower off-diagonal sum. According to Tables \ref{Tab:ResultsCBFree}, \ref{Tab:ResultsCBRest}, \ref{Tab:ResultsFJFree} and \ref{Tab:ResultsFJRest} the models that mimic the real opinion evolution at best are (in order): \emph{Free} CB, \emph{Free} FJ, \emph{Constrained} FJ, and \emph{Constrained} CB. Instead, according to the qualitative evaluation emerging from the transition tables, the models that are more faithful to the real opinion evolution are: \emph{Free} CB, \emph{Free} FJ, \emph{Constrained} CB, and \emph{Constrained} FJ.}





%



\subsection{ Model Outcome Capabilities}

 \CAbnew{Through the \emph{agreement plot}, we explore the model's capacity to produce a wide range of predicted opinions and gain insight into the dynamics that allow these predictions.} 

\begin{figure}[!h]
\centering
\includegraphics[width=\textwidth]{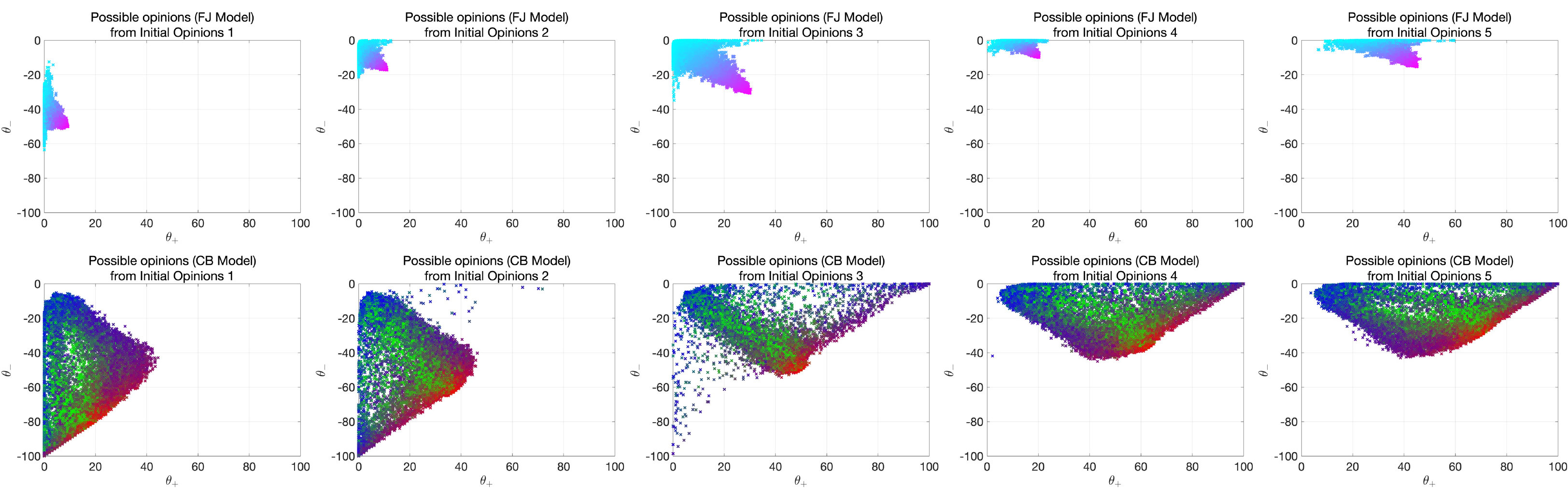}
\caption{ \CAbnew{Row 1 (respectively, row 2) shows the agreement plot of the potential opinion vectors predicted by the FJ (respectively, CB)  model, starting from the initial opinion vectors shown in Figure \ref{Fig:AgPlotExpl}. The marker colour encodes the average traits of the agents producing the final opinion: for the FJ model, cyan and magenta represent susceptibility and prejudice respectively; for the CB model, blue, red and green represent conformism, radicalism and stubbornness respectively. All simulations evolved over 50 time steps. 
}}
\label{Fig:AgrPlots}
\end{figure}

 \CAbnew{Given a single initial opinion vector $x_o$ and a set of inner traits assignations $\SetAA$ and networks $\SetNN$, let $X = X(x_o, \SetAA, \SetNN)$ be the set of all final opinion vectors produced by evolving the initial opinions $x_o$ with inner traits assignation $\innertraits\in\SetAA$ over a network $\Weights\in\SetNN$. The agreement plot of set $X(x_o, \SetAA, \SetNN)$ gives a visual representation of the range of opinions that the model is able to produce starting from $x_o$. Figure \ref{Fig:AgrPlots} shows the agreement plot for five different initial opinion vectors achieved by two models: the classification-based (CB) model and the Friedkin-Johnsen (FJ) model. In the agreement plot, each dot is colour-coded: for the CB model, the colour represents the average inner traits (blue for average conformist trait $\overline{\alpha}$, red for average radical trait $\overline{\beta}$, and green for average stubborn trait $\overline{\gamma}$); for the FJ model, it represents the average susceptibility $\overline{a}$ (cyan represents complete susceptibility $\overline{a} = 1$ and magenta complete stubbornness $\overline{a} = 0$). We consider the same set of traits as for the optimisation problem, i.e., $\SetAA=\tilde{\SetA}$, and a set of networks $\SetNN$ including networks 1 to 5 in $\tilde{\Weights}$, selected because they represent networks with the same topology and varying ratio of negative to positive edges.}

 \CAbnew{Figure \ref{Fig:AgrPlots} highlights that the CB model can produce a wide range of predicted opinions, primarily thanks to the interaction of the three complementary inner traits: conformism brings opinions together, yielding an effect that is similar to that of the FJ model, i.e., moving opinions along the $3\pi/4$ diagonal; radicalism allows the opinions to move in the other 3 diagonal directions ($\pi/4$, $5\pi/4$ and $7\pi/4$); and stubbornness makes opinions stay near the initial opinion. The combined effect of multiple combinations of these traits leads to an agreement plot of the possible predicted opinions that has a greater range than that produced by the FJ model.}

 \CAbnew{Susceptibility or conformism could be considered as a trait that moves all the opinions towards a common opinion in the middle of the interval $[-1,1]$, whereas radicalism moves all the opinions to their corresponding extremes, either $-1$ or $1$. The combined effect moves the opinions across the interval $[-1,1]$, and the degree of stubbornness determines the speed of the change. Interestingly, if the initial opinion is above the line $\theta_+ = -\theta_-$, then the predicted opinions also tend to stay above the line: if the overall population agrees more than it disagrees with a statement, this opinion balance is likely to be preserved. However, a change in opinion balance is not impossible: a considerable fraction of predicted opinions may approach the line $\theta_+ = -\theta_-$, and from there a slight change in population traits may move the predicted opinions to the other side of the line (initial opinion 2 can produce an opinion vector near initial opinion 1 and from there the next opinions can be near initial opinion 3). It is also interesting to note that, although initial opinions 1 are slightly below the $\theta_+ = -\theta_-$ line, most of the predicted opinions are above that line, an effect probably caused by the network topology or the edge signs.}

\section{Summary and Conclusions}\label{Sec:Con}

We have proposed a novel agent-based opinion formation model that has two fundamental distinctive features. First, the model drops the unrealistic assumption that agents can measure the opinion of their neighbours with infinite precision, which drastically affects the opinion evolution, and introduces a novel classification-based approach that more realistically replicates the way individuals assess and evaluate the opinions of their neighbours, by classifying them as agreeing much less, less, comparably, more or much more.
Second, the model captures the complexity of the behaviour of individuals by introducing three different internal traits, associated with conformism, radicalism, and stubbornness. Instead of considering agents of different types, the model allows all these tendencies to coexist in each agent, thus representing multifaceted psychological and sociological phenomena in action within each individual.

Five types of simulation analyses were carried out: ($i$) simulations over simple digraphs and agent parameters to gain insight into the model behaviour; ($ii$) simulations with varying model parameters to perform a parameter sensitivity analysis; ($iii$) simulations with parameters chosen through the approximate solution of two optimization problems to assess the model's potential to predict opinions similar to those seen in real life; ($iv$) comparison with the Friedkin-Johnsen model; and ($v$) exploration of the model's capability to generate different opinion predictions.

We used real data from the World Values Survey to assess the capability of our classification-based model to  \CAbnew{mimic actual opinion evolutions} seen in real life. 
Despite its simplicity, the model  \CAbnew{can yield opinions similar to the ones in survey results and can also produce a rich and wide variety of collective behaviours, without the need of introducing bounded confidence, randomness, or more complex mechanics.}

Possible further directions for future work include a more detailed study of the effects of different network topologies on the opinion evolution, and an investigation of what happens if the opinion evolution parameters $\Omega$ are agent-dependent. 

 \section{  Appendix A: Network Metrics}

   The signed digraph is represented by the weight matrix $\Weights\in\{-1,0,1\}^{\numag\times\numag}$, where $w_{ij}$ is associated with the edge going from vertex $j$ to vertex $i$. We consider six network metrics: average path length (APL), clustering coefficient (CC), positive edges (PE), negative edges (NE), diameter (D), and balance index (BI). This appendix explains how these metrics are computed.
  
A directed path is a $K$-tuple of vertices $(p_1, p_2, \dots, p_i, p_{i+1}, \dots , p_K)$ such that there is an edge from vertex $p_i$ to vertex $p_{i+1}$ for $i=1, \dots, K-1$. The length $|p|$ of a directed path $p$ is the number of edges that it crosses. Let $P(i,j)$ be the set of all directed paths from vertex $i$ to vertex $j$ (if there are none, then $P(i,j) = \emptyset$). Denote by $d(i,j)$ the length of the shortest directed path from $i$ to $j$, i.e., $d(i,j)\coloneqq \min_{p\in P(i,j)}|p|$. Let $C(\Weights)$ be the set of vertex pairs $(i,j)$ such that there exists a direct path from $i$ to $j$ and $i \neq j$, i.e. $C(\Weights) = \{(i,j)\mid P(i,j) \neq \emptyset \text{ and } i \neq j \}$. Then the average path length and diameter of the digraph $\Weights$ are:
  
\begin{equation}
APL = \frac{1}{|C(\Weights)|}\sum_{(i,j)\in C(\Weights)}d(i,j) \qquad\mbox{and}\qquad D = \max_{(i,j)\in C(\Weights)}d(i,j)
\end{equation}

Note that, because all the networks are strongly connected, $|C(\Weights)| = \numag(\numag-1)$.

To compute the clustering coefficient, consider agent $i$, with $k_i$ in-neighbours excluding itself: $k_i = |\tilde{\Neig_i}|$, where $\tilde{\Neig_i} \coloneqq \{ j\in\Vertices \mid w_{ij}\neq0, i \neq j\}$. Then there are at most $k_i(k_i-1)$ directed edges between these neighbours. The fraction $c_i$ of these edges that is actually present is the clustering coefficient of agent $i$. If agent $i$ has only one in-neighbour, then its clustering coefficient is 1, and if it has no in-neighbour but itself $c_i$ is not defined:

\begin{equation}
c_i = 
\begin{cases}
\frac{|\{ (j,k) \mid j \neq k \text{ and } i,k\in\tilde{\Neig_i} \}|}{k_i(k_i-1)} & \text{ if $k_i>1$} \\
1 & \text{ if $k_i=1$} \\ 
nan & \text{ if $k_i=0$} \\ 
\end{cases}
\end{equation}

The clustering coefficient of the network (defined by extending to digraphs the definition for undirected graphs by \cite{Watts1998Collective}) is thus the average of the clustering coefficients of all agents with at least one in-neighbour excluding themselves:

\begin{equation}
CC = \frac{1}{|\{i\in\Vertices \mid k_i > 0\}|} \sum_{i:k_i > 0 }{c_i} 
\end{equation}

The number of positive and negative edges are computed as

\begin{equation}
PE = \sum_{i,j\in\Vertices: w_{ij}>0}1 \qquad  \mbox{and} \qquad 	NE = \sum_{i,j\in\Vertices: w_{ij}<0}1		
\end{equation}

Finally, the balance index is computed as

\begin{equation}
BI = \frac{tr(exp(\Weights))}{tr(exp(D))}
\end{equation}

where $tr(\cdot)$ is the trace operator, $exp(\cdot)$ is the matrix exponential, and $D=|\Weights|$ component-wise. This formula is a direct extension of the balance index for undirected graphs proposed by \cite{Estrada2019Rethinking,Estrada2014Walk}.


 \section{ Appendix B: Simulation Process}
 
The \textbf{free optimisation problem} in Equation \eqref{Eq:LargeOP} with sets $\SetN = \tilde{\SetN}$ and $\SetA = \tilde{\SetA}$ was solved using the algorithm:

\begin{enumerate} 
\setlength\itemsep{-0.4em}
\item \textbf{Input:} survey answers for waves 5 and 6 for a given country.
\item Set $w_0 = \infty$; this will be the minimum cost across all networks
\item For network $\Weights\in\tilde{\SetN}$
	\begin{itemize} 
	\item For question $q\in\{1, 2, \dots, 30\}$
		\begin{itemize} 
		\item Set $v_q = \infty$ to be the minimum cost for question $q$
		\item For inner traits assignation $\innertraits^{(\newj)}\in\tilde{\SetA}$
			\begin{itemize} 
			\item Compute the predicted opinions $\tilde{y}_q$ after $K$ iterations evolving over the network $\Weights$ with inner traits assignation $\innertraits^{(\newj)}$ starting with initial opinions $x_q$. These initial opinions are the survey results to question $q$ in wave 5.

			\begin{equation*}
			\tilde{y}_q = \FancyF_\Omega(x_q, \Weights, \innertraits^{(\newj)}, K)
			\end{equation*}

			\item Compute the mismatch $\Cost$ (Equation \eqref{Eq:CostFunction}) between these predicted opinions $\tilde{y}_\newj$ and the real opinions $y_\newj$ given by survey results of question $q$ in wave 6.

			\item if $\Cost(\tilde{y}_q, y_q)<v_q$
				\begin{itemize} 
				\item Set $v_q = \Cost(\tilde{y}_q, y_q)$ as the current minimum cost across all inner traits assignations.
				\item Set $\widehat{\innertraits^{(q)}} = \innertraits^{(\newj)}$ as the inner traits assignation that gives the lowest cost for question $q$.
				\end{itemize} 

			\end{itemize} 
		\item Add all the minimum costs to obtain the minimum cost for the network $\Weights$
		
		\begin{equation*}
		\Cost_\text{Total} = \sum_{q=1}^{30} v_q
		\end{equation*}
		
		\end{itemize} 
		\item if $\Cost_\text{Total}<w_0$
			\begin{itemize} 
			\item Set $w_0 = \Cost_\text{Total}$ as the current minimum cost across all networks for this country.
			\item Set $\widehat{\Weights} = \Weights$ as the network that produces the minimum cost for this country.
			\end{itemize} 
	\end{itemize} 
\item \textbf{Output:} network $\widehat{\Weights}$ and set of inner traits assignations $(\widehat{\innertraits^{(\newj)}})_{\newj=1}^{30}$ that give the minimum total cost across all questions.
\end{enumerate} 

In the algorithm used to solve the \textbf{constrained optimisation problem} in Equation \eqref{Eq:LargeOP_mod}, both the network and the inner traits assignations are the same for each question:

\begin{enumerate} 
\setlength\itemsep{-0.4em}
\item \textbf{Input:} survey answers for waves 5 and 6 for a given country.
\item Set $w_0 = \infty$; this will be the minimum cost across all networks and inner traits assignations
\item For network $\Weights\in\tilde{\SetN}$
	\begin{itemize} 
	\item For inner traits assignation $\innertraits\in\tilde{\SetA}$
			\begin{itemize} 
			\item For question $q\in\{1, 2, \dots, 30\}$
				\begin{itemize} 
				\item Compute the predicted opinions $\tilde{y}_q$ after $K$ iterations evolving over the network $\Weights$ with inner traits assignation $\innertraits$ starting with initial opinions $x_q$. These initial opinions are the survey results to question $q$ in wave 5.

				\begin{equation*}
				\tilde{y}_q = \FancyF_\Omega(x_q, \Weights, \innertraits, K)
				\end{equation*}

				\item Compute the mismatch $\Cost$ (Equation \eqref{Eq:CostFunction}) between the predicted opinions $\tilde{y}_q$ and the real opinions $y_q$ given by survey results of question $q$ in wave 6.

				\end{itemize} 
			\item Add all the costs to obtain the cost for the network $\Weights$ and the inner traits assignation $\innertraits$.
		
			\begin{equation*}
			\Cost_\text{Total} = \sum_{q=1}^{30} v_q
			\end{equation*}
		
			\item if $\Cost_\text{Total}<w_0$
				\begin{itemize} 
				\item Set $w_0 = \Cost_\text{Total}$ as the current minimum cost across all networks and inner traits assignations for this country.
				\item Set $\widehat{\Weights} = \Weights$ as the network that produces the minimum cost for this country.
				\item Set $\hat{\innertraits} = \innertraits$ as the inner traits assignation that gives the minimum cost for this country.
				\end{itemize} 
		
		\end{itemize} 

	\end{itemize} 
\item \textbf{Output:} network $\widehat{\Weights}$ and inner traits assignations $\widehat{\innertraits}$ that give the minimum total cost across all questions.
\end{enumerate} 

The data sets used to produce the results shown in the paper can be downloaded from the following link: https://giuliagiordano.dii.unitn.it/docs/papers/OpinionModel.zip, together with the corresponding code and instructions on how to use the code.

 \section{Appendix C: Countries and Questions}
\label{App:D}

We report here the list of countries and the list of questions we considered, from the real data collected by the World Values Survey.

\begin{table}[!h]
	\centering
	\begin{tabular}{|| p{0.2in} | p{1.5in} || p{0.2in} | p{1.5in} ||  p{0.2in} | p{1.5in} || }
	\hline
  C1 & Australia &  C2 & Brazil&  C3 & Chile\\
	\hline    
  C4 & China &  C5 & Cyprus&  C6 & Georgia\\
	\hline    
  C7 & Ghana &  C8 & India&  C9 & Jordan\\
	\hline    
  C10 & Japan &  C11 & Malaysia&  C12 & Mexico\\
	\hline    
  C13 & Poland &  C14 & Romania&  C15 & Slovenia\\
	\hline    
  C16 & South Africa &  C17 & Spain&  C18 & Sweden\\
	\hline    
  C19 & South Korea &  C20 & Thailand&  C21 & Trinidad\\
	\hline    
  C22 & Taiwan &  C23 & Turkey&  C24 & Ukraine\\
	\hline    
  C25 & United States &  C26 & Uruguay& \multicolumn{2}{ c ||}{}  \\
	\hline    
	\end{tabular}
	\caption{Countries}
	\label{tab:countries}	
\end{table}

\begin{table}[!h]
	\centering
   \resizebox{\textwidth}{!}{%
	\begin{tabular}{| p{0.2in} | p{8in} | }
	\toprule
 Q1 &  Some people feel they have completely free choice and control over their lives, while other people feel that what they do has no real effect on what happens to them. Please use this scale where 1 means \textbf{no choice at all} and 10 means \textbf{a great deal of choice} to indicate how much freedom of choice and control you feel you have over the way your life turns out \\
	\hline		
 Q2 &  All things considered, how satisfied are you with your life as a whole these days? Using this card on which 1 means you are \textbf{completely dissatisfied} and 10 means you are \textbf{completely satisfied} where would you put your satisfaction with your life as a whole? \\
	\hline		
 Q3 &  How satisfied are you with the financial situation of your household? Please use this card again to help with your answer (1 is completely dissatisfied, 10 is completely satisfied) \\
	\hline		
 Q4 &  How would you place your views on this scale? 1 means you completely agree with the statement \textbf{Incomes should be made more equal}; 10 means you completely agree with the statement \textbf{We need larger income differences as incentives for individual effort}. And if your views fall somewhere in between, you can choose any number in between. \\
	\hline		
 Q5 &  How would you place your views on this scale? 1 means you completely agree with the statement \textbf{Private ownership of business and industry should be increased}; 10 means you completely agree with the statement \textbf{Government ownership of business and industry should be increased}. And if your views fall somewhere in between, you can choose any number in between. \\
	\hline		
 Q6 &  How would you place your views on this scale? 1 means you completely agree with the statement \textbf{The government should take more responsibility to ensure that everyone is provided for}; 10 means you completely agree with the statement \textbf{People should take more responsibility to provide for themselves}. And if your views fall somewhere in between, you can choose any number in between. \\
	\hline		
 Q7 &  How would you place your views on this scale? 1 means you completely agree with the statement \textbf{Competition is good. It stimulates people to work hard and develop new ideas}; 10 means you completely agree with the statement \textbf{Competition is harmful. It brings out the worst in people}. And if your views fall somewhere in between, you can choose any number in between. \\
	\hline		
 Q8 &  How would you place your views on this scale? 1 means you completely agree with the statement \textbf{In the long run, hard work usually brings a better life}; 10 means you completely agree with the statement \textbf{Hard work doesn’t generally bring success—it’s more a matter of luck and connections}. And if your views fall somewhere in between, you can choose any number in between. \\
	\hline		
 Q9 &  How much you agree or disagree with the statement \textbf{Science and technology are making our lives healthier, easier, and more comfortable.}. For this questions, a 1 means that you “completely disagree” and a 10 means that you “completely agree.” \\
	\hline		
 Q10 &  How much you agree or disagree with the statement \textbf{Because of science and technology, there will be more opportunities for the next generation.}. For this questions, a 1 means that you "completely disagree” and a 10 means that you “completely agree.” \\
	\hline		
 Q11 &  How much you agree or disagree with the statement \textbf{We depend too much on science and not enough on faith.}. For this questions, a 1 means that you "completely disagree” and a 10 means that you “completely agree.” \\
	\hline		
 Q12 &  All things considered, would you say that the world is better off, or worse off, because of science and technology? 1 means that "the world is a lot worse off,” and 10 means that “the world is a lot better off.” \\
	\hline		
 Q13 &  How important is God in your life? 10 means "very important” and 1 means “not at all important.” \\
	\hline		
 Q14 &  Indicate if the action of \textbf{Claiming government benefits to which you are not entitled} can be never justified (1); always justified (10); or something in between in a scale from 1 to 10. \\
	\hline		
 Q15 &  Indicate if the action of \textbf{Cheating on taxes if you have a chance} can be never justified (1); always justified (10); or something in between in a scale from 1 to 10. \\
	\hline		
 Q16 &  Indicate if the action of \textbf{Someone accepting a bribe in the course of their duties} can be never justified (1); always justified (10); or something in between in a scale from 1 to 10. \\
	\hline		
 Q17 &  Indicate if the action of \textbf{Homosexuality} can be never justified (1); always justified (10); or something in between in a scale from 1 to 10. \\
	\hline		
 Q18 &  Indicate if the action of \textbf{Abortion} can be never justified (1); always justified (10); or something in between in a scale from 1 to 10. \\
	\hline		
 Q19 &  Indicate if the action of \textbf{Divorce} can be never justified (1); always justified (10); or something in between in a scale from 1 to 10. \\
	\hline		
 Q20 &  Indicate if the action of \textbf{Suicide} can be never justified (1); always justified (10); or something in between in a scale from 1 to 10. \\
	\hline		
 Q21 &  Indicate if the action of \textbf{For a man to beat his wife} can be never justified (1); always justified (10); or something in between in a scale from 1 to 10. \\
	\hline		
 Q22 &  \textbf{Governments tax the rich and subsidize the poor.} an essential characteristic of democracy? Use this scale where 1 means "not at all an essential characteristic of democracy” and 10 means it definitely is “an essential characteristic of democracy” \\
	\hline		
 Q23 &  \textbf{Religious authorities interpret the laws.} an essential characteristic of democracy? Use this scale where 1 means "not at all an essential characteristic of democracy” and 10 means it definitely is “an essential characteristic of democracy” \\
	\hline		
 Q24 &  Is \textbf{People choose their leaders in free elections.} an essential characteristic of democracy? Use this scale where 1 means "not at all an essential characteristic of democracy” and 10 means it definitely is “an essential characteristic of democracy” \\
	\hline		
 Q25 &  Is \textbf{People receive state aid for unemployment.} an essential characteristic of democracy? Use this scale where 1 means "not at all an essential characteristic of democracy” and 10 means it definitely is “an essential characteristic of democracy” \\
	\hline		
 Q26 &  Is \textbf{The army takes over when government is incompetent.} an essential characteristic of democracy? Use this scale where 1 means "not at all an essential characteristic of democracy” and 10 means it definitely is “an essential characteristic of democracy” \\
	\hline		
 Q27 &  Is \textbf{Civil rights protect people7s liberty against oppression.} an essential characteristic of democracy? Use this scale where 1 means “not at all an essential characteristic of democracy” and 10 means it definitely is “an essential characteristic of democracy” \\
	\hline		
 Q28 &  Is \textbf{Women have the same rights as men.} an essential characteristic of democracy? Use this scale where 1 means "not at all an essential characteristic of democracy” and 10 means it definitely is “an essential characteristic of democracy” \\
	\hline		
 Q29 &  How important is it for you to live in a country that is governed democratically? On this scale where 1 means it is "not at all important” and 10 means “absolutely important" what position would you choose? \\
	\hline		
 Q30 &  And how democratically is this country being governed today? Again using a scale from 1 to 10, where 1 means that it is "not at all democratic" and 10 means that it is "completely democratic,” what position would you choose? \\
	\hline		
	\bottomrule		
	\end{tabular}}
	\caption{Questions}
	\label{tab:questions}	
\end{table}

\newpage

\bibliography{Model_Paper.bib} 
\bibliographystyle{ieeetr}

\end{document}